\documentclass[11pt,a4paper]{article}
\usepackage{jheppub}

\newcommand{\Tr}{\mathop{\rm Tr}}

\newcommand{\Group}[2]{{ \hbox{{\itshape{#1}}($#2$)} }}
\newcommand{\U}[1]{\Group{U\kern0.05em}{#1}}
\newcommand{\SU}[1]{\Group{SU\kern0.1em}{#1}}
\newcommand{\SL}[1]{\Group{SL\kern0.05em}{#1}}
\newcommand{\Sp}[1]{\Group{Sp\kern0.05em}{#1}}
\newcommand{\SO}[1]{\Group{SO\kern0.1em}{#1}}

\newcommand{\mybar}[1]%
    {{\kern 0.8pt\overline{\kern -0.8pt#1\kern -0.8pt}\kern 0.8pt}}
\newcommand{\sla}[1]%
    {{\raise.15ex\hbox{$/$}\kern-.57em #1}}
\newcommand{\roughly}[1]%
    {{ \mathrel{\raise.3ex\hbox{ $#1$\kern-.75em\lower1ex\hbox{$\sim$}} } }}

\newcommand{\nop}[1]{:\kern-.3em#1\kern-.3em:}

\newcommand{\del}{\partial}

\newcommand{\delfb}{\overleftrightarrow{\partial}}
\newcommand{\al}{\ensuremath{\alpha}}
\newcommand{\be}{\ensuremath{\beta}}

\newcommand{\de}{\ensuremath{\delta}}

\newcommand{\ep}{\ensuremath{\epsilon}}

\renewcommand{\th}{\ensuremath{\theta}}

\newcommand{\la}{\ensuremath{\lambda}}
\newcommand{\La}{\ensuremath{\Lambda}}
\newcommand{\rh}{\ensuremath{\rho}}
\newcommand{\si}{\ensuremath{\sigma}}

\newcommand{\Ph}{\ensuremath{\Phi}}

\newcommand{\GeV}{ \ensuremath{\mathrm{~GeV}} }

\newcommand{\dof}{DOF}
\newcommand{\nlsm}{\ensuremath{\mathrm{NL}\Sigma \mathrm{M}}}
\newcommand{\hc}{\mathrm{H.c.}} 
\newcommand{\n}{\notag \\}
\newcommand{\mcl}[1]{\mathcal{#1}}
\newcommand{\fb}[1]{\overleftrightarrow{#1}}
\newcommand{\Z}{\phantom{-1}}
\title{{\boldmath 
Structure of dimension-six derivative interactions 
in pseudo Nambu-Goldstone $N$ Higgs doublet models
} }

\author{Yohei Kikuta,}
\author{Yasuhiro Okada}
\author{and Yasuhiro Yamamoto}
\affiliation{KEK Theory Center, 
Institute of Particle and Nuclear Studies, KEK, 1-1 Oho,}
\affiliation{Tsukuba, Ibaraki 305-0801, Japan}
\affiliation{Department of Particle and Nuclear Physics, 
Graduate University for Advanced Studies}
\affiliation{(Sokendai), 1-1 Oho, Tsukuba, Ibaraki 305-0801, Japan}
\emailAdd{kikuta@post.kek.jp}
\emailAdd{yasuhiro.okada@kek.jp}
\emailAdd{yamayasu@post.kek.jp}
\abstract{ 
We derive the general structure of dimension-six derivative interactions 
in the $N$ Higgs doublet models, where Higgs fields arise as 
pseudo Nambu-Goldstone modes of a strongly interacting sector.
We show that 
there are several relations among the dimension-six operators, 
and therefore the number of independent operators 
decreases compared with models 
on which only $SU(2)_L \times U(1)_Y$ invariance is imposed.
As an explicit example, we derive scattering amplitudes 
and cross sections of 
longitudinal gauge bosons and 
Higgs bosons at high energy 
on models involving two Higgs doublets, 
and compare them with 
those of one Higgs doublet.
}


\subheader{KEK-TH-1505}

\begin{document}
\maketitle
\section{Introduction} 
Recent progress in the Higgs boson searches at the Tevatron and 
the Large Hadron Collider (LHC) are remarkable~\cite{Nisati:2011}, 
and we are likely to observe the Higgs boson soon 
if it exists in the mass range favored in the standard model (SM). 
When the Higgs boson is discovered, 
precise study of its properties is an important step to understand 
the dynamics behind the electroweak symmetry breaking (EWSB). 
In many models, a SM-like Higgs boson may be observed at first, 
even if the Higgs sector has complicated structure.
For example, there may be a new symmetry principle like supersymmetry or 
a new strongly interacting sector like little Higgs models. 

The strongly-interacting light Higgs (SILH) model was proposed as 
a framework describing an effective theory of one Higgs doublet models with 
a light physical Higgs boson based on 
a new strongly interacting sector~\cite{Giudice:2007fh}.
In this model, the Higgs doublet is identified as 
a composite field corresponding to pseudo Nambu-Goldstone bosons (PNGBs) of
a global symmetry breaking at some high energy scale. 
The model introduces two new scales, $f$ and $m_\rh =g_\rh f$, 
where $f$ is the decay constant describing the global symmetry breaking and 
$m_\rh$ is the mass scale of new resonances. 
The new coupling constant $g_\rh$ is considered to be in the range of 
$g_{SM} \lesssim g_\rh \lesssim 4\pi$, where
$g_{SM}$ indicates the weak gauge coupling 
or the top Yukawa coupling.
Explicit examples of this kind of structure can be found in 
little Higgs models~\cite{ArkaniHamed:2001nc}
and models with large extra 
dimension~\cite{Contino:2003ve}. 
In Ref.~\cite{Giudice:2007fh}, 
the general form of the effective Lagrangian was introduced, 
and its phenomenological implications were discussed. 
They studied high energy behavior of the scattering amplitudes 
for longitudinal modes of massive gauge bosons and the Higgs boson.
It was shown that the longitudinal gauge bosons and 
the physical Higgs boson production cross sections 
at the LHC satisfy a simple relation at high energy 
because these quantities are determined 
by the same dimension-six derivative coupling of the effective Lagrangian. 
This formulation was further investigated in details in 
Ref.~\cite{Low:2009di}.

In this paper, 
we generalize the SILH model to the model including $N$ Higgs doublets. 
It is found that the number of independent dimension-six  
derivative interactions is strongly constrained by 
requiring that Higgs fields are generated as PNGBs 
of some global symmetry breaking. 
We derive the scattering amplitudes and cross sections of 
the longitudinal gauge bosons and the Higgs bosons at high energy 
for the two Higgs doublet model (2HDM). 
These scattering amplitudes are controlled 
by the dimension-six derivative interactions, 
and therefore study of these cross sections 
at the LHC and future Linear Colliders (LC)
is important to identify this model experimentally.

This paper is organized as follows:
In Sec.~\ref{sec:single}, we review the SILH model.
In Sec.~\ref{sec:multi}, we discuss the extension of the framework to 
the $N$ Higgs doublet model (NHDM).
Then we study phenomenological features of 
the 2HDM in Sec.~\ref{sec:phenomenology}.
Section \ref{sec:conclusion} is conclusion of our results.

\section{A brief review of the strongly interacting light Higgs model} 
\label{sec:single}

We briefly review the SILH model investigated 
in Ref.~\cite{Giudice:2007fh,Low:2009di}.

In order to study scatterings 
of the longitudinal modes and the Higgs boson,
we focus on dimension-six derivative interactions.
The Lagrangian of the derivative interactions 
invariant under the $SU(2)_L \times U(1)_Y$ symmetry is given by
\begin{align} \label{eq:four_independent}
 {\cal L}^6 =& 
   \frac{\la^H}{2\La^2} \del_\mu (H^\dag H) \del^\mu (H^\dag H) 
  +\frac{\la^r}{ \La^2} H^\dag H (\del^\mu H^\dag )(\del_\mu H) \n &
  +\frac{\la^T}{2\La^2} (H^\dag \fb{\del}_\mu  H)(H^\dag \fb{\del}^\mu H) 
  +i\frac{\la^{HT}}{\La^2} \del_\mu (H^\dag H)(H^\dag \fb{\del}^\mu H) \\
 =&
   \frac{\la^H}{2\La^2} O^H 
  +\frac{\la^r}{ \La^2} O^r 
  +\frac{\la^T}{2\La^2} O^T 
  +i\frac{\la^{HT}}{\La^2} O^{HT} ,
  \label{eq:one_Higgs_der_int}
\end{align}
where $\la^H$, $\la^r$, $\la^T$ and $\la^{HT}$ 
are real coefficients,
$H$ is an $SU(2)_L$ doublet and
$H^\dag \fb{\del}_\mu H = H^\dag (\del_\mu H) -(\del_\mu H )^\dag H$. 
We have introduced $\La$ as an appropriate cutoff scale.
In this paper, the SM gauge symmetry, $SU(2)_L \times U(1)_Y$, is 
treated as if it is global symmetry
since we focus only scalar four point interactions with two derivatives.
Full gauge invariant Lagrangian can be recovered by replacing
a partial derivative with the covariant derivative.
We study the effects of these derivative interactions with the requirement that 
the Higgs doublet is considered 
as a pseudo Nambu-Goldstone (NG) field.

In composite Higgs theories,
the nonlinear realization~\cite{Coleman:1969sm}
generates the derivative interactions 
like Eq.~\eqref{eq:one_Higgs_der_int}.
For a nonlinear $\si$ model (\nlsm) where
a global symmetry is broken from $G$ to $H$,
the Lagrangian of NG fields is constructed with
fields parametrizing the degenerate vacua of 
the quotient manifold $G/H$:
\begin{align}
  \xi = e^{i \Pi(x) / f}, \quad \Pi(x) = \Pi^a (x) X^a,
\end{align}
where the fields $\Pi^a(x)$ represent the NG fields and $\{ X^{a} \}$ are 
generators of the broken symmetry $G/H$.
Generators of the unbroken symmetry are denoted by $\{ T^i \}$.
Commutation relations among these generators are 
\begin{align}
 [T^i , T^j ] = if^{ijk} T^k, \qquad 
 [T^i , X^a ] = if^{iab} X^b, \qquad 
 [X^a , X^b ] = if^{abi} T^i + if^{abc} X^c .
\label{eq:com_rel}
\end{align}
If the second term of the right hand side vanishes 
for the third commutator ($f^{abc} = 0$), 
$G/H$ is called symmetric space.
In order to construct the nonlinear realization, 
the Maurer-Cartan one form, $\al_\mu (\Pi)$, is a fundamental constituent:
\begin{align}
 \al_\mu (\Pi) =& 
  -i\xi^{-1} (\Pi)\, \del_\mu \xi (\Pi) \\
 =&
   \frac{1}{f} \del_\mu \Pi -\frac{i}{2f^2} [\Pi ,\del_\mu \Pi] 
  -\frac{1}{6f^3} \left[\Pi,[\Pi ,\del_\mu \Pi]\right] 
  +{\cal O} \left( ( \Pi/f )^4 \right) \\
 =&
   \al_{\perp \mu}^a (\Pi)\, X^a 
  +\al_{\parallel \mu}^i (\Pi)\, T^i \\
 =&
   \al_{\perp \mu} +\al_{\parallel \mu}.
\end{align}
Under a transformation $g\, (\in G)$, we obtain 
\begin{align}
 \al_{\perp \mu} (\Pi) \to \, &
   \al_{\perp \mu} (\Pi') 
 = 
   h(\Pi,g)\, \al_{\perp \mu} (\Pi)\, h^{-1}(\Pi,g), 
\label{eq:MC_perp} \\
 \al_{\parallel \mu} (\Pi) \to \, & 
   \al_{\parallel \mu} (\Pi') 
 = 
   h(\Pi,g)\, \al_{\perp \mu} (\Pi)\, h^{-1}(\Pi,g) 
  -ih(\Pi,g) \partial_{\mu}  h^{-1}(\Pi,g),
\end{align}
where $h \in H$.
Using Eqs.~\eqref{eq:com_rel} and~\eqref{eq:MC_perp}, 
it is straightforward to calculate $\al_{\perp \mu}$:
\begin{align}
 \al_{\perp \mu}(\Pi) =& 
   \frac{1}{f} \del_\mu \Pi 
  -\frac{i}{2f^2} [\Pi ,\del_\mu \Pi]_X 
  -\frac{1}{6f^3} \left[\Pi,[\Pi , \del_\mu \Pi]\right]_X 
  +{\cal O} \left( ( \Pi/f )^4 \right) \n 
 = &
   X^a \left( 
	  \frac{1}{f} \del_\mu \Pi^a 
	 +\frac{1}{2f^2} f^{abc} \Pi^b \del_\mu \Pi^c 
    +\frac{1}{6f^3} (f^{abe} f^{cde} +f^{abi} f^{cdi}) 
	   \Pi^b \Pi^c \del_\mu \Pi^c 
   \right) 
  +{\cal O} \left( ( \Pi/f )^4 \right),
\end{align}
where $[X^a ,X^b]_X$ is the projection of the commutator into 
the broken generator. 
Then the $G$-invariant two-derivative term of the NG fields is given by
\begin{align} 
 \frac{f^2}{2} \Tr[ \al_{\perp \mu} \al_\perp^\mu ] =&
   \frac{1}{2} (\del_\mu \Pi^a)(\del^\mu \Pi^a) \n &
  -\left( 
    \frac{1}{6f^2} f^{aci} f^{bdi} +\frac{1}{24f^2} f^{ace} f^{bde} 
  \right)  
  \Pi^a \Pi^b (\del_\mu \Pi^c)(\del^\mu \Pi^d)
  +{\cal O} \left( ( \Pi/f )^4 \right).
\label{eq:dim6_derivative_int_w/ff}
\end{align}
The normalizations of generators are given by
$\Tr [ T^i T^j ] = \de^{ij}$ and  
$\Tr [ X^a X^b ] = \de^{ab} $.

We assume that the $SO(4)$ multiplet,
$h^a$ $(a\in \{1, ..., 4\})$,
is embedded in the NG fields, $\Pi^a$.
This multiplet corresponds to the $SU(2)$ doublet Higgs field. 
In this paper, 
we treat the Higgs doublet as a NG field
and ignore possible other fields.
The second term of Eq.~\eqref{eq:dim6_derivative_int_w/ff} leads to
dimension-six derivative interactions of the Higgs fields.
We write the derivative interaction as
\begin{align}
 {\cal L}^\text{6NL} =& 
   \frac{1}{f^2}{\cal T}_{abcd} h^{a} h^{b} \del_\mu h^c \del^\mu h^d, 
   \label{eq:dim6_der_int_w/T} \\
 {\cal T}_{abcd} =&
  -\frac{1}{6} f^{aci} f^{bdi} -\frac{1}{24} f^{ace} f^{bde}.
  \label{eq:4th_tensor}
\end{align}
Since the structure constant is totally antisymmetric, 
$f^{aci}$ is  a $4\times 4$ antisymmetric matrix for each $i$.
Since the Lagrangian must be the $SU(2)_L \times U(1)_Y$ invariant, 
it is useful to use the generators of $SO(4) \simeq SU(2)_L \times SU(2)_R$
as a complete set of antisymmetric matrices.
Our definitions of the generators, $T^{L\al} $ and $ T^{R\be}$ 
$(\al, \be \in \{1, 2, 3 \})$, are listed 
in Appendix~\ref{app:so4N}.
In particular, the hyper charge generator corresponds to $ T^{R3}$.

We parametrize the $SU(2)_L$ Higgs doublet as
\begin{align}
 H = 
   \frac{1}{\sqrt{2}}
   \begin{pmatrix}
      h^{1} + i h^{2}  \\
      h^{3} + i h^{4} 
   \end{pmatrix}.
\end{align}
In order to see the property of the $SU(2)_R$ symmetry,
it is useful to use the bidoublet notation:
\begin{align}
 \Ph =& 
   \left( i\si^2 H^\ast \, H \right) 
   \label{eq:bidoublet_for_one_Higgs}  \\
 =&
   \frac{1}{\sqrt{2}} \left(
	 {\bf 1}_2 \, h^3 +i( \si^1 h^2 +\si^2 h^1 -\si^3 h^4 )
	\right),
\end{align}
where ${\bf 1}_2$ is the $2\times 2$ unit matrix.
With the bidoublet, 
$SO(4)$ transformation for $h^a$ is represented as
\begin{align}
 \Ph \to L \Phi R^\dag ,
\end{align}
where $L \in SU(2)_L$ and $R \in SU(2)_R$.
We can show the following relations to connect 
the $SO(4)$ multiplet with the bidoublet:
\begin{align}
 h^a \left( T^{L\al} \right)_{ac} \del_\mu h^c =& 
   \frac{1}{4} \Tr \left[ \Ph^\dag \si^\al \fb{\del}_\mu \Ph \right],
  \label{eq:one_Higgs_bidoublet_TL} \\
 h^a \left( T^{R\be} \right)_{ac} \del_\mu h^c =& 
   \frac{1}{4} \Tr \left[ \Ph \si^\be \fb{\del}_\mu \Ph^\dag \right]. 
  \label{eq:one_Higgs_bidoublet_TR}
\end{align}
Hence, the following form of ${\cal T}_{abcd}$ is invariant
under the $SU(2)_L \times U(1)_Y$ symmetry:
\begin{align}
 {\cal T}_{abcd} = 
   a^L \left( T^{L\al} \right)_{ac} \left( T^{L\al} \right)_{bd} 
  +a^R \left( T^{R\be} \right)_{ac} \left( T^{R\be} \right)_{bd} 
  +a^Y \left( T^{R3}   \right)_{ac} \left( T^{R3}   \right)_{bd}.
  \label{eq:one_Higgs_invariant_T}
\end{align}
With the explicit representation given in Appendix~\ref{app:so4N},
the tensor, $\mcl{T}_{abcd}$, can be written as
\begin{align}
 \mcl{T}_{abcd} =&
  -\frac{a^L + a^R}{4}(\de^{ab} \de^{cd} -\de^{ad} \de^{bc})
  +\frac{a^L - a^R}{4}\ep_{abcd} \n &
  -\frac{a^Y}{4} 
  ( \de^{1a} \de^{2c} -\de^{1c} \de^{2a} 
   +\de^{3a} \de^{4c} -\de^{3c} \de^{4a} )
  ( \de^{1b} \de^{2d} -\de^{1d} \de^{2b} 
   +\de^{3b} \de^{4d} -\de^{3d} \de^{4b} ).
\end{align}
According to Eq.~\eqref{eq:dim6_der_int_w/T}, 
the second term of the above equation does not 
contribute to the Lagrangian due to the Bose symmetry.
This means that degree of freedom (\dof) 
is decreased by the \nlsm \ structure.
After expanding it in terms of $O^H$, $O^r$, $O^T$ and $O^{HT}$,
the following Lagrangian is obtained:
\begin{align}
 {\cal L}^\text{6NL} =
   \frac{a^L +a^R}{4f^2} \del_\mu (H^\dag H) \del^\mu (H^\dag H)
  -\frac{a^L +a^R}{ f^2} H^\dag H (\del_\mu H^\dag )(\del^\mu H) 
  +\frac{a^Y}{4f^2} (H^\dag \fb{\del}_\mu H)(H^\dag \fb{\del}^\mu H).
  \label{eq:one_Higgs_nlsm_Lagrangian}
\end{align}
In the SILH model, the number of independent coefficients is two 
while there are four \dof \  in the Lagrangian preserving only 
the $SU(2)_L \times U(1)_Y$ symmetry.

In the rest of this section, 
the scattering amplitudes of the longitudinal gauge bosons 
and the Higgs boson are studied 
as phenomenological consequences of the Lagrangian.
The given Lagrangian~\eqref{eq:one_Higgs_nlsm_Lagrangian} 
can be simplified with the field redefinition:
\begin{align}
 H^a \to H^a +\frac{a}{f^2} H^a (H^\dag H).
\end{align}
To study the scatterings of gauge boson longitudinal modes, 
it is enough to see the effect of the redefinition in the kinetic term:
\begin{align}
 (\del_\mu H)^\dag (\del^\mu H) \to 
   (\del_\mu H)^\dag (\del^\mu H)
  +\frac{ a}{f^2} \del_\mu (H^\dag H) \del^\mu (H^\dag H) 
  +\frac{2a}{f^2} H^\dag H(\del_\mu H^\dag )(\del^\mu H) 
  +{\cal O} \left( ( H/f )^4  \right).
\end{align}
This indicates that we can always 
choose the coefficient, $a$, so as to eliminate $O^r$.
The Yukawa interaction and 
the Higgs potential are also changed by the redefinition.
However, these $\mcl{O}(v^2/f^2)$ corrections can be neglected
in leading contributions of the derivative interactions.
After the redefinition, the Lagrangian becomes 
\begin{align}
 {\cal L}^\text{6NL} = 
   \frac{3(a^L +a^R)}{4f^2} O^H +\frac{a^Y}{4f^2} O^T .
  \label{eq:L_6NL}
\end{align}
The term $O^T$ breaks the custodial symmetry so that 
the coefficient $a^Y$ is severely constrained 
by electroweak precision measurements.
The term $O^H$ affects high energy behavior in the scattering amplitudes
of the longitudinal modes and the Higgs boson.
If we keep only $O^H$, using the equivalence theorem~\cite{Chanowitz:1985hj},
these amplitudes in high energy region
$(s, t, u \gg m_h^2, m_W^2 \ \ \text{and} \ \ s + t + u = 0)$ are given as follows:
\begin{align} 
 {\cal M} (W_L^+ W_L^- \to W_L^+ W_L^-) =& 
   \frac{3(s +t)}{2f^2} (a^L +a^R), \label{eq:amp1} \\
 {\cal M} (W_L^+ W_L^- \to Z_L Z_L) =& 
 {\cal M} (W_L^+ W_L^- \to h h ) = 
   \frac{3s}{2f^2} (a^L +a^R), \label{eq:amp2} \\
 {\cal M} (W_L^+ W_L^- \to Z_L h ) =& 0.
\end{align}
In the SILH model,
the amplitudes increase as $E^2$ in high energy region
whereas such behavior does not appear in the SM case ($f\to \infty$).
The amplitudes are supposed to grow up to the scale, $m_\rh$,
where new resonances are produced.
It is, therefore, important to precisely measure 
vector boson fusion processes in high energy region
and see correlations among them
at the LHC and LC.
Studies of the scatterings at the LHC are done 
in Refs.~\cite{Han:2009em,Contino,Grober}.
Above the scale $m_{\rh}$, we need to include the effects of heavy resonances, 
see, e.g.~\cite{Contino:2011np}.
\section{Generalization to $N$ Higgs doublet models} 
\label{sec:multi}
In this section, we present the generalizations of the SILH model which
includes only one Higgs doublet to the case with $N$ Higgs doublets.

The $SU(2)_L \times U(1)_Y$ invariant Lagrangian for
dimension-six derivative interactions consists of four kinds of operators:
\begin{align}
 \mcl{O}^H : O^H_{ijkl} &= 
   \del_\mu ( H_i^\dag H_j ) \del^\mu ( H_k^\dag H_l ), \\
 \mcl{O}^T : O^T_{ijkl} &= 
   ( H_i^\dag \delfb_\mu H_j )( H_k^\dag \delfb^\mu H_l ), \\
 \mcl{O}^r : O^r_{ijkl} &=  
   H_i^\dag H_j  ( \del_\mu  H_k )^\dag ( \del^\mu H_l ),  \\
 \mcl{O}^{HT} : O^{HT}_{ijkl} &= 
   \del_\mu ( H_i^\dag H_j )( H_k^\dag \delfb^\mu H_l ),
\end{align}
where $i, j, k, l \in \{1, ..., N\}$ stand for the species of Higgs doublets.

In order to study how the nonlinear nature of the Higgs sector 
reduces the number of independent coefficients,
we classify above operators into the following five types:
\begin{description}
 \item[Type I:]
   All species are same 
	(e.g. $\del_\mu (H_i^\dag H_i) \del^\mu (H_i^\dag H_i)$).
 \item[Type II:]
   Only one species is different from the others
	(e.g. $\del_\mu (H_i^\dag H_i) \del^\mu (H_i^\dag H_j)$).
 \item[Type III:]
   Two pairs of different species appear
	(e.g. $\del_\mu (H_i^\dag H_i) \del^\mu (H_j^\dag H_j)$).
 \item[Type IV:]
   Three species are included
	(e.g. $\del_\mu (H_i^\dag H_i) \del^\mu (H_j^\dag H_k)$).
 \item[Type V:]
   All species are deferent
	(e.g. $\del_\mu (H_i^\dag H_j) \del^\mu (H_k^\dag H_l)$).
\end{description}

On the other hand, 
dimension-six derivative interactions of the NG fields are 
written as Eq.~\eqref{eq:dim6_derivative_int_w/ff}.
In this section,
we regard the NG fields, $\Pi^a$, as not an $SO(4)$ multiplet 
but an $SO(4N)$ multiplet, $h^a \, (a\in \{1, ..., 4N\})$, and 
possible other NG fields are neglected.
In this case, the structure constant, $f^{aci}$, in Eq.~\eqref{eq:4th_tensor}
is considered as $4N\times 4N$ antisymmetric matrices for given $i$. 
Namely it can be expressed using generators of the $SO(4N)$.
The multiplet corresponds to $N$ species of Higgs doublets:
\begin{align}
 H_i =\frac{1}{\sqrt{2}}
  \begin{pmatrix}
   h^{1+4(i-1)} +ih^{2+4(i-1)} \\
   h^{3+4(i-1)} +ih^{4+4(i-1)}
  \end{pmatrix},
\end{align}
where $i \ (i=1,...,N)$ is also the index of the Higgs species.

To see how the $SU(2)_L \times U(1)_Y$ symmetry is imposed on the theory,
it is convenient to use the bidoublet notation
like Eq.~\eqref{eq:bidoublet_for_one_Higgs}.
In the case of $N=1$, 
generators which correspond to $SU(2)_L$, $T^{L\al}$, and
$SU(2)_R$, $T^{R\be}$, can be used
as generators of the $SO(4)$. 
See Eqs.~\eqref{eq:one_Higgs_bidoublet_TL} and~\eqref{eq:one_Higgs_bidoublet_TR}.
They are the ({\bf 3}, {\bf 1}) and ({\bf 1}, {\bf 3}) representations of
$SU(2)_L \times SU(2)_R$.
In the case of $N \geq 2$,
other kind of generator appears, namely, 
the ({\bf 1}, {\bf 1}) and ({\bf 3}, {\bf 3}) representations of 
the $SU(2)_L \times SU(2)_R$ written as $U$ and $S^{\al\be}$, respectively.
Explicit forms of them are shown in Appendix~\ref{app:so4N}.
With the generators, 
the following relations are obtained between 
the multiplet of the $SO(4N)$ and the bidoublet:
\begin{align}
  h^a \left( T_{(i,i)}^{L\al} \right)_{ac} \del_\mu h^c &= 
    \frac{1}{4} \Tr \left[
	  \Ph_{ii}^\dag \si^\al  \fb{\del}_\mu \Ph_{ii}
   \right], \\
  h^a \left( T_{(i,i)}^{R\be} \right)_{ac} \del_\mu h^c &=  
    \frac{1}{4} \Tr \left[ 
	  \Ph_{ii} \si^\be \fb{\del}_\mu \Ph_{ii}^\dag  
	\right], \\
  h^a \left( T_{(i,j)}^{L\al} \right)_{ac} \del_\mu h^c &= 
    \frac{1}{4} \Tr \left[ 
	  \Ph_{ii}^\dag \si^\al \fb{\del}_\mu \Ph_{jj}
	 +\Ph_{jj}^\dag \si^\al \fb{\del}_\mu \Ph_{ii}
	\right], \\
  h^a \left( T_{(i,j)}^{R\be} \right)_{ac} \del_\mu h^c &=
    \frac{1}{4} \Tr \left[ 
	  \Ph_{ii} \si^\be \fb{\del}_\mu \Ph_{jj}^\dag 
    +\Ph_{jj} \si^\be \fb{\del}_\mu \Ph_{ii}^\dag
	\right], \\
  h^a \left( U_{(i,j)} \right)_{ac} \del_\mu h^c &= 
    \frac{i}{4} \Tr \left[ 
      \Ph_{jj}^\dag \fb{\del}_\mu \Ph_{ii}
	  +\Ph_{jj} \fb{\del}_\mu \Ph_{ii}^\dag
	\right], \label{eq:U} \\
  h^a \left( S^{\al\be}_{(i,j)} \right)_{ac} \del_\mu h^c &= 
    \frac{i}{4} \Tr \left[ 
	   \Ph_{ii}^\dag \si^\al \fb{\del}_\mu \Ph_{jj} \si^\be  
	  +\Ph_{ii} \si^\be \fb{\del}_\mu \Ph_{jj}^\dag \si^\al 
	\right], 
\label{eq:S}
\end{align}
where
\begin{align}
 \Ph_{ii} &= \left( i\si^2 H_i^\ast \, H_i \right).
\end{align}
The $SU(2)_{L} \times SU(2)_{R}$ transformation is defined by 
\begin{align}
 \Ph_{ii} &\rightarrow L \Ph_{ii} R^{\dagger}.
 \label{eq:N_bidoublet_tr}
\end{align}
If the Higgs sector is invariant under the 
transformation~\eqref{eq:N_bidoublet_tr},
it preserves the $SO(4)$ symmetry which is 
directly connected with the custodial symmetry.
We simply refer the $SO(4)$ invariance as the custodial invariance.
For details of the bidoublet notation.
See Appendix \ref{app:bidoublet}.

The $SU(2)_L \times U(1)_Y$ invariant forms can be 
constructed in two ways.
One is to contract the indices of $SU(2)_L$ or $SU(2)_R$,
which gives $SO(4)$ invariant terms.
The other is to keep only $\si^3$ term of $T^{R3}$,
which corresponds to $SO(4)$ breaking 
terms\footnote{
There is a special case where the $SO(4)$ invariance is 
preserved with combination of $T^{R3}$ terms.
See Sec.~\ref{subsec:amp}.
}.

In the following subsections,
we show that the number of independent derivative interactions 
needed to describe $N$ Higgs doublets in the \nlsm \ 
is reduced compared to that of all operators
invariant under $SU(2)_L \times U(1)_Y$.
\subsection{Type I}
\label{subsec:sym_based}

The Lagrangian for type I can be constructed in the same way
as the SILH model reviewed in the previous section.
In order to show our notation for the NHDM,
we rewrite several equations.

The Lagrangian of derivative interactions invariant under 
$SU(2)_L \times U(1)_Y$ is
\begin{align}
 \mcl{L}^6_\text{I} =&
   \frac{\la^H_{iiii}}{2\La^2} O^H_{iiii}
  +\frac{\la^r_{iiii}}{ \La^2} O^r_{iiii}
  +\frac{\la^T_{iiii}}{2\La^2} O^T_{iiii}
  +i\frac{\la^{HT}_{iiii}}{\La^2} O^{HT}_{iiii},
\end{align}
where all of the coefficients, $\la^H$, $\la^r$, $\la^T$ and $\la^{HT}$, are real
so that there are $3N$ real and $N$ imaginary \dof \  in the general NHDM.

For the case of the \nlsm, as is given in Eq.~\eqref{eq:dim6_der_int_w/T},
the Lagrangian corresponding to the above can be parametrized as
\begin{align}
 \mcl{L}^\text{6NL}_\text{I} =&
   \frac{1}{f^2} \mcl{T}_{abcd}^\text{I}\,
	 h^a h^b (\del_\mu h^c )(\del^\mu h^d ),
	 \label{eq:der_int_typeI}
\end{align}
where
\begin{align}
 {\cal T}_{abcd}^\text{I} =& 
   a^L_{iiii} 
   \left( T^{L\al}_{(i,i)} \right)_{ac} \left( T^{L\al}_{(i,i)} \right)_{bd}
  +a^R_{iiii} 
   \left( T^{R\be}_{(i,i)} \right)_{ac} \left( T^{R\be}_{(i,i)} \right)_{bd}
  +2a^Y_{iiii} 
   \left( T^{R3}_{(i,i)} \right)_{ac} \left( T^{R3}_{(i,i)} \right)_{bd}.
   \label{eq:typeI_T}
\end{align}
The coefficients, $a^L_{iiii}$, $a^R_{iiii}$ and $a^Y_{iiii}$ 
are real.
Substituting Eq.~\eqref{eq:typeI_T} into Eq.~\eqref{eq:der_int_typeI},
the following Lagrangian is obtained:
\begin{align}
 \mcl{L}^\text{6NL}_\text{I} =&
   \frac{c^{\text{I}1}_{[i]}}{4f^2} O^H_{iiii}
  -\frac{c^{\text{I}1}_{[i]}}{ f^2} O^r_{iiii}
  +\frac{c^{\text{I}2}_{[i]}}{2f^2} O^T_{iiii},
  \label{eq:type_I_Lagrangian}
\end{align}
\begin{align}
 c^{\text{I}1}_{[i]} = & 
  a^L_{iiii} +a^R_{iiii} ,
  \label{eq:typeI_coeffA} \\
 c^{\text{I}2}_{[i]} = &
  a^Y_{iiii}.
  \label{eq:typeI_coeffB}
\end{align}
Note that the factor of 2 in Eq.~\eqref{eq:typeI_T}
is introduced; this is different from Eq.~\eqref{eq:one_Higgs_invariant_T}.

In NHDM,
there are $3N$ real and $N$ imaginary \dof.
However, assuming the Higgs doublets are embedded in the \nlsm,
$2N$ real \dof \  remain (Table~\ref{tb:typeI}).
For example, in the case of $N=1$, 
three real and one imaginary \dof \  decrease to two real \dof.
It is consistent with the result in the previous section.
\begin{table}[!h] 
\centering
 \begin{displaymath}
  \begin{array}{l|ll}
                & \text{Re} & \text{Im} \\ \hline \hline
   \text{General} & 3N &  N \\ \hline
   \mcl{O}^H    &         N &         0 \\
   \mcl{O}^T    &         N &         0 \\
   \mcl{O}^r    &         N &         0 \\
   \mcl{O}^{HT} &         0 &         N \\ \hline \hline
   \text{Nonlinear} & 2N &  0 \\
  \end{array} 
 \end{displaymath}
 \caption{
  The number of the independent dimension-six derivative interactions for type I.
  The first and second columns mean real and imaginary \dof, respectively.
  We show the number of \dof \ for the case 
  where $SU(2)_L \times U(1)_Y$ invariance is imposed (General)
  and the case with nonlinear realization (Nonlinear). 
  In the case of General, the \dof \ of each type of operators is also shown.
 } 
 \label{tb:typeI}
\end{table}
\subsection{Type II}

For type II, the Lagrangian of the dimension-six derivative interactions is
\begin{align}
 \mcl{L}^6_\text{II} =&
   \frac{\la_{iiij}^H}{\La^2} O^H_{iiij}
  +\frac{\la_{iiij}^r}{\La^2} O^r_{iiij}
  +\frac{\la_{ijii}^r}{\La^2} O^r_{ijii}
  +\frac{\la_{iiij}^T}{\La^2} O^T_{iiij}
  +\frac{\la_{iiij}^{HT}}{\La^2} O^{HT}_{iiij}
  +\frac{\la_{ijii}^{HT}}{\La^2} O^{HT}_{ijii}
  +\hc \ .
\end{align}
All of the coefficients can be complex
so that there are $6N(N-1)$ real and imaginary \dof, respectively.

Assuming the Higgs doublets are described as the NG fields,
the Lagrangian can be written as
\begin{align}
 \mcl{L}^\text{6NL}_\text{II} =&
   \frac{1}{f^2} \mcl{T}_{abcd}^\text{II}\,
	 h^a h^b (\del_\mu h^c )(\del^\mu h^d ),
\end{align}
where, using real coefficients,
\begin{align}
 {\cal T}_{abcd}^\text{II} =& 
   2a^L_{iiij} 
   \left( T^{L\al}_{(i,i)} \right)_{ac} \left( T^{L\al}_{(i,j)} \right)_{bd}
  +2a^R_{iiij} 
   \left( T^{R\be}_{(i,i)} \right)_{ac} \left( T^{R\be}_{(i,j)} \right)_{bd}
  +2a^{LS}_{iiij} 
   \left( T^{L\al}_{(i,i)} \right)_{ac} \left( S^{\al 3}_{(i,j)} \right)_{bd}
 \n &
  +4a^Y_{iiij} 
   \left( T^{R3}_{(i,i)} \right)_{ac} \left( T^{R3}_{(i,j)} \right)_{bd}
  +4a^{YU}_{iiij} 
   \left( T^{R3}_{(i,i)} \right)_{ac} \left( U_{(i,j)} \right)_{bd} .
\end{align}

The Lagrangian, $\mcl{L}^\text{6NL}_\text{II}$, can be expanded in terms of 
the $SU(2)_L$ doublets, $H_i$.
The normalization of the coefficients are chosen to make the following
coefficients simple:
\begin{align}
 \mcl{L}^\text{6NL}_\text{II} =&
   \frac{c^{\text{II}1}_{[ij]}}{2f^2} O^H_{iiij}
  -\frac{c^{\text{II}1}_{[ij]}}{ f^2} O^r_{iiij} 
  -\frac{c^{\text{II}1}_{[ij]}}{ f^2} O^r_{ijii}
  +\frac{c^{\text{II}2}_{[ij]}}{ f^2} O^T_{iiij}
  +\hc \ ,
\end{align}
\begin{align}
 c^{\text{II}1}_{[ij]} =& 
  a^L_{iiij} +a^R_{iiij} -ia^{LS}_{iiij}, \\
 c^{\text{II}2}_{[ij]} =&
  a^Y_{iiij} -ia^{YU}_{iiij}.
\end{align}

The original Lagrangian, $\mcl{L}^6_\text{II}$, has
$6N(N-1)$ real and imaginary \dof.
However, 
the corresponding Lagrangian consists of $2N(N-1)$ real and imaginary 
\dof, respectively, if the Higgs doublets
are considered as NG fields (Table~\ref{tb:typeII}).
\begin{table}[!h] 
 \centering
  \begin{displaymath}
  \begin{array}{l|ll}
                & \text{Re} & \text{Im} \\ \hline \hline
   \text{General} & 6N(N-1) & 6N(N-1) \\ \hline
   \mcl{O}^H    &    N(N-1) &    N(N-1) \\ 
   \mcl{O}^T    &    N(N-1) &    N(N-1) \\ 
   \mcl{O}^r    &   2N(N-1) &   2N(N-1) \\
   \mcl{O}^{HT} &   2N(N-1) &   2N(N-1)  \\ \hline \hline
   \text{Nonlinear} & 2N(N-1) & 2N(N-1)
  \end{array} 
  \end{displaymath}
 \caption{
  The number of the independent dimension-six derivative interactions for type II.
  The first and second columns mean real and imaginary \dof, respectively.
  We show the number of \dof \ for the case 
  where $SU(2)_L \times U(1)_Y$ invariance is imposed (General)
  and the case with nonlinear realization (Nonlinear). 
  In the case of General, the \dof \ of each type of operators is also shown.
 } 
\label{tb:typeII}
\end{table}
\subsection{Type III}

The Lagrangian of type III derivative interactions is
\begin{align}
 \mcl{L}^6_\text{III} =&
   \frac{\la_{iijj}^H}{2\La^2} O^H_{iijj}
  +\frac{\la_{ijji}^H}{2\La^2} O^H_{ijji}
  +\frac{\la_{ijij}^H}{2\La^2} O^H_{ijij} \n &
  +\frac{\la_{iijj}^r}{2\La^2} O^r_{iijj}
  +\frac{\la_{jjii}^r}{2\La^2} O^r_{jjii}
  +\frac{\la_{ijij}^r}{ \La^2} O^r_{ijij}
  +\frac{\la_{ijji}^r}{ \La^2} O^r_{ijji} \n &
  +\frac{\la_{iijj}^T}{2\La^2} O^T_{iijj}
  +\frac{\la_{ijji}^T}{2\La^2} O^T_{ijji}
  +\frac{\la_{ijij}^T}{2\La^2} O^T_{ijij} \n &
  +i\frac{\la_{iijj}^{HT}}{2\La^2} O^{HT}_{iijj}
  +i\frac{\la_{jjii}^{HT}}{2\La^2} O^{HT}_{jjii}
  + \frac{\la_{ijij}^{HT}}{ \La^2} O^{HT}_{ijij}
  + \frac{\la_{ijji}^{HT}}{ \La^2} O^{HT}_{ijji}
  +\hc \ .
\end{align}
In the above operators, $O^H_{iijj}$, $O^H_{ijji}$, $O^r_{iijj}$, 
$O^r_{jjii}$, $O^T_{iijj}$ and $O^T_{ijji}$, are Hermitian operators while
$O^{HT}_{iijj}$ and $O^{HT}_{jjii}$ are skew Hermitian operators.
The number of independent real and imaginary coefficients are 
$6N(N-1)$ and $4N(N-1)$, respectively.

If the Higgs doublets are embedded in the NG fields,
Lagrangian can be written as follows:
\begin{align}
 \mcl{L}^\text{6NL}_\text{III} =&
   \frac{1}{f^2} \mcl{T}_{abcd}^\text{III}\,
	 h^a h^b (\del_\mu h^c )(\del^\mu h^d ).
\end{align}
Parametrization of the tensor, $\mcl{T}_{abcd}^\text{III}$, is
\begin{align}
 {\cal T}_{abcd}^\text{III} =& 
   2a^L_{iijj} 
   \left( T^{L\al}_{(i,i)} \right)_{ac} 
	\left( T^{L\al}_{(j,j)} \right)_{bd}
  +a^L_{ijij} 
   \left( T^{L\al}_{(i,j)} \right)_{ac} 
	\left( T^{L\al}_{(i,j)} \right)_{bd}
  +2a^R_{iijj} 
   \left( T^{R\be}_{(i,i)} \right)_{ac}  
	\left( T^{R\be}_{(j,j)} \right)_{bd} \n &
  +a^R_{ijij} 
   \left( T^{R\be}_{(i,j)} \right)_{ac} 
	\left( T^{R\be}_{(i,j)} \right)_{bd}
  +a^S_{ijij} 
   \left( S^{\al 3}_{(i,j)} \right)_{ac} 
	\left( S^{\al 3}_{(i,j)} \right)_{bd}
  +a^{SS}_{ijij} 
   \left( S^{\al \be}_{(i,j)} \right)_{ac} 
	\left( S^{\al \be}_{(i,j)} \right)_{bd} \n &
  +a^{LS}_{ijij} 
   \left( T^{L\al }_{(i,j)} \right)_{ac} 
	\left( S^{\al 3}_{(i,j)} \right)_{bd}
  +2a^Y_{iijj} 
   \left( T^{R3}_{(i,i)} \right)_{ac} 
	\left( T^{R3}_{(j,j)} \right)_{bd}
  +a^Y_{ijij} 
   \left( T^{R3}_{(i,j)} \right)_{ac} 
	\left( T^{R3}_{(i,j)} \right)_{bd} \n &
  +a^U_{ijij} 
   \left( U_{(i,j)} \right)_{ac} 
	\left( U_{(i,j)} \right)_{bd} 
  +a^{YU}_{ijij} 
   \left( T^{R3}_{(i,j)} \right)_{ac} 
	\left( U_{(i,j)} \right)_{bd},
\end{align}
where the coefficients are real.
The derivative interactions embedded in \nlsm \ can be written as
\begin{align}
 \mcl{L}^\text{6NL}_\text{III} =&
   \frac{c^{\text{III}1}_{[ij]}}{2f^2} O^H_{iijj}
  +\frac{c^{\text{III}2}_{[ij]}}{2f^2} O^H_{ijji}
  -\frac{c^{\text{III}1}_{[ij]}}{ f^2} ( O^r_{iijj} +O^r_{jjii} ) \n &
  -\frac{c^{\text{III}2}_{[ij]}}{ f^2} ( O^r_{ijji} +O^r_{jiij} )
  +\frac{c^{\text{III}3}_{[ij]}}{2f^2} O^T_{ijji}
  +\frac{c^{\text{III}4}_{[ij]}}{2f^2} O^T_{iijj} \n &
  +\left(
   \frac{c^{\text{III}5}_{[ij]}}{4f^2} O^H_{ijij}
  -\frac{c^{\text{III}5}_{[ij]}}{ f^2} O^r_{ijij}
  +\frac{c^{\text{III}6}_{[ij]}}{4f^2} O^T_{ijij} +\hc 
  \right),
\end{align}
\begin{align}
 c^{\text{III}1}_{[ij]} =&
   a^L_{ijij} +a^R_{ijij} +a^S_{ijij} +2a^{SS}_{ijij},\\
 c^{\text{III}2}_{[ij]} =&
   a^L_{iijj} +a^R_{ijij} -a^{SS}_{ijij},\\
 c^{\text{III}3}_{[ij]} =&
  -a^L_{ijij} +a^L_{iijj} -a^S_{ijij} +a^Y_{ijij} +a^U_{ijij},\\
 c^{\text{III}4}_{[ij]} =&
  -a^L_{iijj} +a^L_{ijij} -a^R_{ijij} +a^R_{iijj}
  +a^S_{ijij} +a^Y_{iijj},\\
 c^{\text{III}5}_{[ij]} =&
   a^L_{ijij} +a^R_{iijj} -a^S_{ijij} -a^{SS}_{ijij} -ia^{LS}_{ijij},\\
 c^{\text{III}6}_{[ij]} =&
   a^R_{ijij} -a^R_{iijj} +a^Y_{ijij} -a^U_{ijij} -ia^{YU}_{ijij}.
\end{align}

Because of the nature of the nonlinear dynamics, 
the number of independent \dof \  decreases from
$6N(N-1)$ real \dof \  and $4N(N-1)$ imaginary \dof \  to
$3N(N-1)$ real \dof \  and $N(N-1)$ imaginary \dof \  (Table~\ref{tb:typeIII}).
\begin{table}[!h] 
 \centering
  \begin{displaymath}
  \begin{array}{l|ll}
                & \text{Re}   & \text{Im}   \\ \hline \hline
   \text{General} & 6N(N-1) & 4N(N-1) \\ \hline
   \mcl{O}^H    & (3/2)N(N-1) & (1/2)N(N-1) \\ 
   \mcl{O}^T    & (3/2)N(N-1) & (1/2)N(N-1) \\ 
   \mcl{O}^r    &     2N(N-1) &      N(N-1) \\
   \mcl{O}^{HT} &      N(N-1) &     2N(N-1) \\ \hline \hline
   \text{Nonlinear} & 3N(N-1) & N(N-1)
  \end{array} 
  \end{displaymath}
 \caption{
  The number of the independent dimension-six derivative interactions for type III.
  The first and second columns mean real and imaginary \dof, respectively.
  We show the number of \dof \ for the case 
  where $SU(2)_L \times U(1)_Y$ invariance is imposed (General)
  and the case with nonlinear realization (Nonlinear). 
  In the case of General, the \dof \ of each type of operators is also shown.
 } 
 \label{tb:typeIII}
\end{table}

\subsection{Type IV}

The Lagrangian of type IV derivative interactions is
\begin{align}
 \mcl{L}^6_\text{IV} =&
   \frac{\la_{iijk}^H}{\La^2} O^H_{iijk}
  +\frac{\la_{ijik}^H}{\La^2} O^H_{ijik}
  +\frac{\la_{ijki}^H}{\La^2} O^H_{ijki} \n &
  +\frac{\la_{iijk}^r}{\La^2} O^r_{iijk}
  +\frac{\la_{jkii}^r}{\La^2} O^r_{jkii}
  +\frac{\la_{ijik}^r}{\La^2} O^r_{ijik}
  +\frac{\la_{ikij}^r}{\La^2} O^r_{ikij}
  +\frac{\la_{ijki}^r}{\La^2} O^r_{ijki}
  +\frac{\la_{kiij}^r}{\La^2} O^r_{kiij} \n &
  +\frac{\la_{iijk}^T}{\La^2} O^T_{iijk}
  +\frac{\la_{ijik}^T}{\La^2} O^T_{ijik}
  +\frac{\la_{ijki}^T}{\La^2} O^T_{ijki} \n &
  +\frac{\la_{iijk}^{HT}}{\La^2} O^{HT}_{iijk}
  +\frac{\la_{jkii}^{HT}}{\La^2} O^{HT}_{jkii}
  +\frac{\la_{ijik}^{HT}}{\La^2} O^{HT}_{ijik}
  +\frac{\la_{ikij}^{HT}}{\La^2} O^{HT}_{ikij}
  +\frac{\la_{ijki}^{HT}}{\La^2} O^{HT}_{ijki}
  +\frac{\la_{kiij}^{HT}}{\La^2} O^{HT}_{kiij}
  +\hc \ ,
  \label{eq:der_int_typeIV}
\end{align}
where $j<k$.
Since all of the coefficients can be complex,
the Lagrangian has $9N(N-1)(N-2)$ \dof \  for real and imaginary coefficients,
respectively.
In the case that the Higgs sector is governed by the nonlinear dynamics, 
the derivative interactions can be written as follows:
\begin{align}
 \mcl{L}^\text{6NL}_\text{IV} =&
   \frac{1}{f^2} \mcl{T}_{abcd}^\text{IV}\,
	 h^a h^b (\del_\mu h^c )(\del^\mu h^d ),
	 \label{eq:der_int_nlsm_typeIV}
\end{align}
where
\begin{align}
 {\cal T}_{abcd}^\text{IV} =& 
   2a^L_{iijk} 
   \left( T^{L\al}_{(i,i)} \right)_{ac} 
	\left( T^{L\al}_{(j,k)} \right)_{bd}
  +2a^L_{ijik} 
   \left( T^{L\al}_{(i,j)} \right)_{ac} 
	\left( T^{L\al}_{(i,k)} \right)_{bd}
  +2a^R_{iijk} 
   \left( T^{R\be}_{(i,i)} \right)_{ac} 
	\left( T^{R\be}_{(j,k)} \right)_{bd} \n &
  +2a^R_{ijik} 
   \left( T^{R\be}_{(i,j)} \right)_{ac} 
	\left( T^{R\be}_{(i,k)} \right)_{bd}
  +2a^S_{ijik} 
   \left( S^{\al 3}_{(i,j)} \right)_{ac} 
	\left( S^{\al 3}_{(i,k)} \right)_{bd}
  +2a^{SS}_{ijik} 
   \left( S^{\al \be}_{(i,j)} \right)_{ac} 
	\left( S^{\al \be}_{(i,k)} \right)_{bd} \n &
  +2a^{LS}_{iijk} 
   \left( T^{L \al}_{(i,i)} \right)_{ac} 
	\left( S^{\al 3}_{(j,k)} \right)_{bd}
  +2a^{LS}_{ijik} 
   \left( T^{L \al}_{(i,j)} \right)_{ac} 
	\left( S^{\al 3}_{(i,k)} \right)_{bd}
  +2a^{LS}_{ikij} 
   \left( T^{L \al}_{(i,k)} \right)_{ac} 
	\left( S^{\al 3}_{(i,j)} \right)_{bd} \n &
  +2a^Y_{iijk} 
   \left( T^{R3}_{(i,i)} \right)_{ac} 
	\left( T^{R3}_{(j,k)} \right)_{bd}
  +2a^Y_{ijik} 
   \left( T^{R3}_{(i,j)} \right)_{ac} 
	\left( T^{R3}_{(i,k)} \right)_{bd}
  +2a^U_{ijik} 
   \left( U_{(i,j)} \right)_{ac} 
	\left( U_{(i,k)} \right)_{bd}  \n &
  +2a^{YU}_{iijk} 
   \left( T^{R3}_{(i,i)} \right)_{ac} 
	\left( U_{(j,k)} \right)_{bd} 
  +2a^{YU}_{ijik} 
   \left( T^{R3}_{(i,j)} \right)_{ac} 
	\left( U_{(i,k)} \right)_{bd} 
  +2a^{YU}_{ikij} 
   \left( T^{R3}_{(i,k)} \right)_{ac} 
	\left( U_{(i,j)} \right)_{bd}.
\end{align}
Equation~\eqref{eq:der_int_nlsm_typeIV} can be expanded 
in terms of the $SU(2)$ doublets:
\begin{align}
 \mcl{L}^\text{6NL}_\text{IV} =&
   \frac{c^{\text{IV}1}_{[ijk]}}{2f^2} O^H_{iijk}
  +\frac{c^{\text{IV}2}_{[ijk]}}{2f^2} O^H_{ijik}
  +\frac{c^{\text{IV}3}_{[ijk]}}{2f^2} O^H_{ijki} \n &
  -\frac{c^{\text{IV}1}_{[ijk]}}{ f^2} ( O^r_{iijk} +O^r_{jkii} )
  -\frac{c^{\text{IV}2}_{[ijk]}}{ f^2} ( O^r_{ijik} +O^r_{ikij} )
  -\frac{c^{\text{IV}3}_{[ijk]}}{ f^2} ( O^r_{ijki} +O^r_{kiij} ) \n &
  +\frac{c^{\text{IV}4}_{[ijk]}}{2f^2} O^T_{iijk}
  +\frac{c^{\text{IV}5}_{[ijk]}}{2f^2} O^T_{ijik} 
  +\frac{c^{\text{IV}6}_{[ijk]}}{2f^2} O^T_{ijki}
 +\hc \ ,
\end{align}
\begin{align}
 c^{\text{IV}1}_{[ijk]} =&
   a^L_{ijik} +a^R_{ijik} +a^S_{ijik} +2a^{SS}_{ijik}
  +i\left(-a^{LS}_{ijik} +a^{LS}_{ikij} \right),\\ 
 c^{\text{IV}2}_{[ijk]} =&
   a^L_{ijik} +a^R_{iijk} -a^S_{ijik} -a^{SS}_{ijik}
  -i\left( a^{LS}_{ijik} +a^{LS}_{ikij} \right),\\
 c^{\text{IV}3}_{[ijk]} =&
   a^L_{iijk} +a^R_{ijik} -a^{SS}_{ijik} +ia^{LS}_{iijk},\\
 c^{\text{IV}4}_{[ijk]} =&
  -a^L_{iijk} +a^L_{ijik} +a^R_{iijk} -a^R_{ijik} +a^S_{ijik} +a^Y_{iijk}
  +i\left( a^{LS}_{iijk} -a^{LS}_{ijik} 
          +a^{LS}_{ikij} -a^{YU}_{iijk} \right),\\
 c^{\text{IV}5}_{[ijk]} =&
  -a^R_{iijk} +a^R_{ijik} +a^Y_{ijik} -a^U_{ijik}
  -i\left(  a^{YU}_{ijik} +a^{YU}_{ikij} \right),\\
 c^{\text{IV}6}_{[ijk]} =&
   a^L_{iijk} -a^L_{ijik} -a^S_{ijik} +a^Y_{ijik} +a^U_{ijik}
  +i\left( a^{LS}_{iijk} -a^{LS}_{ijik} +a^{LS}_{ikij} 
          +a^{YU}_{ijik} -a^{YU}_{ikij} \right).
\end{align}
The last Lagrangian has $3N(N-1)(N-2)$ real and $3N(N-1)(N-2)$ imaginary \dof.
Namely, each \dof \ is reduced to one third of the original one 
in the Lagrangian~\eqref{eq:der_int_typeIV} (Table~\ref{tb:typeIV}).
\begin{table}[!h] 
 \centering
  \begin{displaymath}
  \begin{array}{l|ll}
                & \text{Re}        & \text{Im}        \\ \hline \hline
   \text{General} & 9N(N-1)(N-2) & 9N(N-1)(N-2) \\ \hline 
   \mcl{O}^H    & (3/2)N(N-1)(N-2) & (3/2)N(N-1)(N-2) \\ 
   \mcl{O}^T    & (3/2)N(N-1)(N-2) & (3/2)N(N-1)(N-2) \\ 
   \mcl{O}^r    &     3N(N-1)(N-2) &     3N(N-1)(N-2) \\
   \mcl{O}^{HT} &     3N(N-1)(N-2) &     3N(N-1)(N-2) \\ \hline \hline
   \text{Nonlinear} & 3N(N-1)(N-2) & 3N(N-1)(N-2) 
  \end{array} 
  \end{displaymath}
 \caption{
  The number of the independent dimension-six derivative interactions for type IV.
  The first and second columns mean real and imaginary \dof, respectively.
  We show the number of \dof \ for the case 
  where $SU(2)_L \times U(1)_Y$ invariance is imposed (General)
  and the case with nonlinear realization (Nonlinear). 
  In the case of General, the \dof \ of each type of operators is also shown.
 } 
 \label{tb:typeIV}
\end{table}
\subsection{Type V}

Type V derivative interactions are described by the Lagrangian below:
\begin{align}
 \mcl{L}^6_\text{V} =&
   \frac{\la_{ijkl}^H}{\La^2} O^H_{ijkl}
  +\frac{\la_{ijlk}^H}{\La^2} O^H_{ijlk}
  +\frac{\la_{ikjl}^H}{\La^2} O^H_{ikjl}
  +\frac{\la_{iklj}^H}{\La^2} O^H_{iklj}
  +\frac{\la_{iljk}^H}{\La^2} O^H_{iljk}
  +\frac{\la_{ilkj}^H}{\La^2} O^H_{ilkj} \n &
  +\frac{\la_{ijkl}^r}{\La^2} O^r_{ijkl}
  +\frac{\la_{ijlk}^r}{\La^2} O^r_{ijlk}
  +\frac{\la_{ikjl}^r}{\La^2} O^r_{ikjl}
  +\frac{\la_{iklj}^r}{\La^2} O^r_{iklj}
  +\frac{\la_{iljk}^r}{\La^2} O^r_{iljk}
  +\frac{\la_{ilkj}^r}{\La^2} O^r_{ilkj} \n &
  +\frac{\la_{jkil}^r}{\La^2} O^r_{jkil}
  +\frac{\la_{jkli}^r}{\La^2} O^r_{jkli}
  +\frac{\la_{jlik}^r}{\La^2} O^r_{jlik}
  +\frac{\la_{jlki}^r}{\La^2} O^r_{jlki}
  +\frac{\la_{klij}^r}{\La^2} O^r_{klij}
  +\frac{\la_{klji}^r}{\La^2} O^r_{klji} \n &
  +\frac{\la_{ijkl}^T}{\La^2} O^T_{ijkl}
  +\frac{\la_{ijik}^T}{\La^2} O^T_{ijik}
  +\frac{\la_{ikjl}^T}{\La^2} O^T_{ikjl}
  +\frac{\la_{iklj}^T}{\La^2} O^T_{iklj}
  +\frac{\la_{iljk}^T}{\La^2} O^T_{iljk}
  +\frac{\la_{ilkj}^T}{\La^2} O^T_{ilkj} \n &
  +\frac{\la_{ijkl}^{HT}}{\La^2} O^{HT}_{ijkl}
  +\frac{\la_{ijlk}^{HT}}{\La^2} O^{HT}_{ijlk}
  +\frac{\la_{ikjl}^{HT}}{\La^2} O^{HT}_{ikjl}
  +\frac{\la_{iklj}^{HT}}{\La^2} O^{HT}_{iklj}
  +\frac{\la_{iljk}^{HT}}{\La^2} O^{HT}_{iljk}
  +\frac{\la_{ilkj}^{HT}}{\La^2} O^{HT}_{ilkj} \n &
  +\frac{\la_{jkil}^{HT}}{\La^2} O^{HT}_{jkil}
  +\frac{\la_{jkli}^{HT}}{\La^2} O^{HT}_{jkli}
  +\frac{\la_{jlik}^{HT}}{\La^2} O^{HT}_{jlik}
  +\frac{\la_{jlki}^{HT}}{\La^2} O^{HT}_{jlki}
  +\frac{\la_{klij}^{HT}}{\La^2} O^{HT}_{klij}
  +\frac{\la_{klji}^{HT}}{\La^2} O^{HT}_{klji}
  +\hc \ , 
\end{align}
where $i<j<k<l$.
The Lagrangian has $3N(N-1)(N-2)(N-3)$ \dof, and 
real (imaginary) \dof \ is a half of them.
Similarly to previous sections,
let us study the derivative interactions with the constraint of 
the nonlinear realization:
\begin{align}
 \mcl{L}^\text{6NL}_\text{V} =&
   \frac{1}{f^2} \mcl{T}_{abcd}^\text{V}\,
	 h^a h^b (\del_\mu h^c )(\del^\mu h^d ),
\end{align}
where
\begin{align}
 {\cal T}_{abcd}^\text{V} =& 
	2a^L_{ijkl} 
	\left( T^{L\al}_{(i,j)} \right)_{ac} 
	\left( T^{L\al}_{(k,l)} \right)_{bd}
  +2a^L_{ikjl} 
   \left( T^{L\al}_{(i,k)} \right)_{ac} 
	\left( T^{L\al}_{(j,l)} \right)_{bd}
  +2a^L_{iljk} 
   \left( T^{L\al}_{(i,l)} \right)_{ac} 
	\left( T^{L\al}_{(j,k)} \right)_{bd} \n &
  +2a^R_{ijkl} 
   \left( T^{R\be}_{(i,j)} \right)_{ac} 
	\left( T^{R\be}_{(k,l)} \right)_{bd}
  +2a^R_{ikjl} 
   \left( T^{R\be}_{(i,k)} \right)_{ac} 
	\left( T^{R\be}_{(j,l)} \right)_{bd}
  +2a^R_{iljk} 
   \left( T^{R\be}_{(i,l)} \right)_{ac} 
	\left( T^{R\be}_{(j,k)} \right)_{bd} \n &
  +2a^S_{ijkl} 
   \left( S^{\al 3}_{(i,j)} \right)_{ac} 
	\left( S^{\al 3}_{(k,l)} \right)_{bd}
  +2a^S_{ikjl} 
   \left( S^{\al 3}_{(i,k)} \right)_{ac} 
	\left( S^{\al 3}_{(j,l)} \right)_{bd}
  +2a^S_{iljk} 
   \left( S^{\al 3}_{(i,l)} \right)_{ac} 
	\left( S^{\al 3}_{(j,k)} \right)_{bd} \n &
  +2a^{SS}_{ijkl} 
   \left( S^{\al \be}_{(i,j)} \right)_{ac} 
	\left( S^{\al \be}_{(k,l)} \right)_{bd}
  +2a^{SS}_{ikjl}     
   \left( S^{\al \be}_{(i,k)} \right)_{ac} 
	\left( S^{\al \be}_{(j,l)} \right)_{bd}
  +2a^{SS}_{iljk}     
   \left( S^{\al \be}_{(i,l)} \right)_{ac} 
	\left( S^{\al \be}_{(j,k)} \right)_{bd} \n &
  +2a^{LS}_{ijkl} 
   \left( T^{L \al}_{(i,j)} \right)_{ac} 
	\left( S^{\al 3}_{(k,l)} \right)_{bd} 
  +2a^{LS}_{ikjl} 
   \left( T^{L \al}_{(i,k)} \right)_{ac} 
	\left( S^{\al 3}_{(j,l)} \right)_{bd} 
  +2a^{LS}_{iljk} 
   \left( T^{L \al}_{(i,l)} \right)_{ac} 
	\left( S^{\al 3}_{(j,k)} \right)_{bd} \n &
  +2a^{LS}_{klij} 
   \left( T^{L \al}_{(k,l)} \right)_{ac} 
	\left( S^{\al 3}_{(i,j)} \right)_{bd} 
  +2a^{LS}_{jlik} 
   \left( T^{L \al}_{(j,l)} \right)_{ac} 
	\left( S^{\al 3}_{(i,k)} \right)_{bd} 
  +2a^{LS}_{jkil} 
   \left( T^{L \al}_{(j,k)} \right)_{ac} 
	\left( S^{\al 3}_{(i,l)} \right)_{bd} \n &
  +2a^Y_{ijkl} 
   \left( T^{R3}_{(i,j)} \right)_{ac} 
	\left( T^{R3}_{(k,l)} \right)_{bd}
  +2a^Y_{ikjl} 
   \left( T^{R3}_{(i,k)} \right)_{ac} 
	\left( T^{R3}_{(j,l)} \right)_{bd}
  +2a^Y_{iljk} 
   \left( T^{R3}_{(i,l)} \right)_{ac} 
	\left( T^{R3}_{(j,k)} \right)_{bd} \n &
  +2a^U_{ijkl} 
   \left( U_{(i,j)} \right)_{ac} 
	\left( U_{(k,l)} \right)_{bd}
  +2a^U_{ikjl} 
   \left( U_{(i,k)} \right)_{ac} 
	\left( U_{(j,l)} \right)_{bd}
  +2a^U_{iljk} 
   \left( U_{(i,l)} \right)_{ac} 
	\left( U_{(j,k)} \right)_{bd}  \n &
  +2a^{YU}_{ijkl} 
   \left( T^{R3}_{(i,j)} \right)_{ac} 
	\left( U_{(k,l)} \right)_{bd} 
  +2a^{YU}_{ikjl} 
   \left( T^{R3}_{(i,k)} \right)_{ac} 
	\left( U_{(j,l)} \right)_{bd} 
  +2a^{YU}_{iljk} 
   \left( T^{R3}_{(i,l)} \right)_{ac} 
	\left( U_{(j,k)} \right)_{bd} \n &
  +2a^{YU}_{klij} 
   \left( T^{R3}_{(k,l)} \right)_{ac} 
	\left( U_{(i,j)} \right)_{bd} 
  +2a^{YU}_{jlik} 
   \left( T^{R3}_{(j,l)} \right)_{ac} 
	\left( U_{(i,k)} \right)_{bd} 
  +2a^{YU}_{jkil} 
   \left( T^{R3}_{(j,k)} \right)_{ac} 
	\left( U_{(i,l)} \right)_{bd}.
\end{align}
Finally, we obtain the derivative interactions of the Higgs bosons 
realized as the NG bosons below:
\begin{align}
 \mcl{L}^\text{6NL}_\text{V} =&
   \frac{c^{\text{V}1}_{[ijkl]}}{2f^2} O^H_{ijkl}
  +\frac{c^{\text{V}2}_{[ijkl]}}{2f^2} O^H_{ijlk}
  +\frac{c^{\text{V}3}_{[ijkl]}}{2f^2} O^H_{ikjl}
  +\frac{c^{\text{V}4}_{[ijkl]}}{2f^2} O^H_{iklj}
  +\frac{c^{\text{V}5}_{[ijkl]}}{2f^2} O^H_{iljk}
  +\frac{c^{\text{V}6}_{[ijkl]}}{2f^2} O^H_{ilkj} \n &
  -\frac{c^{\text{V}1}_{[ijkl]}}{ f^2} ( O^r_{ijkl} +O^r_{klij} )
  -\frac{c^{\text{V}2}_{[ijkl]}}{ f^2} ( O^r_{ijlk} +O^r_{klji} )
  -\frac{c^{\text{V}3}_{[ijkl]}}{ f^2} ( O^r_{ikjl} +O^r_{jlik} ) \n &
  -\frac{c^{\text{V}4}_{[ijkl]}}{ f^2} ( O^r_{iklj} +O^r_{jlki} )
  -\frac{c^{\text{V}5}_{[ijkl]}}{ f^2} ( O^r_{iljk} +O^r_{jkil} )
  -\frac{c^{\text{V}6}_{[ijkl]}}{ f^2} ( O^r_{ilkj} +O^r_{jkli} ) \n &
  +\frac{c^{\text{V}7}_{[ijkl]}}{2f^2} O^T_{ijkl}
  +\frac{c^{\text{V}8}_{[ijkl]}}{2f^2} O^T_{ijlk}
  +\frac{c^{\text{V}9}_{[ijkl]}}{2f^2} O^T_{ikjl}
  +\frac{c^{\text{V}10}_{[ijkl]}}{2f^2} O^T_{iklj}
  +\frac{c^{\text{V}11}_{[ijkl]}}{2f^2} O^T_{iljk}
  +\frac{c^{\text{V}12}_{[ijkl]}}{2f^2} O^T_{ilkj} +\hc \ ,
\end{align}
\begin{align}
 c^{\text{V}1}_{[ijkl]} =&
   a^L_{iljk} +a^R_{ikjl} +a^S_{iljk} +a^{SS}_{ikjl} +a^{SS}_{iljk} 
  +i\left( a^{LS}_{iljk} -a^{LS}_{jkil} \right),\\
 c^{\text{V}2}_{[ijkl]} =&
   a^L_{ikjl} +a^R_{iljk} +a^S_{ikjl} +a^{SS}_{ikjl} +a^{SS}_{iljk} 
  +i\left( a^{LS}_{ikjl} -a^{LS}_{jlik} \right),\\
 c^{\text{V}3}_{[ijkl]} =&
   a^L_{iljk} +a^R_{ijkl} -a^S_{iljk} +a^{SS}_{ijkl} -a^{SS}_{iljk}
  -i\left( a^{LS}_{iljk} +a^{LS}_{jkil} \right),\\
 c^{\text{V}4}_{[ijkl]} =&
   a^L_{ijkl} +a^R_{iljk} +a^S_{ijkl} +a^{SS}_{ijkl} -a^{SS}_{iljk}
  +i\left( a^{LS}_{ijkl} -a^{LS}_{klij} \right),\\
 c^{\text{V}5}_{[ijkl]} =&
   a^L_{ikjl} +a^R_{ijkl} -a^S_{ikjl} -a^{SS}_{ijkl} -a^{SS}_{ikjl}
  -i\left( a^{LS}_{ikjl} +a^{LS}_{jlik} \right),\\
 c^{\text{V}6}_{[ijkl]} =&
   a^L_{ijkl} +a^R_{ikjl} -a^S_{ijkl} -a^{SS}_{ijkl} -a^{SS}_{ikjl}
  -i\left( a^{LS}_{ijkl} +a^{LS}_{klij} \right),\\
 c^{\text{V}7}_{[ijkl]} =&
  -a^L_{ijkl} +a^L_{iljk} +a^R_{ijkl} -a^R_{ikjl} +a^Y_{ijkl} -a^U_{ijkl} 
  +a^S_{ijkl} +a^S_{iljk} +a^{SS}_{ijkl} -a^{SS}_{ikjl} +a^{SS}_{iljk} \n &
  +i\left( -a^{YU}_{ijkl} -a^{YU}_{klij} +a^{LS}_{ijkl} +a^{LS}_{iljk} 
           +a^{LS}_{klij} -a^{LS}_{jkil} \right),\\
 c^{\text{V}8}_{[ijkl]} =&
  -a^L_{ijkl} +a^L_{ikjl} +a^R_{ijkl} -a^R_{iljk} +a^Y_{ijkl} +a^U_{ijkl} 
  -a^S_{ijkl} +a^S_{ikjl} -a^{SS}_{ijkl} +a^{SS}_{ikjl} -a^{SS}_{iljk} \n &
  +i\left( a^{YU}_{ijkl} -a^{YU}_{klij} -a^{LS}_{ijkl} -a^{LS}_{ikjl} 
		    +a^{LS}_{klij} -a^{LS}_{jlik} \right),\\
 c^{\text{V}9}_{[ijkl]} =&
  -a^L_{ikjl} +a^L_{iljk} -a^R_{ijkl} +a^R_{ikjl} +a^Y_{ikjl} -a^U_{ikjl} 
  +a^S_{ikjl} -a^S_{iljk} -a^{SS}_{ijkl} +a^{SS}_{ikjl} -a^{SS}_{iljk} \n &
  +i\left( -a^{YU}_{ikjl} -a^{YU}_{jlik} +a^{LS}_{ikjl} -a^{LS}_{iljk} 
		     +a^{LS}_{jlik} -a^{LS}_{jkil} \right),\\
 c^{\text{V}10}_{[ijkl]} =&
   a^L_{ijkl} -a^L_{ikjl} +a^R_{ikjl} -a^R_{iljk} +a^Y_{ikjl} +a^U_{ikjl} 
  -a^S_{ijkl} -a^S_{ikjl} +a^{SS}_{ijkl} -a^{SS}_{ikjl} +a^{SS}_{iljk} \n &
  +i\left( a^{YU}_{ikjl} -a^{YU}_{jlik} +a^{LS}_{ijkl} -a^{LS}_{ikjl} 
			 -a^{LS}_{klij} +a^{LS}_{jlik} \right),\\
 c^{\text{V}11}_{[ijkl]} =&
   a^L_{ikjl} -a^L_{iljk} -a^R_{ijkl} +a^R_{iljk} +a^Y_{iljk} -a^U_{iljk} 
  -a^S_{ikjl} +a^S_{iljk} +a^{SS}_{ijkl} -a^{SS}_{ikjl} +a^{SS}_{iljk} \n &
  +i\left( -a^{YU}_{iljk} -a^{YU}_{jkil} -a^{LS}_{ikjl} +a^{LS}_{iljk} 
			  -a^{LS}_{jlik} +a^{LS}_{jkil} \right),\\
 c^{\text{V}12}_{[ijkl]} =&
   a^L_{ijkl} -a^L_{iljk} -a^R_{ikjl} +a^R_{iljk} +a^Y_{iljk} +a^U_{iljk} 
  -a^S_{ijkl} -a^S_{iljk} -a^{SS}_{ijkl} +a^{SS}_{ikjl} -a^{SS}_{iljk} \n &
  +i\left( a^{YU}_{iljk} -a^{YU}_{jkil} -a^{LS}_{ijkl} -a^{LS}_{iljk} 
			 -a^{LS}_{klij} +a^{LS}_{jkil} \right).
\end{align}
There are $(1/2)N(N-1)(N-2)(N-3)$ real \dof, and
the imaginary \dof \  is the same as the real one.
The \dof \  becomes one third of original one 
in either case (Table~\ref{tb:typeV}).
\begin{table}[!h] 
 \centering
  \begin{displaymath}
  \begin{array}{l|ll}
                & \text{Re}             & \text{Im}          \\ \hline \hline 
   \text{General} & (3/2)N(N-1)(N-2)(N-3) & (3/2)N(N-1)(N-2)(N-3) \\ \hline
   \mcl{O}^H    & (1/4)N(N-1)(N-2)(N-3) & (1/4)N(N-1)(N-2)(N-3) \\ 
   \mcl{O}^T    & (1/4)N(N-1)(N-2)(N-3) & (1/4)N(N-1)(N-2)(N-3) \\ 
   \mcl{O}^r    & (1/2)N(N-1)(N-2)(N-3) & (1/2)N(N-1)(N-2)(N-3) \\
   \mcl{O}^{HT} & (1/2)N(N-1)(N-2)(N-3) & (1/2)N(N-1)(N-2)(N-3) \\ \hline \hline
   \text{Nonlinear} & (1/2)N(N-1)(N-2)(N-3) & (1/2)N(N-1)(N-2)(N-3)
  \end{array} 
  \end{displaymath}
 \caption{
  The number of the independent dimension-six derivative interactions for type V.
  The first and second columns mean real and imaginary \dof, respectively.
  We show the number of \dof \ for the case 
  where $SU(2)_L \times U(1)_Y$ invariance is imposed (General)
  and the case with nonlinear realization (Nonlinear). 
  In the case of General, the \dof \ of each type of operators is also shown.
 } 
 \label{tb:typeV}
\end{table}

As a conclusion of this section, 
we discuss the number of independent derivative interactions 
in the case of the NHDM.
Summing up the \dof \  given by this section, 
there are $(1/2)N^2(N^2 +3)$ real and $(1/2)N^2(N^2-1)$ imaginary
\dof \  for models where Higgs doublets are generated by nonlinear dynamics,
while general NHDM have $(3/2)N^2(N^2 +1)$ real and $(1/2)N^2(3N^2 -1)$
imaginary coefficients (Table~\ref{tb:summary}).
If $N$ is large enough, 
the \dof \  is about one third compared with the original one.

\begin{table}[!h] 
 \centering
 \begin{tabular}{l|ll}
             & Re                 & Im                     \\ \hline \hline
   General   & $(3/2)N^2(N^2 +1)$ & $(1/2)N^2(3N^2-1)$     \\ 
   Nonlinear & $(1/2)N^2(N^2 +3)$ & $(1/2)N^2( N^2-1)$ 
 \end{tabular} 
 \caption{
  The number of the independent dimension-six derivative operators 
  for NHDMs.
  The first and second columns mean real and imaginary \dof, respectively. 
  The first and second raws correspond to the models 
  imposing only $SU(2)_L \times U(1)_Y$ and nonlinear dynamics, respectively.
 } 
 \label{tb:summary}
\end{table}
\section{Application to two Higgs doublet models} 
\label{sec:phenomenology}

In the previous section, we have derived the general expression 
of the dimension-six derivative interactions for the NHDM, 
assuming the Higgs bosons as PNGBs.
Its phenomenological consequences are studied 
for the 2HDM in this section as an explicit example.
Many models which include the two Higgs doublets 
as NG fields are proposed~\cite{
ArkaniHamed:2002qx,Mrazek:2011iu}.
In particular, $SO(4)$ invariant dimension-six derivative interactions 
are examined in Ref.~\cite{Mrazek:2011iu}.

We discuss the scalar four point interactions with two derivatives,
$\mcl{O}^H,\mcl{O}^T$ and $\mcl{O}^r$.
The operator, $\mcl{O}^{HT}$, does not appear 
in the case of the nonlinear dynamics.
Since the Higgs doublets include 
the longitudinal modes of the gauge bosons 
and the physical Higgs bosons, 
the interactions contribute to the vector boson fusion processes 
in high energy region.
As we show in the following, the rising behavior of
the amplitudes is determined by these derivative 
operators in high energy region 
if we neglect $\mcl{O}(v^2/f^2)$ corrections from the EWSB.

Finally, 
cross sections of vector boson fusion processes are presented 
in the case where the custodial symmetry is imposed.
We show that a relation between cross sections of 
$W_L^+ W_L^- \to W_L^+ W_L^-$ and $W_L^+ W_L^- \to h h$ 
in the SILH is violated by additional parameters.
\subsection{Dimension-six derivative interactions} 

The derivative interactions on the 2HDM governed by nonlinear dynamics can 
be written using interactions of type I, II and III:
\begin{align}
  {\cal T}^{abcd}_\text{2HDM} &= 
   a^L_{1111} 
	\left( T_{(1,1)}^{L\al} \right)_{ac} 
	\left( T_{(1,1)}^{L\al} \right)_{bd} 
  +2a^L_{1112}
   \left( T_{(1,1)}^{L\al} \right)_{ac} 
	\left( T_{(1,2)}^{L\al} \right)_{bd}  
  +2a^L_{1122}
   \left( T_{(1,1)}^{L\al} \right)_{ac} 
	\left( T_{(2,2)}^{L\al} \right)_{bd} \n &
  +a^L_{1212} 
   \left( T_{(1,2)}^{L\al} \right)_{ac} 
	\left( T_{(1,2)}^{L\al} \right)_{bd}
  +2a^L_{2221}
   \left( T_{(2,2)}^{L\al} \right)_{ac} 
	\left( T_{(2,1)}^{L\al} \right)_{bd}
  +a^L_{2222} 
   \left( T_{(2,2)}^{L\al} \right)_{ac} 
	\left( T_{(2,2)}^{L\al} \right)_{bd} \n &  
  +a^R_{1111} 
	\left( T_{(1,1)}^{R\al} \right)_{ac} 
	\left( T_{(1,1)}^{R\al} \right)_{bd} 
  +2a^R_{1112}
   \left( T_{(1,1)}^{R\al} \right)_{ac} 
	\left( T_{(1,2)}^{R\al} \right)_{bd}  
  +2a^R_{1122}
   \left( T_{(1,1)}^{R\al} \right)_{ac} 
	\left( T_{(2,2)}^{R\al} \right)_{bd} \n &
  +a^R_{1212} 
   \left( T_{(1,2)}^{R\al} \right)_{ac} 
	\left( T_{(1,2)}^{R\al} \right)_{bd}
  +2a^R_{2221}
   \left( T_{(2,2)}^{R\al} \right)_{ac} 
	\left( T_{(2,1)}^{R\al} \right)_{bd}
  +a^R_{2222} 
   \left( T_{(2,2)}^{R\al} \right)_{ac} 
	\left( T_{(2,2)}^{R\al} \right)_{bd} \n &  
  +a^S_{1212}
   \left( S_{(1,2)}^{\al 3} \right)_{ac} 
	\left( S_{(1,2)}^{\al 3} \right)_{bd}
  +a^{SS}_{1212}
   \left( S_{(1,2)}^{\al \be} \right)_{ac} 
	\left( S_{(1,2)}^{\al \be} \right)_{bd}
  +2a^{LS}_{1112}  
   \left( T_{(1,1)}^{L\al} \right)_{ac} 
	\left( S_{(1,2)}^{\al 3} \right)_{bd} \n &
  +a^{LS}_{1212} 
   \left( T_{(1,2)}^{L\al} \right)_{ac} 
	\left( S_{(1,2)}^{\al 3} \right)_{bd}
  +2a^{LS}_{2221} 
   \left( T_{(2,2)}^{L\al} \right)_{ac} 
	\left( S_{(2,1)}^{\al 3} \right)_{bd} 
  +2a^Y_{1111} 
   \left( T_{(1,1)}^{R3} \right)_{ac} 
	\left( T_{(1,1)}^{R3} \right)_{bd} \n &
  +4a^Y_{1112} 
   \left( T_{(1,1)}^{R3} \right)_{ac} 
	\left( T_{(1,2)}^{R3} \right)_{bd}
  +2a^Y_{1122} 
   \left( T_{(1,1)}^{R3} \right)_{ac} 
	\left( T_{(2,2)}^{R3} \right)_{bd}
  +a^Y_{1212} 
   \left( T_{(1,2)}^{R3} \right)_{ac} 
	\left( T_{(1,2)}^{R3} \right)_{bd} \n & 
  +4a^Y_{2212} 
   \left( T_{(2,2)}^{R3} \right)_{ac} 
	\left( T_{(1,2)}^{R3} \right)_{bd}
  +2a^Y_{2222} 
   \left( T_{(2,2)}^{R3} \right)_{ac} 
	\left( T_{(2,2)}^{R3} \right)_{bd}
  +a^U_{1212} 
   \left( U_{(1,2)} \right)_{ac} 
	\left( U_{(1,2)} \right)_{bd} \n &
  +4a^{YU}_{1112}   
   \left( T_{(1,1)}^{R3} \right)_{ac} 
	\left( U_{(1,2)} \right)_{bd}
  +a^{YU}_{1212} 
   \left( T_{(1,2)}^{R3} \right)_{ac} 
	\left( U_{(1,2)} \right)_{bd}
  +4a^{YU}_{2212}  
   \left( T_{(2,2)}^{R3} \right)_{ac} 
	\left( U_{(1,2)} \right)_{bd}.
\label{eq:2HDM_T}
\end{align}
In the following, 
we assume that the Lagrangian possesses the custodial invariance
which automatically ensures 
the $CP$ invariance~\cite{Grzadkowski:2010dj}\footnote{
In Ref.~\cite{Grzadkowski:2010dj},
it is pointed out the custodial symmetry of the Higgs potential 
is violated by the imaginary part of $H^\dag_i H_j$. 
Since the symmetry in the derivative interactions is violated 
by the imaginary part of $H^\dag_i \del H_j$, 
their discussion can be applied to this case 
by the replacement of 
$\partial H_k$ with $H_l$ ($k \neq l$).
}.
From the phenomenological standpoint, 
the custodial symmetry breaking terms 
are severely constrained. 
For the case with custodial symmetry violating terms, 
see Appendix \ref{app:amp}. 
The $SO(4)$ invariance is achieved by the following conditions:
\begin{align}
 a^{YU}_{1112} &= 0, \quad
 a^{YU}_{1212}  = 0, \quad
 a^{YU}_{2212} = 0, \\
 a^{LS}_{1112} &= 0, \quad
 a^{LS}_{1212} = 0, \quad
 a^{LS}_{2212} = 0, \\
 a^Y_{1111} &= 0, \quad
 a^Y_{1112} = 0, \quad
 a^Y_{2212} = 0, \quad
 a^Y_{2222} = 0, \\
 a^S_{1212} &= a^Y_{1212} = - \frac{a^Y_{1122}}{2}.
 \label{eq:SO(4)_condition}
\end{align}
Note that there exists a nontrivial realization of the $SO(4)$ invariance 
even if each term corresponding to the coefficient appearing in
Eq.~\eqref{eq:SO(4)_condition} 
violates the $SO(4)$ symmetry. 

In terms of the $SU(2)_L$ doublet,
since $\mcl{O}^r$ can be eliminated by the field redefinition 
(see Appendix~\ref{app:eom}),
we obtain the following Lagrangian:
\begin{align}
 {\cal L}^6_\text{2HDM} =& 
   \frac{c^H_{1111}}{2f^2}  O^H_{1111} 
  +\frac{c^H_{1112}}{ f^2}( O^H_{1112} +O^H_{1121} ) 
  +\frac{c^H_{1122}}{ f^2}  O^H_{1122}
  +\frac{c^H_{1221}}{ f^2}  O^H_{1221} \n &
  +\frac{c^H_{1212}}{2f^2}( O^H_{1212} +O^H_{2121} )
  +\frac{c^H_{2221}}{ f^2}( O^H_{2212} +O^H_{2221} )
  +\frac{c^H_{2222}}{2f^2}  O^H_{2222}  \n &
  +\frac{c^T_{1122}}{ f^2}  O^T_{1122}
  +\frac{c^T_{1221}}{ f^2}  O^T_{1221} 
  +\frac{c^T_{1212}}{2f^2}( O^T_{1212} +O^T_{2121} ).
  \label{eq:2HDM_nlsm_Lagrangian}
\end{align}
The relations between the above coefficients and 
the ones in Eq.~\eqref{eq:2HDM_T} are
\begin{align}
 c^H_{1111} =&
   \frac{3}{2} (a^L_{1111} +a^R_{1111}) ,\\
 c^H_{1112} =&
   \frac{3}{2} (a^L_{1112} +a^R_{1112}) ,\\
 c^H_{1122} =&
   \frac{3}{2} (a^L_{1212} +a^R_{1212} +a^S_{1212} +2a^{SS}_{1212}) ,\\
 c^H_{1221} =&
   \frac{3}{2} (a^L_{1122} +a^R_{1212} -a^{SS}_{1212}) ,\\
 c^H_{1212} =&
   \frac{3}{2} (a^L_{1212} +a^R_{1122} -a^S_{1212} -a^{SS}_{1212}) ,\\
 c^H_{2221} =&
   \frac{3}{2} (a^L_{2221} +a^R_{2221}) ,\\
 c^H_{2222} =&
   \frac{3}{2} (a^L_{2222} +a^R_{2222}) ,\\
 c^T_{1122} =&
   \frac{1}{2} (-a^L_{1122} +a^L_{1212} 
   +a^R_{1122} -a^R_{1212} -a^S_{1212}) ,\\
 c^T_{1221} =&
   \frac{1}{2} (-a^L_{1212} +a^L_{1122} +a^U_{1212}) ,\\
 c^T_{1212} =& 
   \frac{1}{2} (a^R_{1212} -a^R_{1122} +a^S_{1212} -a^U_{1212} ).
\end{align}
The other coefficients vanish.
In the above equations, the following relations are obtained:
\begin{align}
  c^T_{1122} = - (c^T_{1221} + c^T_{1212}) 
   = - \frac{1}{3} (c^H_{1221} -c^H_{1212}).
 \label{eq:relation_of_coeff}
\end{align}
Therefore, the Lagrangian discussed here 
has eight real \dof.
This is equivalent to a result in Ref.~\cite{Mrazek:2011iu}.
On the other hand,
using equations in Table~\ref{tb:summary},
the general 2HDM realized by nonlinear dynamics 
without the $SO(4)$ symmetry 
has 14 real and six imaginary \dof.
For the following discussions, 
we eliminate $c^T_{1122}$ and $c^T_{1212}$ 
using Eq.~\eqref{eq:relation_of_coeff}.

We first discuss the kinetic term mixing 
induced by the dimension-six derivative interactions 
and show that this mixing can be neglected 
in deriving leading behavior of the scattering amplitudes 
at high energy.

After the EWSB, the kinetic term mixing 
appears in the Lagrangian for neutral Higgs bosons 
at $\mcl{O}(v^2/f^2)$.
Since we assume the $CP$ invariance, the mixing appears 
for the $CP$ even and odd sector separately.
We show the prescription for the mixing 
with the $CP$ even sector as an example.
The kinetic term and the mass term of the $CP$ even sector 
after the EWSB can be written as
\begin{align}
 \mcl{L}_K =&
  \frac{1}{2}
  \begin{pmatrix}
   \del^\mu h^3_0 & \del^\mu h^7_0
  \end{pmatrix}
  \left( \mathbf{1}_2 + C_K \right)
  \begin{pmatrix}
   \del_\mu h^3_0 \\ \del_\mu h^7_0
  \end{pmatrix}
 -
  \frac{1}{2}
  \begin{pmatrix}
   h^3_0 & h^7_0
  \end{pmatrix}
  M
  \begin{pmatrix}
   h^3_0 \\ h^7_0
  \end{pmatrix},
 \label{eq:KT&mass}
\end{align}
where $h^3 = v\cos \be +h^3_0$, $h^7 = v\sin \be +h^7_0$ 
and the matrix $C_K$ stands for the $\mcl{O}(v^2/f^2)$ corrections.
The mass matrix, $M$, includes the effects 
from dimension-six Higgs potentials.
The following matrices are introduced to 
make the kinetic term to be the canonical form:
\begin{align}
 V = 
 \begin{pmatrix}
   \cos \th & \sin \th \\
  -\sin \th & \cos \th
 \end{pmatrix},&\qquad
 K = 
 \begin{pmatrix}
  1/\sqrt{1 +k_1} & 0 \\
  0 & 1/\sqrt{1 +k_2} 
 \end{pmatrix},
\end{align}
such that,
\begin{align}
 V C_K V^\dag = 
 \begin{pmatrix}
  k_1 & 0 \\ 0 & k_2
 \end{pmatrix}.
\end{align}
Mass eigenstates of the $CP$ even Higgs bosons, 
$h$ and $H$ ($m_h \leq m_H$), can be written as
\begin{align}
 \begin{pmatrix}
  h \\ H
 \end{pmatrix}
 =&
 U V^\dag K^{-1} V
 \begin{pmatrix}
  h^3_0 \\ h^7_0
 \end{pmatrix},
\end{align}
where $U$ diagonalizes the mass matrix,
$V^\dag KV M V^\dag K V$, in the basis 
giving the canonical kinetic term.
Using the mixing angle $\al$ which diagonalizes 
the mass matrix $M$ in Eq.~\eqref{eq:KT&mass}, 
we can write
\begin{align}
 U =&
 \begin{pmatrix}
   \cos \al  & \sin \al  \\
   -\sin \al  & \cos \al 
 \end{pmatrix}  + U_\xi, 
 \label{eq:diag_U} \\
   V^\dag KV =&
   \mathbf{1}_2 + K_\xi,
 \label{eq:VKV} 
\end{align}
where $U_\xi$ and $K_\xi$ are $\mcl{O}(v^2/f^2)$ corrections 
given by the kinetic mixing.
When we neglect $\mcl{O}(v^2/f^2)$ corrections,
the mass eigenstates for the $CP$ even sector are given 
as follows\footnote{This is a different notation 
from Ref.~\cite{Gunion} 
in which CP-even mass eigenstates are defined as 
$H = \cos \al \, h^3_0 + \sin \al \, h^7_0$ and 
$h = - \sin \al \, h^3_0 + \cos \al \, h^7_0$.}
:
\begin{align}
 \begin{pmatrix}
  h \\ H
 \end{pmatrix}
 =&
 \begin{pmatrix}
   \cos \al & \sin \al \\
  -\sin \al & \cos \al
 \end{pmatrix}
 \begin{pmatrix}
  h^3_0 \\ h^7_0
 \end{pmatrix}.
\end{align}
As we mentioned, the dimension-six derivative interactions,
Eq.~\eqref{eq:2HDM_nlsm_Lagrangian}, 
give the leading behavior of $E^2 / f^2$, and 
the kinetic mixing only induces the correction of $\mcl{O}(v^2/f^2)$.

Mass eigenstates of the $CP$ odd sector, $G^0$ and $A$, can be obtained 
similarly to the $CP$ even sector 
by replacing $\al$ with $\be$ 
for the diagonalization matrix at the leading order:
\begin{align}
 \begin{pmatrix}
  G^0 \\ A
 \end{pmatrix}
 =&
 \begin{pmatrix}
   \cos \be & \sin \be \\
  -\sin \be & \cos \be
 \end{pmatrix}
 \begin{pmatrix}
  h^4 \\ h^8
 \end{pmatrix},
\end{align}
where $G^0$ is the longitudinal mode of $Z$ boson 
and $A$ is the $CP$ odd Higgs boson.
The matrix also gives us the mass eigenstates 
in the charged Higgs sector
even if we take the $\mcl{O}(v^2/f^2)$ corrections into consideration
because the dimension-six derivative interactions never contribute to
the kinetic term of the charged Higgs sector
after the field redefinition.
\subsection{Scattering amplitudes of the longitudinal modes and the Higgs bosons}
\label{subsec:amp}
We derive the scattering amplitudes of the longitudinal modes 
and the Higgs bosons 
from dimension-six two-derivative interactions
in the 2HDM governed by nonlinear dynamics 
with the custodial symmetry. 
Other contributions are neglected 
since they are subleading corrections 
in high energy region.
To study the influence on vector boson fusion processes,
we consider the case that the initial states consist of 
longitudinal modes of the SM gauge bosons, 
$W_L^\pm$ and $Z_L$.

The amplitudes shown below 
are given in terms of the Mandelstam variables.
For a process, $V_1 V_2 \to X_1 X_2$, 
the variables are defined as,
\begin{align}
  s =& (p_1 +p_2)^2 = (q_1 +q_2)^2, 
 \label{eq:mandelstam_s} \\
  t =& (q_1 -p_1)^2 = (q_2 -p_2)^2,
 \label{eq:mandelstam_t} \\
  u =& (q_2 -p_1)^2 = (q_1 -p_2)^2,
 \label{eq:mandelstam_u}
\end{align}
where 
$p_1^\mu$, $p_2^\mu$, $q_1^\mu$ and $q_2^\mu$ 
are momenta of the particles 
$V_1$, $V_2$, $X_1$ and $X_2$, 
respectively.
Since amplitudes are considered in the energy region 
which is much higher than the mass scale 
of the physical Higgs bosons,
we show the results in terms of $s$ and $t$ 
using a relation, $s+t+u=0$.
The results of this subsection are based on the custodial invariance.
In Appendix~\ref{app:amp}, we also calculate 
scattering amplitudes including terms violating the custodial symmetry.

Firstly, amplitudes producing the SM particles, 
$W_L^\pm$, $Z_L$ and $h$, are presented.
These processes also appear in the SILH model.
In the following amplitudes,
coefficients are given in Eq.~\eqref{eq:2HDM_nlsm_Lagrangian}. 
For notational simplicity, 
we define $c_x = \cos x$, $s_x = \sin x$.
The amplitudes are given by 

\begin{align}
 \mcl{M}(W^+_L W^-_L \to W^+_L W^-_L)_{\text{cust}} =& \frac{s+t}{f^2} C_1 (\be), 
 \label{eq:W+W-->WW} \\
 \mcl{M}(W^+_L W^-_L \to h h )_{\text{cust}} =& \frac{s}{f^2} C_2 (\al,\be), 
 \label{eq:WW->hh} \\ 
 \mcl{M}(W^+_L W^-_L \to Z_L Z_L)_{\text{cust}} =& \frac{s}{f^2} C_1 (\be), 
 \label{eq:WW->ZZ} \\
 \mcl{M}(W^+_L W^-_L \to h Z_L)_{\text{cust}} =& 0, 
 \label{eq:WW->hZ}
\end{align}
\begin{align}
 \mcl{M}(Z_L Z_L \to W^+_L W^-_L)_{\text{cust}} =& \frac{s}{f^2} C_1 (\be), 
  \label{eq:ZZ->WW} \\
 \mcl{M}(Z_L Z_L \to h h)_{\text{cust}} =& \frac{s}{f^2} C_2 (\al ,\be), 
 \label{eq:ZZ->hh} \\
 \mcl{M}(Z_L Z_L \to Z_L Z_L)_{\text{cust}} =& 0,
 \label{eq:ZZ->ZZ} \\
 \mcl{M}(Z_L Z_L \to h Z_L)_{\text{cust}} =& 0, 
 \label{eq:ZZ->hZ}
\end{align}
\begin{align}
 \mcl{M}(W^+_L Z_L \to W^+_L h)_{\text{cust}} =& 0,
  \label{eq:WZ->Wh} \\
 \mcl{M}(W^+_L Z_L \to W^+_L Z_L)_{\text{cust}} =& \frac{t}{f^2} C_1 (\be), 
  \label{eq:WZ->WZ} \\
 \mcl{M}(W^+_L W^+_L \to W^+_L W^+_L)_{\text{cust}} =& -\frac{s}{f^2} C_1 (\be), 
  \label{eq:WW->WW}
\end{align}
where 
\begin{align}
C_1(\be) =& \frac{1}{8} \Bigl(
    (3 +4c_{2\be} +c_{4\be})
	c^H_{1111}
  +4(2s_{2\be} +s_{4\be})
   c^H_{1112} \n & \qquad
  +2(1 -c_{4\be})
   (c^H_{1122} +c^H_{1221} +c^H_{1212}) \n & \qquad
  +4(2s_{2\be} -s_{4\be})
   c^H_{2221}
  + (3 -4c_{2\be} +c_{4\be})
   c^H_{2222}
 \Bigr), \\
C_2(\al,\be) =& \frac{1}{8} \Bigl(
    (2(1 +c_{2\be}) +(1 +2c_{2\be} +c_{4\be})c_{2(\al -\be)} 
    -(2s_{2\be} +s_{4\be})s_{2(\al -\be)})
   c^H_{1111} \n & \quad
  +4(s_{2\be} +(s_{2\be} +s_{4\be})c_{2(\al -\be)} 
    +(c_{2\be} +c_{4\be})s_{2(\al -\be)})
   c^H_{1112} \n & \quad
  + 2(2 -(1 +c_{4\be})c_{2(\al -\be)} +s_{4\be}s_{2(\al -\be)}) 
   c^H_{1122} \n &\quad
  + 2((1 -c_{4 \be})c_{2(\al -\be)} + s_{4 \be} s_{2(\al -\be)})
   (c^H_{1221} + c^H_{1212}) \n &\quad
  +4(s_{2\be} +(s_{2\be} -s_{4\be})c_{2(\al -\be)} 
    +(c_{2\be} -c_{4\be})s_{2(\al -\be)})
   c^H_{2221} \n & \quad
  + (2(1 -c_{2\be}) +(1 -2c_{2\be} +c_{4\be})c_{2(\al -\be)} 
    +(2s_{2\be} -s_{4\be})s_{2(\al -\be)})
   c^H_{2222} 
 \Bigr).
\end{align}

\noindent
In the above expressions, 
we can see that 
five amplitudes, Eqs.~\eqref{eq:W+W-->WW}, 
\eqref{eq:WW->ZZ}, \eqref{eq:ZZ->WW},
\eqref{eq:WZ->WZ} and \eqref{eq:WW->WW}, 
are determined by $C_1 (\be)$, and
two amplitudes, Eqs.~\eqref{eq:WW->hh} and
\eqref{eq:ZZ->hh}, are determined by $C_2 (\al ,\be)$.
Among them, amplitudes in Eqs.~\eqref{eq:W+W-->WW} 
and \eqref{eq:WW->WW} 
(Eqs.~\eqref{eq:WW->ZZ}, \eqref{eq:ZZ->WW} and \eqref{eq:WZ->WZ}) 
are related by the crossing symmetry.
If the custodial symmetry is broken, 
the amplitudes of $W^+_L W^-_L \to W^+_L W^-_L$, 
$W^+_L W^+_L \to W^+_L W^+_L$ 
and $Z_L Z_L \to h h$ have 
$c^T$ dependence as shown in Appendix \ref{app:amp}.
Amplitudes in Eqs.~\eqref{eq:WW->hZ} and \eqref{eq:WZ->Wh}
are also connected through the crossing symmetry.
In particular, these amplitudes vanish when 
we require the custodial invariance.

We expand the coefficient $C_2(\al,\be)$ with $\al -\be$:
\begin{align}
  C_2(\al,\be) =  & C_1 (\be) \n
  &+2(\al - \be) \Bigl(
  -(2 s_{2\be} + s_{4\be})c^H_{1111} + 4(c_{2\be} +c_{4\be})c^H_{1112}
  \n & \qquad \qquad \quad
  +2 s_{4\be}(c^H_{1122} +c^H_{1221} + c^H_{1212}) 
  +4 (c_{2\be} -c_{4\be}) c^H_{2221} + (2s_{2\be} -s_{4\be})c^H_{2222}
 \Bigr) \n
  &+ \mcl{O} \left((\al -\be)^2 \right)
\end{align}

\noindent
In the SILH model, 
there is a simple relation among the amplitudes, 
$W_L^+ W_L^- \to W_L^+ W_L^-$, $W_L^+ W_L^- \to Z_L Z_L$ 
and $W_L^+ W_L^- \to hh$, 
as shown in Eqs.~\eqref{eq:amp1} 
and~\eqref{eq:amp2}. 
Such a relation is violated in the 2HDM 
because the process, 
$W_L^+ W_L^- \to hh$, depends on the parameter $\al$ as well. 
However, if the decoupling limit, 
$\al - \be = 0 $\footnote{Note that 
the decoupling limit in our notation, 
$\al - \be = \ep \ (\ep \to 0)$,
corresponds to 
$\al - \be = - \pi/2 \, + \, \ep \ (\ep \to 0)$
in another notation~\cite{Haber:1995}.}, 
is imposed, the relation is recovered.
In other words, 
even if we observe the scatterings including 
only the SM particles in the high energy region, 
the SILH model can be discriminated from the 
models involving several Higgs doublets.

Secondly, we show amplitudes including 
one heavy Higgs boson, $H^\pm$, $H$ or $A$, 
in the final states.
If they are heavier than the SM ones, then 
the following behavior can be derived 
in the energy region much higher than the mass scale of
heavy Higgs bosons.
The amplitudes are given by

\begin{align}
  \mcl{M}(W^+_L W^-_L \to W^+_L H^-)_{\text{cust}} =& \frac{s + t}{f^2} C_3 (\be), 
 \label{eq:WW -> W+H-}\\
 \mcl{M}(W^+_L W^-_L \to h H)_{\text{cust}} =& \frac{s}{f^2} C_4 (\al,\be), 
 \label{eq:WW -> hH} \\
 \mcl{M}(W^+_L W^-_L \to h A)_{\text{cust}} =& -i\frac{s+2t}{3 f^2}
    s_{\al -\be}
   (c^H_{1221} -c^H_{1212}), 
  \label{eq:WW -> hA}\\
  \mcl{M}(W^+_L W^-_L \to Z_L H )_{\text{cust}} =& 0,
 \label{eq:WW -> ZH} \\
  \mcl{M}(W^+_L W^-_L \to Z_L A )_{\text{cust}} =& \frac{s}{f^2} C_3 (\be), 
 \label{eq:WW -> ZA} 
\end{align}
\begin{align}
  \mcl{M}(Z_L Z_L \to W_L^+ H^-)_{\text{cust}} =& \frac{s}{f^2} C_3 (\be), 
 \label{eq:ZZ -> W+H-}\\
 \mcl{M}(Z_L Z_L \to h H)_{\text{cust}} =& \frac{s}{f^2} C_4 (\al,\be), 
 \label{eq:ZZ -> hH} \\
  \mcl{M}(Z_L Z_L \to h A)_{\text{cust}} =& 0,
 \label{eq:ZZ -> hA} \\
  \mcl{M}(Z_L Z_L \to Z_L H)_{\text{cust}} =& 0,
 \label{eq:ZZ -> ZH}\\
 \mcl{M}(Z_L Z_L \to Z_L A)_{\text{cust}} =& 0,
 \label{eq:ZZ -> ZA}
\end{align}
\begin{align}
 \mcl{M}(W^+_L Z_L \to H^+ h)_{\text{cust}} =& - i\frac{s + 2t}{3f^2}
   s_{\al -\be} \left(
	c^H_{1221} -c^H_{1212}
  \right), 
  \label{eq:WZ -> Hh}\\
  \mcl{M}(W^+_L Z_L \to W^+_L H)_{\text{cust}} =& 0,
 \label{eq:WZ->WH} \\
  \mcl{M}(W^+_L Z_L \to W^+_L A)_{\text{cust}} =& \frac{t}{f^2} C_3 (\be), 
 \label{eq:WZ->WA} \\
  \mcl{M}(W^+_L Z_L \to H^+ Z_L)_{\text{cust}} =& \frac{t}{f^2} C_3 (\be), 
 \label{eq:WZ->HZ} \\
  \mcl{M}(W^+_L W^+_L \to W^+_L H^+)_{\text{cust}} =& -\frac{s}{f^2} C_3 (\be),
 \label{eq:WW->WH}
\end{align}
where 
\begin{align}
 C_3 (\be) =& \frac{1}{8} \Bigl(
  - (2s_{2\be} +s_{4\be})
   c^H_{1111}
  +4(c_{2\be} +c_{4\be})
   c^H_{1112} \n & \qquad
  +2s_{4\be}
   (c^H_{1122} +c^H_{1221} +c^H_{1212}) \n & \qquad
  +4(c_{2\be} -c_{4\be})
   c^H_{2221}
  + (2s_{2\be} -s_{4\be})
   c^H_{2222}
 \Bigr), \\
 C_4 (\al,\be) =& \frac{1}{8} \Bigl(
  -((2s_{2\be} +s_{4\be})c_{2(\al -\be)} 
    +(1 +2c_{2\be} +c_{4\be})s_{2(\al -\be)})
   c^H_{1111} \n & \quad
  +4( (c_{2\be} +c_{4\be})c_{2(\al -\be)} -(s_{2\be} +s_{4\be})s_{2(\al -\be)})
   c^H_{1112} \n & \quad
  +2( s_{4\be}c_{2(\al -\be)} +(1 +c_{4\be})s_{2(\al -\be)})
   c^H_{1122} \n & \quad
  +2( s_{4\be}c_{2(\al -\be)} -(1 -c_{4\be})s_{2(\al -\be)})
   (c^H_{1221} +c^H_{1212}) \n & \quad
  +4((c_{2\be} -c_{4\be})c_{2(\al -\be)} -(s_{2\be} -s_{4\be})s_{2(\al -\be)}) 
   c^H_{2221} \n & \quad
  +( (2s_{2\be} -s_{4\be})c_{2(\al -\be)} 
    -(1 -2c_{2\be} +c_{4\be})s_{2(\al -\be)})
   c^H_{2222}
 \Bigr).
\end{align}
As seen from above, 
Eqs.~\eqref{eq:WW -> W+H-}, \eqref{eq:WW -> ZA}, \eqref{eq:ZZ -> W+H-},
\eqref{eq:WZ->WA}, \eqref{eq:WZ->HZ} and \eqref{eq:WW->WH},
are expressed by a common constant, $C_3 (\be)$, 
and Eqs.~\eqref{eq:WW -> hH} and \eqref{eq:ZZ -> hH} 
are expressed by $C_4 (\al,\be)$. 
Among them, 
Eqs.~\eqref{eq:WW -> W+H-} and \eqref{eq:WW->WH}
(Eqs.~\eqref{eq:WW -> ZA} and \eqref{eq:WZ->WA},
Eqs.~\eqref{eq:ZZ -> W+H-} and \eqref{eq:WZ->HZ}) 
are related through the crossing symmetry.
In the case without the custodial symmetry, 
the amplitudes of $W^+_L W^-_L \to W^+_L H^-$ and 
$W^+_L W^+_L \to W^+_L H^+$
are different from the other four amplitudes  
because of $c^T$ dependence.
See Appendix \ref{app:amp}.
Amplitudes in Eqs.~\eqref{eq:WW -> ZH} and 
\eqref{eq:WZ->WH} related by the crossing symmetry 
vanish in the case with the custodial invariance.
We also obtain the equality between 
Eqs.~\eqref{eq:WW -> hA} and \eqref{eq:WZ -> Hh}.
This equality is broken by the custodial symmetry breaking terms 
as shown in Appendix \ref{app:amp}.

The coefficient, $C_4 (\al,\be)$, is expanded 
in terms of $\al -\be$ as follows:
\begin{align}
  C_4 (\al,\be) = &C_3 (\be) \n
   &  + 2(\al - \be) \Bigl(
   (1 +2c_{2\be} + c_{4\be}) c^H_{1111}
   - 4 (s_{2\be} -s_{4\be}) c^H_{1112}
   +2( (1 +c_{4\be}) c^H_{1122} 
  \n & \qquad \qquad \quad
   -(1 -c_{4\be})( c^H_{1221} +c^H_{1212} )
   - 4(s_{2\be} -s_{4\be}) c^H_{2221}
   -(1 -2c_{2\be} +c_{4\be})c^H_{2222}
 \Bigr)
  \n & + \mcl{O}((\al - \be)^2),  
\end{align}

\noindent
If we take the decoupling limit, all of nonzero amplitudes 
can be expressed by one parameter, $C_3 (\be)$.
Hence, the following relations are satisfied:
\begin{align}
 \mcl{M}(W^+_L W^-_L \to W^+_L H^- )_{\text{cust}} =& \frac{s +t}{f^2} C_3 (\be), \\
 \mcl{M}(W^+_L W^-_L \to Z_L A )_{\text{cust}} =& 
 \mcl{M}(W^+_L W^-_L \to h H )_{\text{cust}} = \frac{s}{f^2} C_3 (\be).
\end{align}

Finally, amplitudes of double heavy Higgs boson production are listed:

\begin{align}
  \mcl{M}(W^+_L W^-_L \to H^+ H^-)_{\text{cust}} =& \frac{s+t}{f^2} C_5 (\be) 
  +\frac{t}{f^2} (c^H_{1221} - c^H_{1122}), 
  \label{eq:WW -> H+H-} \\
 \mcl{M}(W^+_L W^-_L \to H H)_{\text{cust}} =& \frac{s}{f^2} C_6 (\al,\be), 
  \label{eq:WW -> HH} \\
 \mcl{M}(W^+_L W^-_L \to A A)_{\text{cust}} =& \frac{s}{f^2} C_5 (\be), 
 \label{eq:WW -> AA} \\
 \mcl{M}(W^+_L W^-_L \to H A)_{\text{cust}} =& -i\frac{s+2t}{3 f^2}
  c_{\al -\be} (c^H_{1221} -c^H_{1212}),
  \label{eq:WW -> HA} 
\end{align}
\begin{align}
 \mcl{M}(Z_L Z_L \to H^+ H^- )_{\text{cust}} =& \frac{s}{f^2} C_5 (\be), 
 \label{eq:ZZ -> H+H-} \\
 \mcl{M}(Z_L Z_L \to H H)_{\text{cust}} =& \frac{s}{f^2} C_6 (\al,\be), 
 \label{eq:ZZ -> HH} \\
 \mcl{M}(Z_L Z_L \to A A)_{\text{cust}} =& \frac{s}{f^2} (
   c^H_{1122} -3c^T_{1221}),
   \label{eq:ZZ -> AA} \\
 \mcl{M}(Z_L Z_L \to H A)_{\text{cust}} =& 0,
  \label{eq:ZZ -> HA}
\end{align}
\begin{align}
 \mcl{M}(W^+_L Z_L \to H^+ H)_{\text{cust}} =& -i\frac{s +2t}{3f^2}
   c_{\al -\be}
  (c^H_{1221} -c^H_{1212}), 
   \label{eq:WZ -> H+ H} \\
  \mcl{M}(W^+_L Z_L \to H^+ A)_{\text{cust}} =& \frac{t}{f^2} C_5 (\be) 
  - \frac{s -t}{3f^2} c^H_{1221} + \frac{s +2t}{3f^2} c^H_{1212} 
  - \frac{t}{f^2} c^H_{1122}
  + \frac{2s +t}{f^2} c^T_{1221}, 
  \label{eq:WZ -> HA} \\
  \mcl{M}(W^+_L W^+_L \to H^+ H^+)_{\text{cust}} =& - \frac{s}{f^2} C_5 (\be),
  \label{eq:WW -> H+H+}
\end{align}
where 
\begin{align}
  C_5 (\be) =& \frac{1}{8} \Bigl(
    (1 -c_{4\be})
	(c^H_{1111} -2c^H_{1221} -2c^H_{1212} +c^H_{2222}) \n & \quad
  -4s_{4\be}
   (c^H_{1112} -c^H_{2221})
  +2(3 +c_{4\be})
   c^H_{1122}
 \Bigr), \\
 C_6 (\al,\be) =& \frac{1}{8} \Bigl(
    ( 2(1 +c_{2\be}) -(1 +2c_{2\be} +c_{4\be})c_{2(\al -\be)} 
	 +(2s_{2\be} +s_{4\be})s_{2(\al -\be)})
	c^H_{1111} \n & \quad
  +4( s_{2\be} -(s_{2\be} +s_{4\be})c_{2(\al -\be)} 
    -(c_{2\be} +c_{4\be})s_{2(\al -\be)})
   c^H_{1112} \n & \quad
  +2( 2 +(1 +c_{4\be})c_{2(\al -\be)} -s_{4\be}s_{2(\al -\be)})
   c^H_{1122} \n & \quad
  -2( (1 -c_{4\be})c_{2(\al -\be)} +s_{4\be}s_{2(\al -\be)})
   (c^H_{1221} +c^H_{1212}) \n & \quad
  +4( s_{2\be} -(s_{2\be} -s_{4\be})c_{2(\al -\be)} 
    -(c_{2\be} -c_{4\be})s_{2(\al -\be)})
	c^H_{2221} \n & \quad
  + ( 2(1 -c_{2\be}) -(1 -2c_{2\be} +c_{4\be})c_{2(\al -\be)} 
	 -(2s_{2\be} -s_{4\be})s_{2(\al -\be)})
   c^H_{2222}
 \Bigr).
\end{align}

\noindent
The amplitudes, 
Eqs.~\eqref{eq:WW -> AA}, \eqref{eq:ZZ -> H+H-} and 
\eqref{eq:WW -> H+H+}, 
are identical up to overall sign, and
the amplitudes, Eqs.~\eqref{eq:WW -> HH} and \eqref{eq:ZZ -> HH}, 
are identical.
Among them, for the process, $W^+_L W^+_L \to H^+ H^-$, 
$c^T$ dependence appears when the custodial symmetry breaking terms exist. 
See Appendix \ref{app:amp}.
Unlike previous cases, 
the processes, $Z_L Z_L \to AA$ and $W^+_L Z_L \to H^+ A$, 
have the $c^T_{1221}$ dependence. 
Equation~\eqref{eq:WW -> HA} is identical to Eq.~\eqref{eq:WZ -> H+ H} 
due to the custodial symmetry.

We present the power series expansion of $\al - \be$ 
for the coefficient, $C_6 (\al,\be)$, as follows:
\begin{align}
  C_6 (\al,\be) = &C_5 (\be) \n
   & + 2(\al -\be) \Bigl(
  (2s_{2\be} + s_{4\be})c^H_{1111} - 4(c_{2\be} +c_{4\be}) c^H_{1112}
  - 2 s_{4\be} (c^H_{1122} +c^H_{1221} +c^H_{1212})
 \n & \qquad \qquad \quad
  - 4(c_{2\be} - c_{4\be}) c^H_{2221} - (2s_{2\be} -s_{4\be}) c^H_{2222}
 \Bigr) \n &
 + \mcl{O}((\al - \be)^2).
\end{align}

\noindent
The amplitude, $\mcl{M}(W_L^+ W_L^- \to HH)$, becomes the same as 
$\mcl{M}(W_L^+ W_L^- \to AA)$
in the decoupling limit.
\subsection{Cross sections and numerical results}
Since we assume that masses of scalar bosons can be neglected, 
a cross section is written as
\begin{align}
  \si (V_1 V_2 \to X_1 X_2) =&
  \frac{s}{32\pi f^4} \left( \frac{(2C_s -C_t)^2}{2} +\frac{C_t^2}{6} \right),
\end{align}
for an amplitude
\begin{align}
  \mcl{M} = \frac{C_s s +C_t t}{f^2}.
\label{EqAmp}
\end{align}
If $X_1 =X_2$,
the cross section must be divided by two.
With the above formula,
cross sections of vector boson fusion subsystems are
given as follows:
\begin{align}
  \si(W^+_L W^-_L \to W^+_L W^-_L)_{\text{cust}} =&
    \frac{s}{32\pi f^4} \frac{2}{3} C_1 (\be)^2 ,\\
  \si(W^+_L W^-_L \to h h)_{\text{cust}} =&
    \frac{s}{32\pi f^4} C_2 (\al, \be)^2 ,\\
  \si(W^+_L W^-_L \to Z_L Z_L)_{\text{cust}} =& 
    \frac{3}{2} \si(W^+_L W^-_L \to W^+_L W^-_L)_{\text{cust}} ,\\
  \si(Z_L Z_L \to W^+_L W^-_L)_{\text{cust}} =& 
    3 \si(W^+_L W^-_L \to W^+_L W^-_L)_{\text{cust}}, \\
  \si(Z_L Z_L \to h h)_{\text{cust}} =& 
    \si(W^+_L W^-_L \to h h)_{\text{cust}}, \\
  \si(W^+_L Z_L \to W^+_L Z_L)_{\text{cust}} =&  
    \si(W^+_L W^-_L \to W^+_L W^-_L)_{\text{cust}}, \\
  \si(W^+_L W^+_L \to W^+_L W^+_L)_{\text{cust}} =&
    \frac{3}{2} \si(W^+_L W^-_L \to W^+_L W^-_L)_{\text{cust}},
\end{align}
\begin{align}
  \si(W^+_L W^-_L \to W^+_L H^-)_{\text{cust}} =&
    \frac{s}{32\pi f^4} \frac{2}{3} C_3 (\be)^2 ,\\
  \si(W^+_L W^-_L \to h H )_{\text{cust}} =&
    \frac{s}{32\pi f^4} 2 C_4 (\al, \be)^2 ,\\
  \si(W^+_L W^-_L \to h A )_{\text{cust}} =&
    \frac{s}{32\pi f^4} \frac{2}{27} 
	 \sin^2 (\al -\be) (c^H_{1221} -c^H_{1212} )^2 ,\\
  \si(W^+_L W^-_L \to Z_L A)_{\text{cust}} =& 
    3\si(W^+_L W^-_L \to W^+_L H^-)_{\text{cust}} ,\\
  \si(Z_L Z_L \to W^+_L H^-)_{\text{cust}} =& 
    3 \si(W^+_L W^-_L \to W^+_L H^-)_{\text{cust}}, \\
  \si(Z_L Z_L \to h H)_{\text{cust}} =& 
    \si(W^+_L W^-_L \to h H )_{\text{cust}}, \\
  \si(W^+_L Z_L \to H^+ h)_{\text{cust}} =& 
    \si(W^+_L W^-_L \to h A )_{\text{cust}}, \\
  \si(W^+_L Z_L \to W^+_L A)_{\text{cust}} =&
    \si(W^+_L W^-_L \to W^+_L H^-)_{\text{cust}} , \\
  \si(W^+_L Z_L \to H^+ Z_L)_{\text{cust}} =& 
    \si(W^+_L W^-_L \to W^+_L H^-)_{\text{cust}}, \\
  \si(W^+_L W^+_L \to W^+_L H^+)_{\text{cust}} =&
    3 \si(W^+_L W^-_L \to W^+_L H^-)_{\text{cust}}, 
\end{align}
\begin{align}
  \si(W^+_L W^-_L \to H^+ H^-)_{\text{cust}} =&
    \frac{s}{32\pi f^4} \frac{2}{3} 
	 ( C_5(\be)^2 -C_5(\be) (c^H_{1221} -c^H_{1122} ) 
	  +(c^H_{1221} -c^H_{1122} )^2 ) ,\\
  \si(W^+_L W^-_L \to H H)_{\text{cust}} =&
    \frac{s}{32\pi f^4} C_6 (\al, \be)^2 ,\\
  \si(W^+_L W^-_L \to A A)_{\text{cust}} =&
    \frac{s}{32\pi f^4} C_5 (\be)^2 ,\\
  \si(W^+_L W^-_L \to H A )_{\text{cust}} =&
    \frac{s}{32\pi f^4} \frac{2}{27} 
	 \cos^2 (\al -\be) (c^H_{1221} -c^H_{1212} )^2 ,\\
 \si(Z_L Z_L \to H^+ H^-)_{\text{cust}} =& 
   2 \si(W^+_L W^-_L \to A A)_{\text{cust}}, \\
 \si(Z_L Z_L \to H H)_{\text{cust}} =& 
   \si(W^+_L W^-_L \to H H)_{\text{cust}}, \\
  \si(Z_L Z_L \to A A )_{\text{cust}} =&
     \frac{s}{32\pi f^4} (c^H_{1122} -3c^T_{1221})^2 ,\\
  \si(W^+_L Z_L \to H^+ H )_{\text{cust}} =&
     \si(W^+_L W^-_L \to H A )_{\text{cust}}, \\
  \si(W^+_L Z_L \to H^+ A )_{\text{cust}} =&
     \frac{s}{32\pi f^4} \frac{1}{2} \biggl(
	    ( C_5 (\be) -c^H_{1122} +c^H_{1221} -3c^T_{1221} )^2 \n & \qquad \quad
		+\frac{1}{3} \left( 
		  C_5 (\be) +\frac{c^H_{1221} +2c^H_{1212}}{3} -c^H_{1122} +c^T_{1221}
		\right)^2
	\biggr), \\
  \si(W^+_L W^+_L \to H^+ H^+)_{\text{cust}} =& 
     \si(W^+_L W^-_L \to A A)_{\text{cust}}.
\end{align}
In these cross sections,
we omitted processes whose amplitudes are zero.
Some of cross sections are proportional to others 
even if their amplitudes are not.
Notice that since we still have ten independent cross sections,
all of parameters, $c^{H,T} / f^2 , \al $ and $ \be$, 
can be fixed by measuring these processes.

\begin{figure}
\centering
 \includegraphics[width=19em]{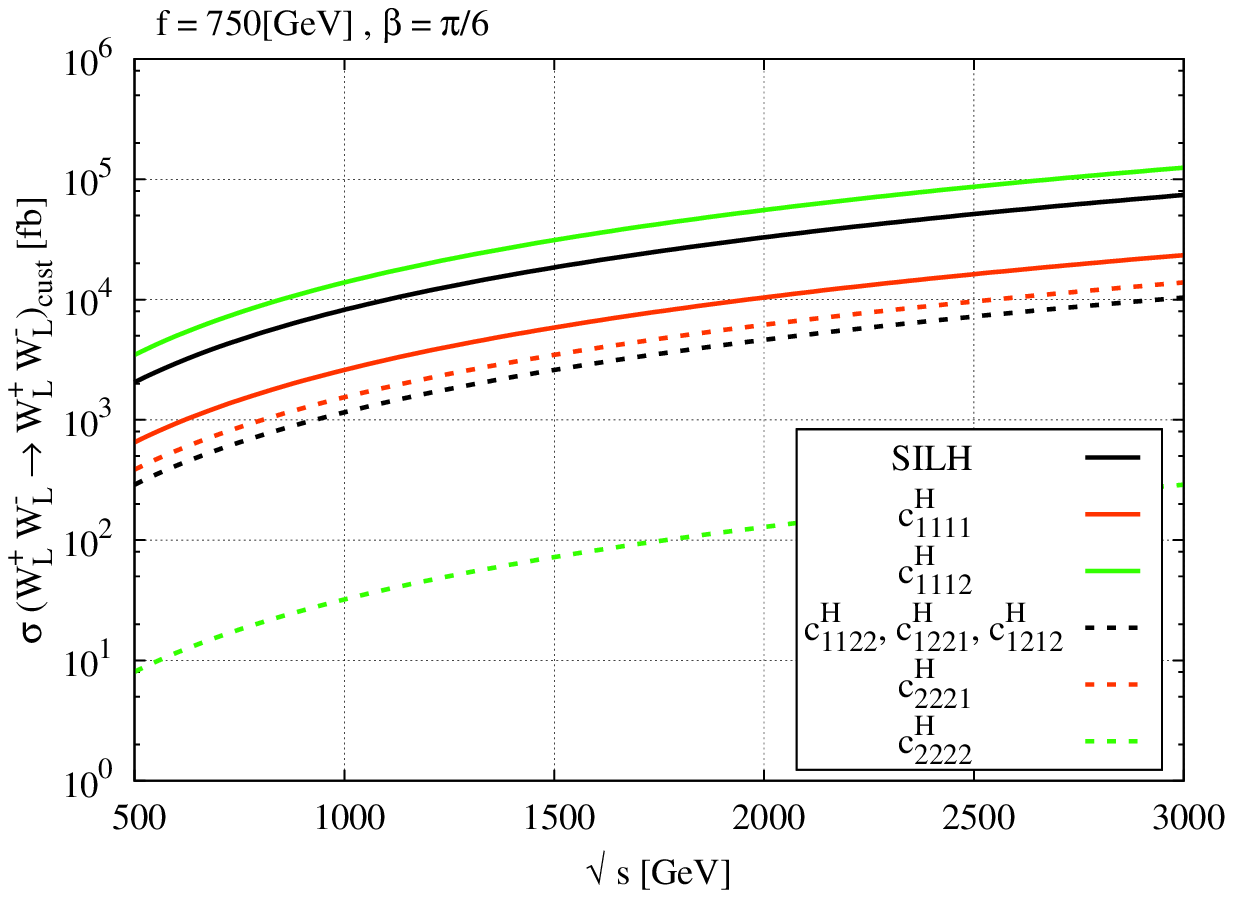}
 \includegraphics[width=19em]{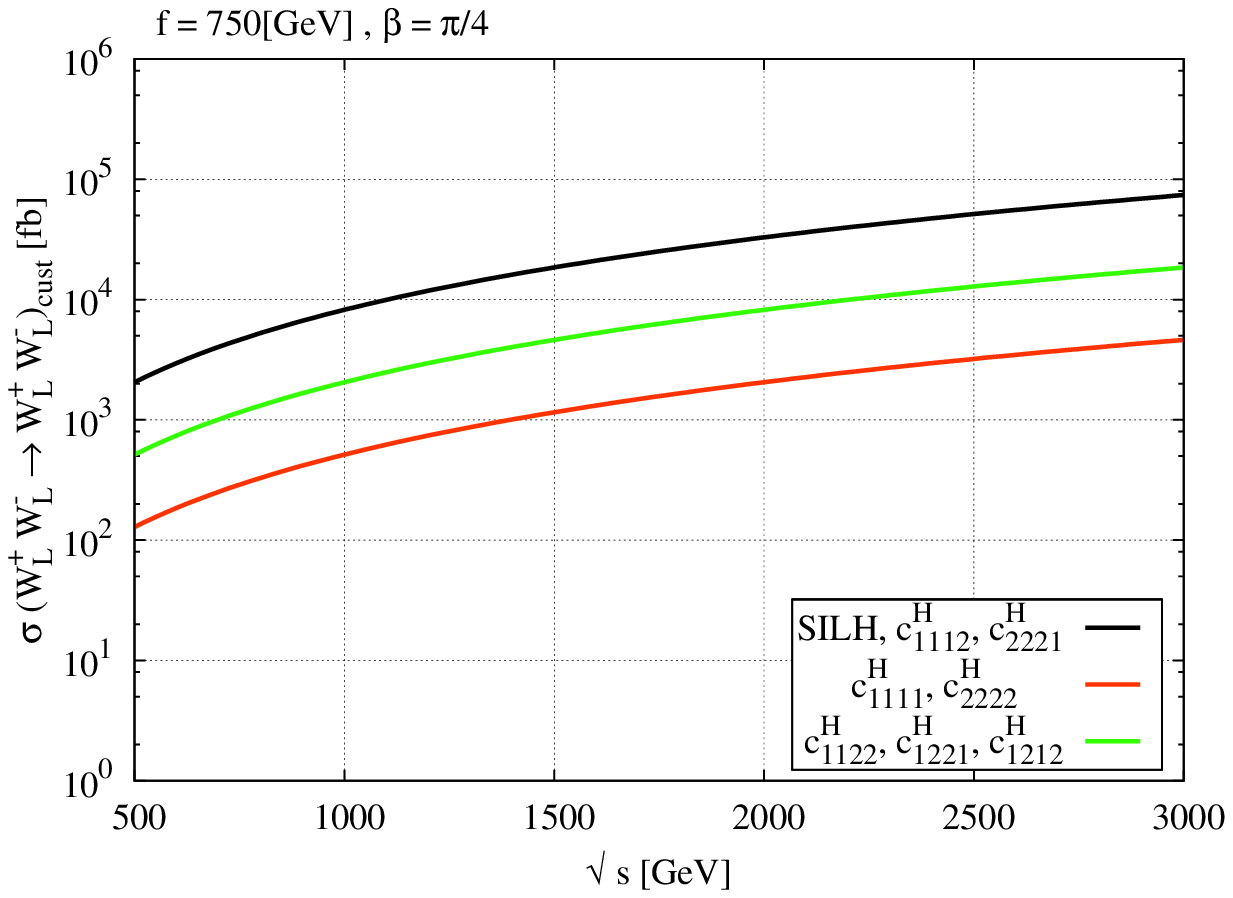}
\caption{
  Cross sections of $W_L^+ W_L^- \to W_L^+ W_L^-$ for $\be = \pi/6$ (left)
  and $\pi/4$ (right).
  The decay constant $f$ is fixed as 750 GeV.
  The line of the SILH means 
  the cross section of $W_L^+ W_L^- \to W_L^+ W_L^-$ with 
  $\al = \be = 0$, $c^H_{1111} = 1$ and the others vanish.
  For the other lines,
  only one of coefficients is unity and the others are zero.
}
\label{FigXsec}
\end{figure}

Cross sections depend on eight coefficients and two angles.
In the following figures, 
one of coefficients is turned on and the others are turned off.
Coefficients turned on are set as unity.
The parameter that $\al = \be = 0$ and only $c^H_{1111} \neq 0$ 
reproduces the SILH, hence we call this case the SILH.
The decay constant of the \nlsm, $f$, is set to be 750 GeV.

In Fig.~\ref{FigXsec}, 
cross sections of $W_L^+ W_L^- \to W_L^+ W_L^-$ are presented, 
where $\be$ is chosen as $\pi/6$ or $\pi/4$ and,
for each line, only one of coefficients is turned on.
Since all lines are linear functions of the squared invariant mass, $s$,
they are parallel to each other.

In Ref.~\cite{Contino} it is shown that 
focusing on the central region, $-1/2 < \cos \th < 1/2 $,
helps us discriminate the derivative interactions from
the other contributions in high energy region.
Without restriction to the central region, 
the cross section of the longitudinal mode production is much smaller than 
that of the transverse mode in the SM 
($\sim 2 \times 10^{6}$ fb)\footnote{
  Our SILH line ($f=750\GeV$, $\al = \be = 0$ and $c^H_{1111} =1$) corresponds 
  to the line of $a^2 \sim 0.9$ in terms of Ref.~\cite{Contino}.
}.
Further investigation with polarization measurement proposed by 
Ref.~\cite{Han:2009em} could be useful to distinguish the longitudinal mode
from the transverse mode at the lower energy region.
Cross sections of the central region are given in 
Appendix~\ref{SecCentral}.

\begin{figure}
\centering
 \includegraphics[width=19em]{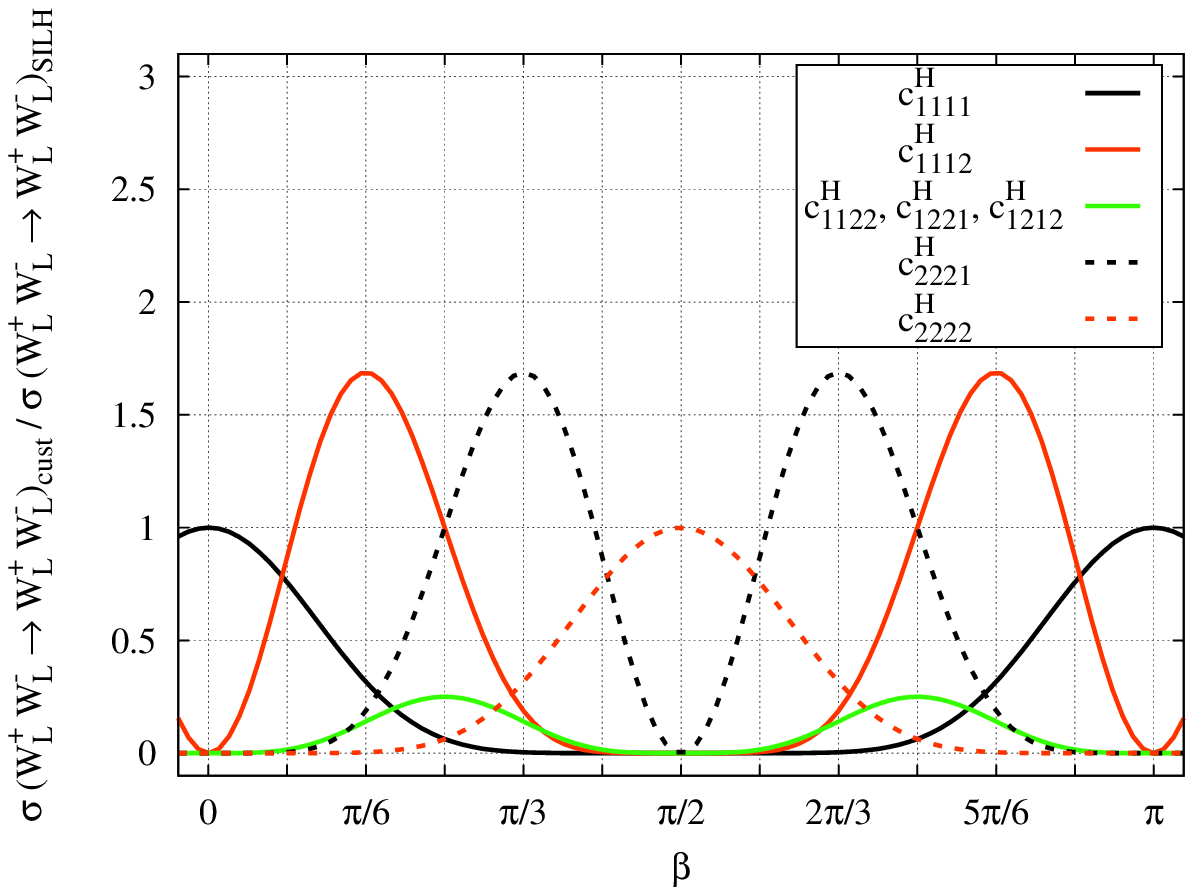}
 \includegraphics[width=19em]{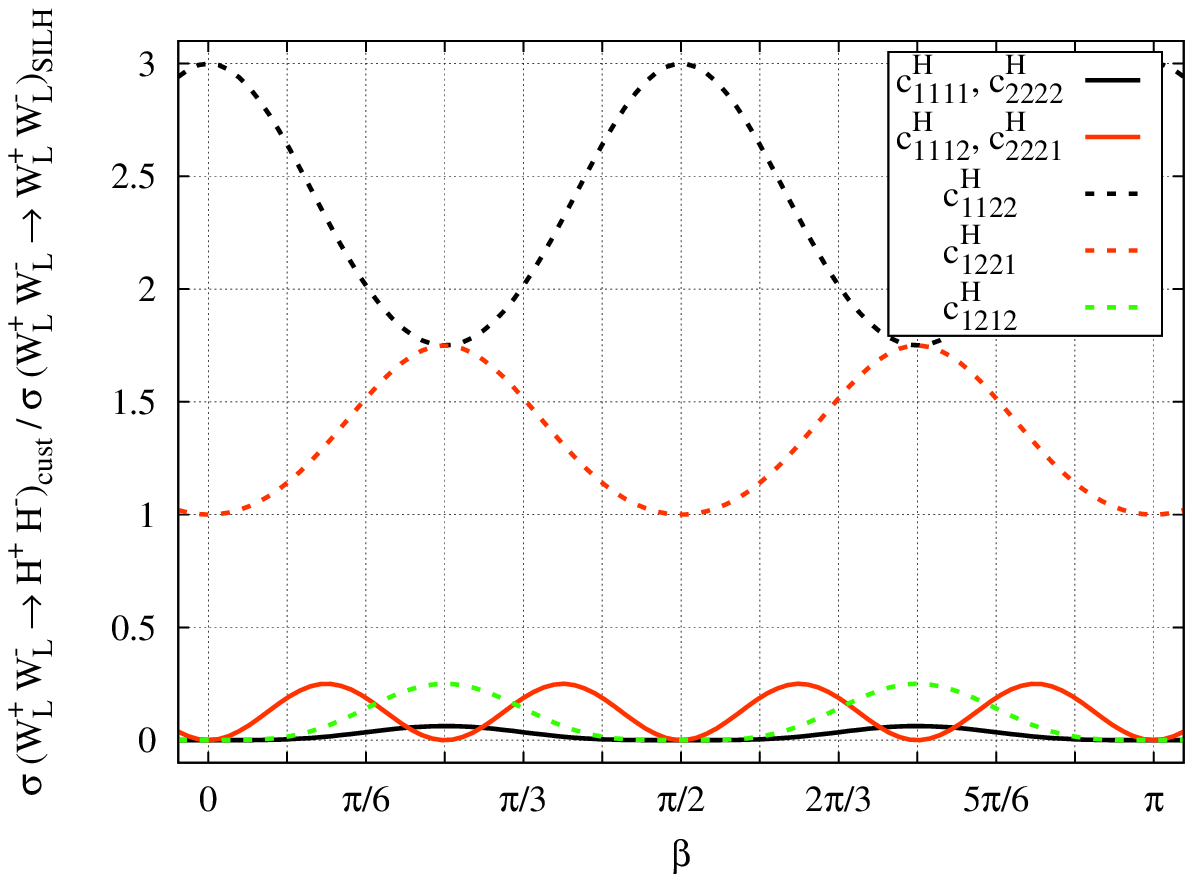}
\caption{
 The $\be$ dependences of $W_L^+ W_L^- \to W_L^+ W_L^-$ (left) and
 $W_L^+ W_L^- \to H^+ H^-$ (right).
 These cross sections are given in the unit of the
 $W_L^+ W_L^- \to W_L^+ W_L^-$ cross section with the SILH parameters.
}
\label{FigAnybeta}
\end{figure}

Cross sections of $W_L^+ W_L^- \to W_L^+ W_L^-$ and 
$W_L^+ W_L^- \to H^+ H^-$ as functions of $\be$ are shown in 
Fig.~\ref{FigAnybeta}.
Each line corresponds to the case where one of coefficients is nonzero.
Cross sections are presented in the unit of the $W_L^+ W_L^- \to W_L^+ W_L^-$ 
cross section with the SILH condition.
For $W_L^+ W_L^- \to W_L^+ W_L^-$,
the largest ratios given by $c^H_{1112}$ and $c^H_{2221}$ are 27/16.
Peaks of small bumps given by $c^H_{1122}$, $c^H_{1221}$ and $c^H_{1212}$
are one fourth comparing to the SILH cross section.
Absence of $c^H_{1111}$ and $c^H_{2222}$, respectively, erases 
the cross section at $\be = 0$ and $\pi/2$.
For $W_L^+ W_L^- \to H^+ H^-$,
$c^H_{1122}$ and $c^H_{1221}$ generate larger cross section than 
that given by $W_L^+ W_L^- \to W_L^+ W_L^-$ in the SILH.
Contributions of $c^H_{1122}$ and $c^H_{1221}$ do not vanish for any $\be$.
In summary, cross sections receive large contribution from
$c^H_{1112}$ and $c^H_{2221}$ for $W_L^+ W_L^- \to W_L^+ W_L^-$
and 
from $c^H_{1122}$ and $c^H_{1221}$ for $W_L^+ W_L^- \to H^+ H^-$.
On the other hand, 
if only $c^H_{1212}$ appears,
the production cross sections of both processes are 
much smaller than 
the $W$ boson pair production cross section in the SM.

\begin{figure}
\centering
 \includegraphics[width=19em]{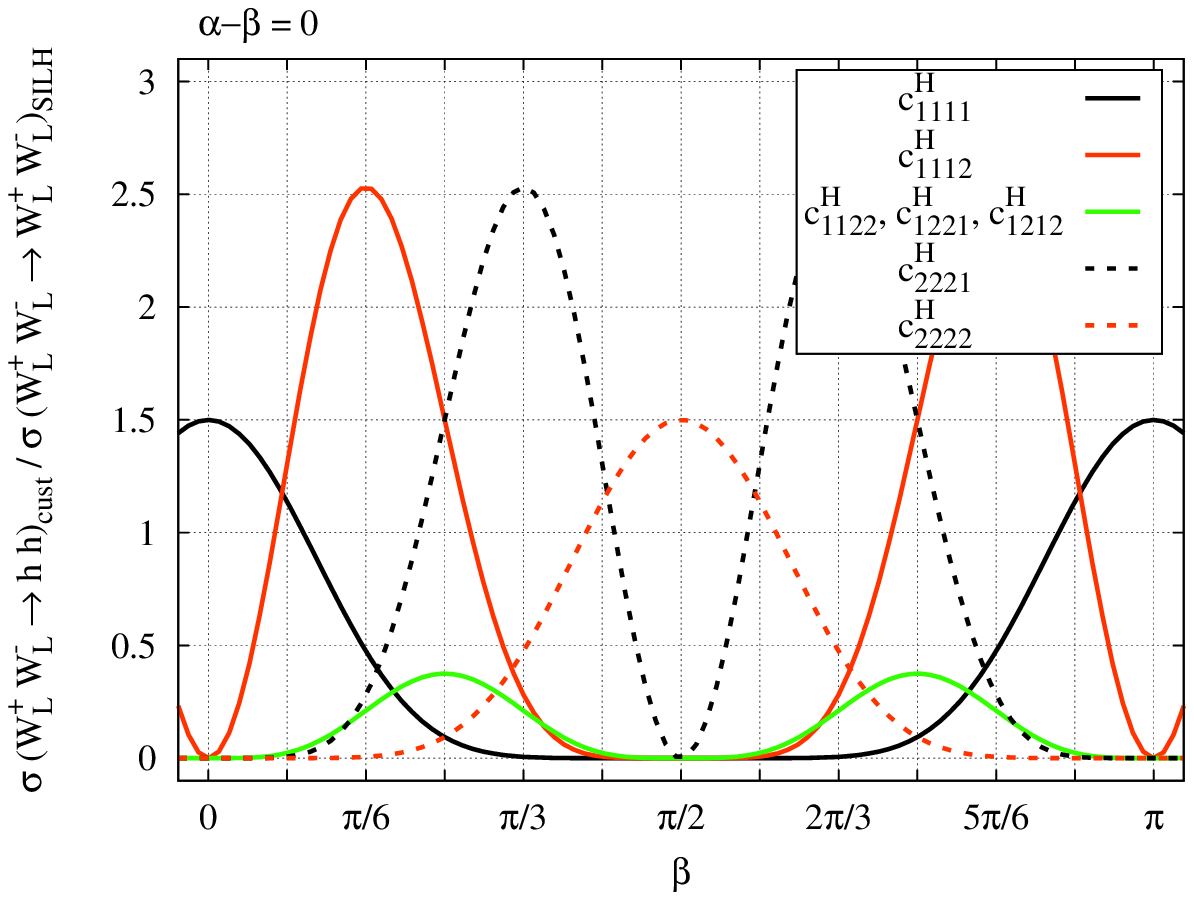}
 \includegraphics[width=19em]{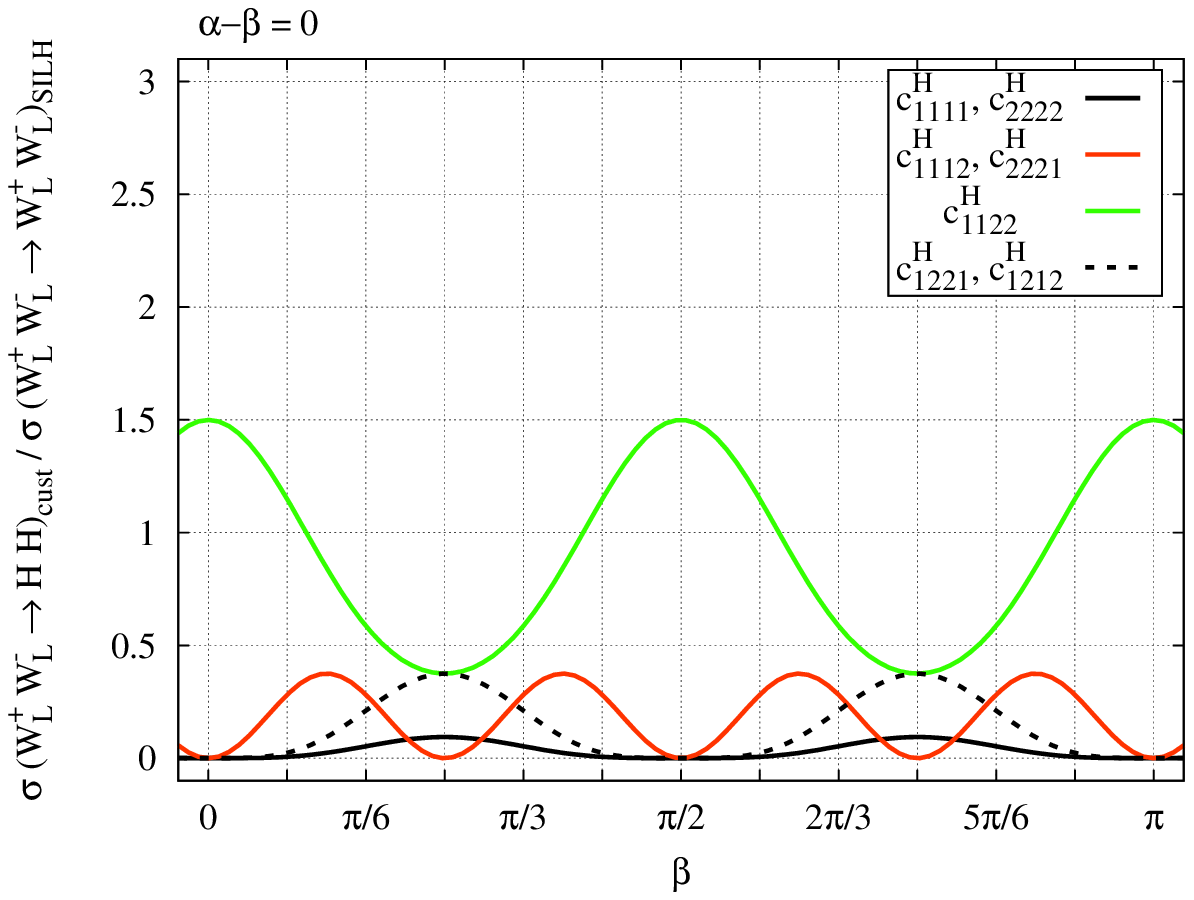}
 \includegraphics[width=19em]{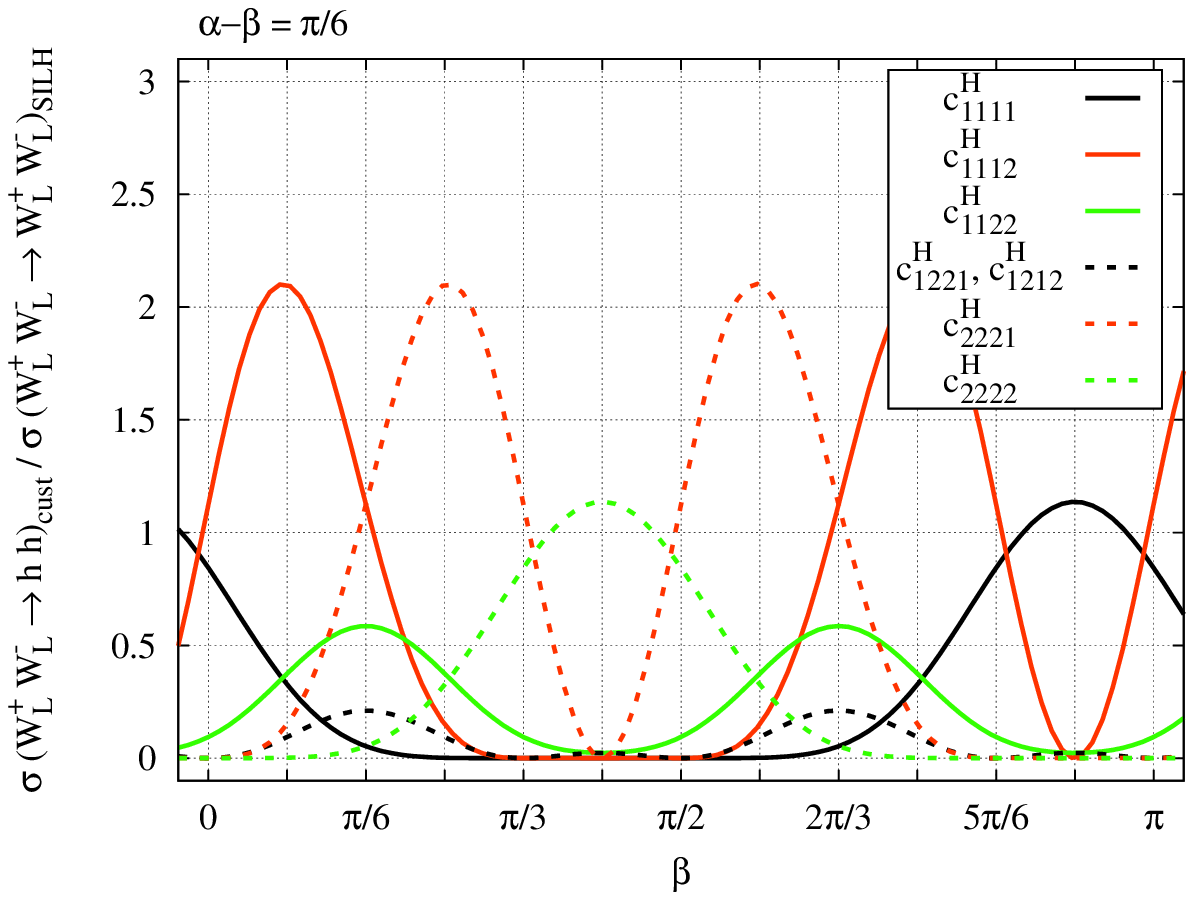}
 \includegraphics[width=19em]{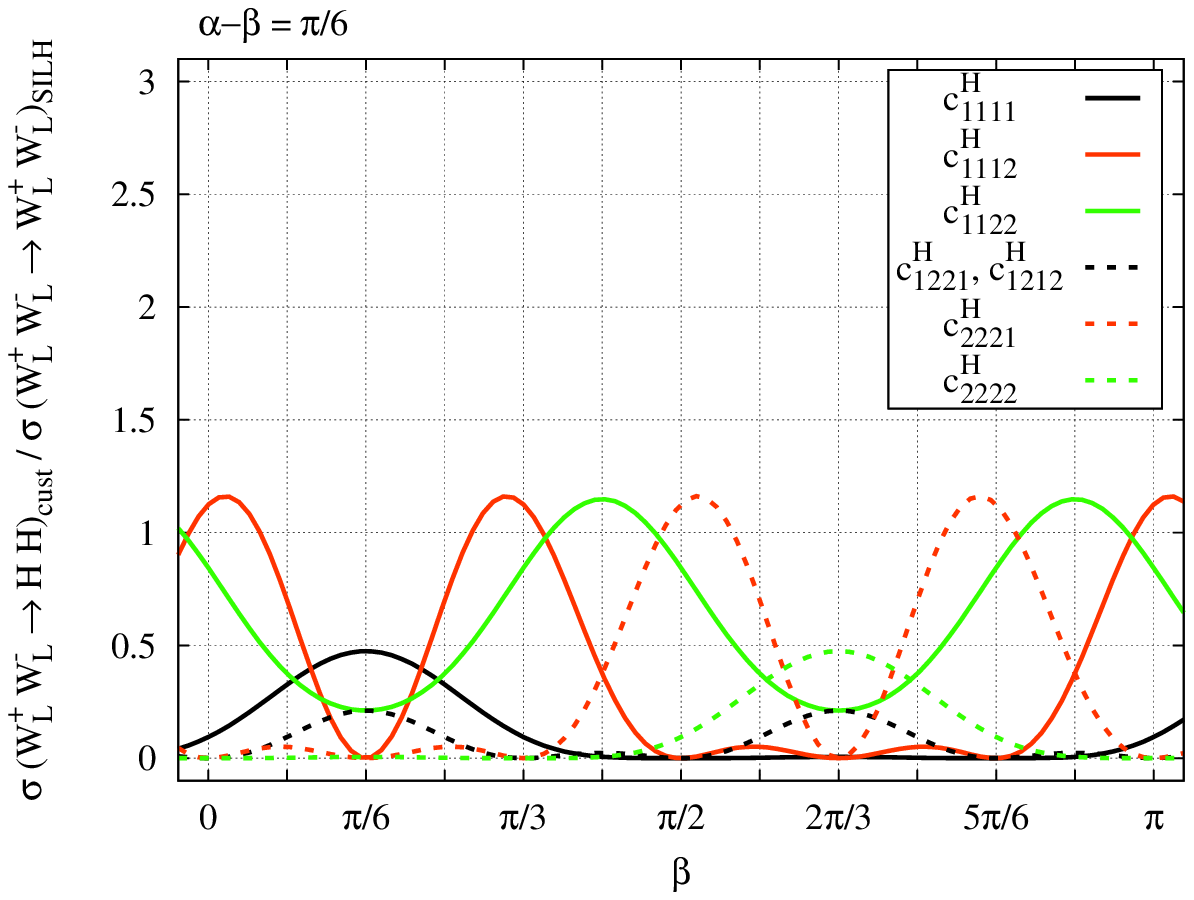}
 \includegraphics[width=19em]{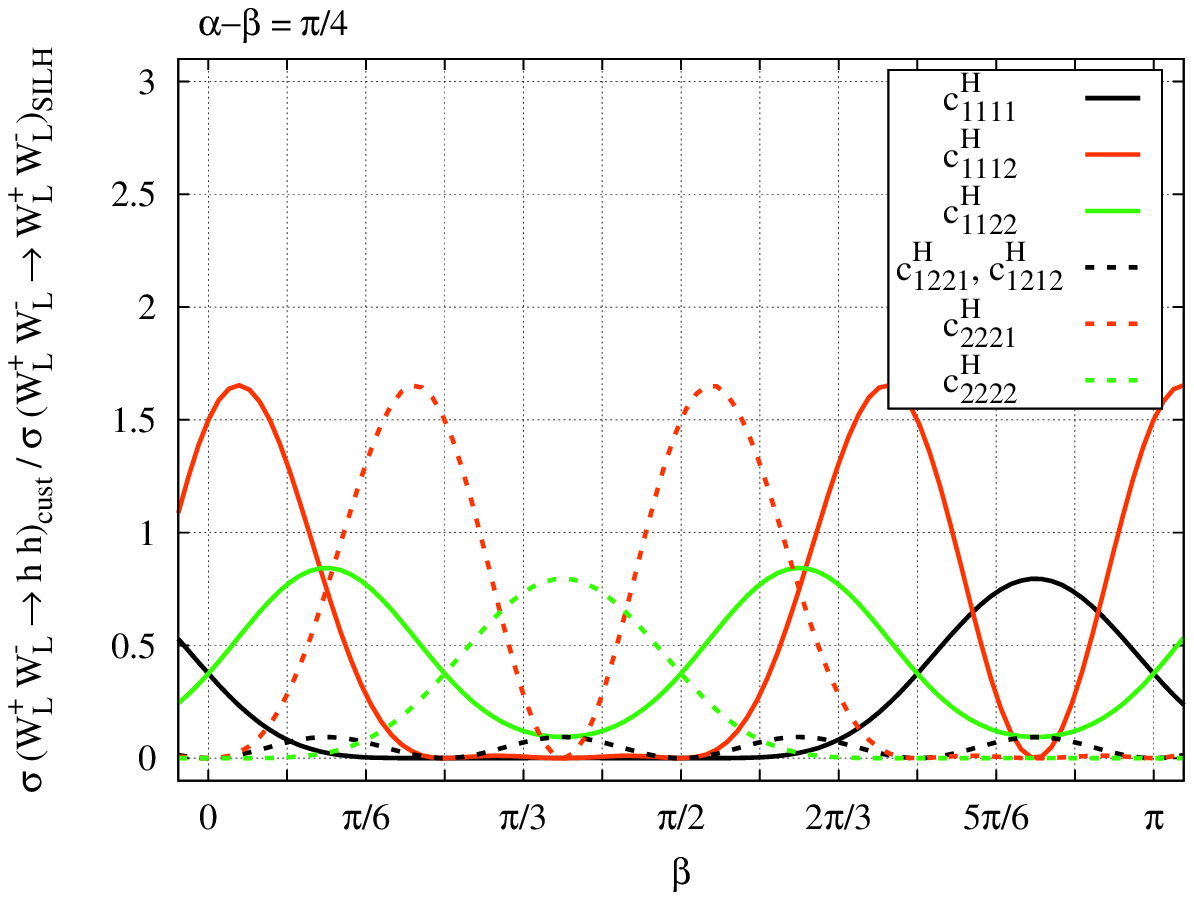}
 \includegraphics[width=19em]{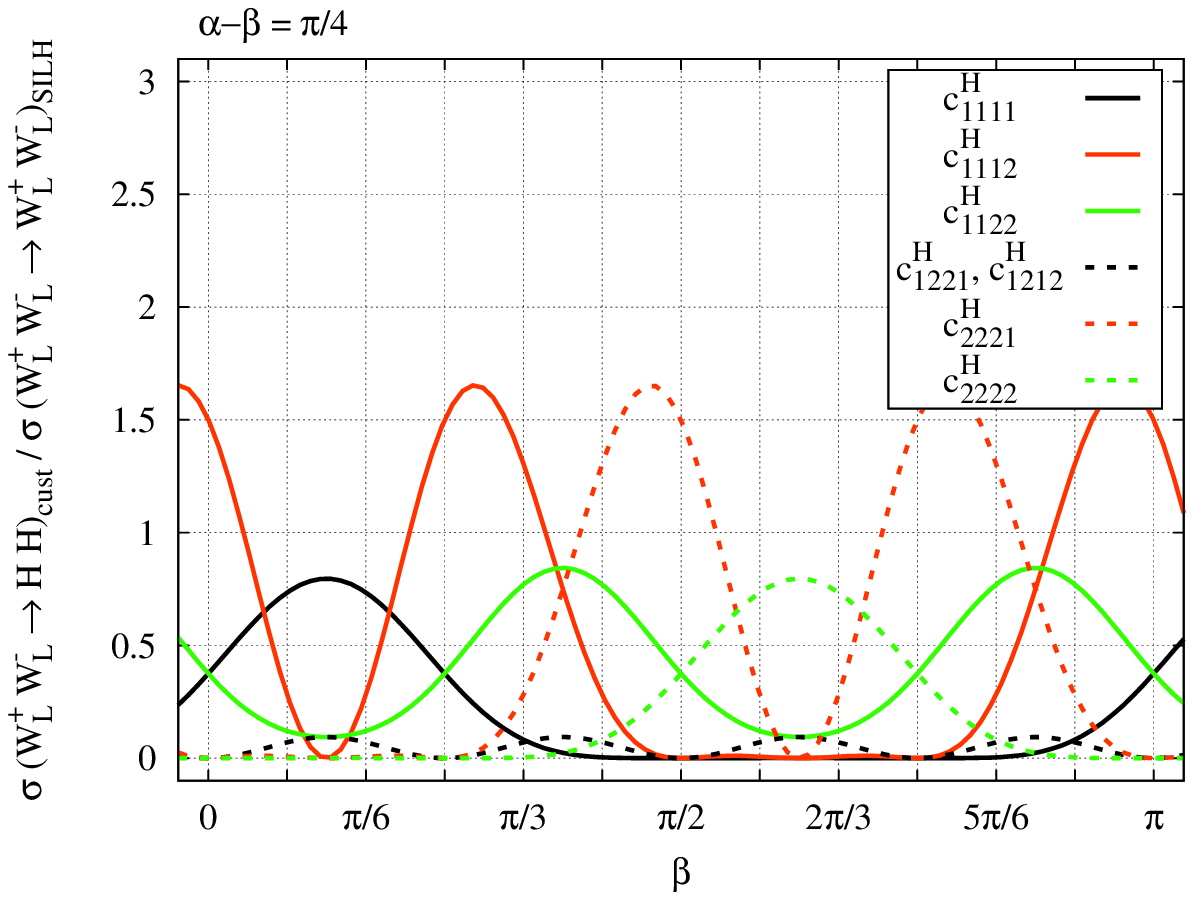}
\caption{
 The $\be$ dependences of $W_L^+ W_L^- \to h h$ (left) and
 $W_L^+ W_L^- \to H H$ (right).
 Each row means $\al -\be = 0$, $\pi/6$ and $\pi/4$.
 These cross sections are given in the unit of the
 $W_L^+ W_L^- \to W_L^+ W_L^-$ cross section with the SILH parameters.
 }
\label{FigAlpha}
\end{figure}

Figure~\ref{FigAlpha} shows
$\be$ dependences of $W_L^+ W_L^- \to h h$ and $W_L^+ W_L^- \to H H$
for $\al -\be = 0$, $\pi/6$ and $\pi/4$\footnote{
  The cross section of $W_L^+ W_L^- \to h h$ with the SILH case
  ($f=750\GeV$, $\al = \be = 0$ and $c^H_{1111}=1$) corresponds to
  that with the line of $a^2 -b \sim 0.9$ in terms of Ref.~\cite{Contino}.
}.
Their cross sections are divided by
the cross section of $W_L^+ W_L^- \to W_L^+ W_L^-$ in the SILH.
These cross sections can exceed the SM $WW \to hh$
production cross section ($  \sim 5 \times 10^4 $ fb)
in a few TeV region without studying on the central region.
By analysis in the central region,
the contributions of derivative interactions can be distinguished 
from the other effects even in the region lower than 1 TeV.

In summary, we found that
the longitudinal mode of $W$ boson and the charged Higgs boson 
possessed quite different parameter dependences.
Because the production cross section of $W^+ W^- \to h h$ is 
smaller than that of $W^+ W^- \to W^+ W^-$  
by 2 orders of magnitude in the SM,
$W_L^+ W_L^- \to h h$ is the promising process to see 
effects of derivative interactions.
There are several possibilities to 
discriminate the NG 2HDM from the SILH 
depending on parameters: the decay constant, coefficients, angles
and mass spectra of heavy Higgs bosons.
For example, since the cross section of $W_L^+ W_L^- \to h h$ 
is three halves larger than
that of $W_L^+ W_L^- \to W_L^+ W_L^-$ in the SILH,
the difference from the SILH could be easily found
if $W_L^+ W_L^- \to W_L^+ W_L^-$ is observed.
In the case that additional Higgs bosons are light,
other processes may be promising to discriminate this model from the SILH.
\section{Conclusions} 
\label{sec:conclusion}

We have extended the treatment of the dimension-six 
derivative interactions in the SILH model 
to the case of NHDM.
The operators are phenomenologically important 
because they determine high energy behavior of the scattering amplitudes 
of the longitudinal gauge bosons and the Higgs bosons.

In the \nlsm, the information of the global symmetry breaking 
determines the form of the operators, and hence
the derivative interactions are governed by 
the structure constant.
Since the $N$ Higgs doublets can be embedded in the NG fields as
the $SO(4N)$ multiplet, 
the derivative interactions can be expressed using the 
$SO(4N)$ generators.
Using the bidoublet notation,
it is easy to impose the $SU(2)_L \times U(1)_Y$ to the interactions.
As a consequence, the number of the derivative interactions,
$(3/2)N^2(N^2 +1)$ real \dof \  and $(1/2)N^2(3N^2 -1)$ imaginary \dof \
in the general NHDM,
is reduced to $(1/2)N^2(N^2 +3)$ real \dof \  
and $(1/2)N^2(N^2 -1)$ imaginary \dof \ 
due to the nature of a strongly interacting dynamics.

We have then applied these results to the 2HDM.
By the phenomenological requirement, we impose the manifest
custodial invariance.
We have calculated the scattering amplitudes of the longitudinal gauge bosons 
and the Higgs bosons by the dimension-six
derivative interactions. 
We have derived various relations among the scattering amplitudes,
and clarified the differences 
between the custodial symmetry violating case and preserving case.
The amplitudes including only 
the SM particles violate the simple relations
found in the SILH model.
These relations are recovered in the decoupling limit, $\al - \be = 0$.
In other words,
the parameter $\al - \be$ is a key parameter to distinguish 
the composite 2HDM from the SILH model.
The high energy behavior of the scattering amplitudes can provide 
an alternative way to study the Higgs sector to 
the Higgs coupling measurements 
through the Higgs boson production and decay processes. 
We have obtained several relations among the amplitudes
involving heavy Higgs bosons, 
which provides us clues to the structure of the 2HDM 
described by the $\nlsm$.
Pair production cross sections of longitudinal gauge bosons and Higgs bosons 
due to dimension-six operators also have been calculated.
We have found that $W_L^+ W_L^- \to h h$ is a promising process 
to see the effect of derivative interactions.
Cross sections of other processes may be large enough to be observed,
so that their observations are useful to distinguish this model from the SILH.

Even if the Higgs boson is discovered,
it is not easy to reveal the structure of the Higgs sector.
By investigation of the strongly interacting NHDM,
we show that precise measurements of the longitudinal gauge boson
scatterings can give us hints for physics beyond the SM;
how many the Higgs doublets exist, 
whether the fundamental interactions are strong or weak, etc.
These phenomena are important physics targets 
in collider experiments at the LHC and LC.
\acknowledgments  
The authors would like to thank P. Posch for useful discussions.
The work of Y.O. is supported in part by the Grant-in-Aid 
for Science Research, Japan Society
for the Promotion of Science (JSPS), No.20244037 and No.22244031.
The work of Y.Y. is supported in part by the Grant-in-Aid 
for Science Research, Japan Society
for the Promotion of Science (JSPS), No.22.3834.
\appendix
\section{Generators of the $SO(4N)$} 
\label{app:so4N}

In order to respect the $ SO(4) \simeq SU(2)_L \times SU(2)_R $ symmetry,
it is convenient to classify the generators of the $SO(4N)$ 
into the irreducible representation of the $SO(4)$.
We can write $2N(4N-1)$ generators in terms of the Kronecker delta:
\begin{align}
  \left( T_{(i,j)}^{ab} \right)_{cd} = 
 -\frac{i}{2} \left( 
    \de^{a+4(i-1),\, c}\, \de^{b+4(j-1),\, d} 
	-\de^{a+4(i-1),\, d}\, \de^{b+4(j-1),\, c} 
  \right),
\end{align}
where $i, j \in \{1, ..., N\}$ $(i\leq j)$, $a, b \in \{1, ..., 4\}$ 
and $c, d \in \{1, ..., 4N\}$.
Namely, $i$ and $j$ mean the indices of $4\times 4$ blocks and 
$a$ and $b$ stand for components in each blocks.

The generators can be classified as follows:
\begin{align}
  T_{(i,i)}^{L1} &= -T_{(i,i)}^{14} +T_{(i,i)}^{23} ,\\
  T_{(i,i)}^{L2} &=  T_{(i,i)}^{13} +T_{(i,i)}^{24} , \\
  T_{(i,i)}^{L3} &= -T_{(i,i)}^{12} +T_{(i,i)}^{21} ,
\end{align}
\begin{align}
  T_{(i,i)}^{R1} &= -T_{(i,i)}^{14} -T_{(i,i)}^{23} , \\
  T_{(i,i)}^{R2} &=  T_{(i,i)}^{13} -T_{(i,i)}^{24} , \\
  T_{(i,i)}^{R3} &= -T_{(i,i)}^{12} +T_{(i,i)}^{21} ,
\end{align}
\begin{align}
  T_{(i,j)}^{L1} &= 
	 -T_{(i,j)}^{14} +T_{(i,j)}^{23} -T_{(i,j)}^{32} +T_{(i,j)}^{41} ,\\
  T_{(i,j)}^{L2} &= 
	  T_{(i,j)}^{13} +T_{(i,j)}^{24} -T_{(i,j)}^{31} -T_{(i,j)}^{42} , \\
  T_{(i,j)}^{L3} &= 
	 -T_{(i,j)}^{12} +T_{(i,j)}^{21} +T_{(i,j)}^{34} -T_{(i,j)}^{43} ,
\end{align}
\begin{align}
  T_{(i,j)}^{R1} &= 
	 -T_{(i,j)}^{14} -T_{(i,j)}^{23} +T_{(i,j)}^{32} +T_{(i,j)}^{41} , \\
  T_{(i,j)}^{R2} &= 
	  T_{(i,j)}^{13} -T_{(i,j)}^{24} +T_{(i,j)}^{31} -T_{(i,j)}^{42} , \\
  T_{(i,j)}^{R3} &= 
	 -T_{(i,j)}^{12} +T_{(i,j)}^{21} -T_{(i,j)}^{34} +T_{(i,j)}^{43} ,
\end{align}
\begin{align}
  U_{(i,j)} = 
	  T_{(i,j)}^{11} +T_{(i,j)}^{22} +T_{(i,j)}^{33} +T_{(i,j)}^{44} ,
\end{align}
\begin{align}
  S_{(i,j)}^{11} &= 
	  T_{(i,j)}^{11} -T_{(i,j)}^{22} -T_{(i,j)}^{33} +T_{(i,j)}^{44} , \\
  S_{(i,j)}^{12} &= 
	 -T_{(i,j)}^{12} -T_{(i,j)}^{21} +T_{(i,j)}^{34}  +T_{(i,j)}^{43} , \\
  S_{(i,j)}^{13} &= 
	  T_{(i,j)}^{13} +T_{(i,j)}^{24} +T_{(i,j)}^{31} +T_{(i,j)}^{42} , \\
  S_{(i,j)}^{21} &= 
	 -T_{(i,j)}^{12} -T_{(i,j)}^{21} -T_{(i,j)}^{34} -T_{(i,j)}^{43} , \\
  S_{(i,j)}^{22} &= 
	 -T_{(i,j)}^{11} +T_{(i,j)}^{22} -T_{(i,j)}^{33} +T_{(i,j)}^{44} , \\
  S_{(i,j)}^{23} &= 
	  T_{(i,j)}^{14} -T_{(i,j)}^{23} -T_{(i,j)}^{32} +T_{(i,j)}^{41} , \\
  S_{(i,j)}^{31} &= 
	 -T_{(i,j)}^{13} +T_{(i,j)}^{24} -T_{(i,j)}^{31} +T_{(i,j)}^{42} , \\
  S_{(i,j)}^{32} &= 
	  T_{(i,j)}^{14} +T_{(i,j)}^{23} +T_{(i,j)}^{32} +T_{(i,j)}^{41} , \\
  S_{(i,j)}^{33} &= 
	  T_{(i,j)}^{11} +T_{(i,j)}^{22} -T_{(i,j)}^{33} -T_{(i,j)}^{44} ,
\end{align}
where $T^{L\al}_{(i,j)}$, $T^{R\be}_{(i,j)}$, $U_{(i,j)}$ and 
$S^{\al \be}_{(i,j)}$ 
are $({\bf 3},\, {\bf 1})$, $({\bf 1},\, {\bf 3})$, $({\bf 1},\, {\bf 1})$
and $({\bf 3}, {\bf 3})$ representations 
of $SU(2)_L \times SU(2)_R$, respectively.
Normalization depends on whether
the generator is in diagonal block or off-diagonal block.
According to the above, all kinds of generators appear for $N=2$.
Therefore, as an example, we show the generators of the $SO(4)$ and 
the generators, $({\bf 1}, {\bf 1})$ and $({\bf 3}, {\bf 3})$, 
in the $SO(8)$ below:
\begin{align}
 T^{L1} =& \frac{i}{2}
  \begin{pmatrix}
      &    &    &  1 \\
      &    & -1 &    \\
      &  1 &    &    \\
   -1 & \Z &    & \Z
  \end{pmatrix},&
 T^{L2} =& \frac{i}{2}
  \begin{pmatrix}
   \Z & \Z & -1 &    \\
      &    &    & -1 \\
    1 &    &    &    \\
      &  1 &    &   
  \end{pmatrix},&
 T^{L3} =& \frac{i}{2}
  \begin{pmatrix}
      &  1 &    &    \\
   -1 & \Z & \Z &    \\
      &    &    & -1 \\
      &    &  1 &   
  \end{pmatrix} \\
 T^{R1} =& \frac{i}{2}
  \begin{pmatrix}
      &    &    &  1 \\
      &    &  1 &    \\
      & -1 & \Z & \Z \\
   -1 &    &    &    
  \end{pmatrix},&
 T^{R2} =& \frac{i}{2}
  \begin{pmatrix}
      &    &  1 &    \\
      & \Z & \Z & -1 \\
   -1 &    &    &    \\
      &  1 &    &    
  \end{pmatrix},&
 T^{R3} =& \frac{i}{2}
  \begin{pmatrix}
      &  1 &    &    \\
   -1 & \Z &    & \Z \\
      &    &    &  1 \\
      &    & -1 &   
  \end{pmatrix}
\end{align}
\begin{align}
 U_{(1,2)} = \frac{i}{2}
  \begin{pmatrix}
   \mathbf{0}_4 & -\mathbf{1}_4 \\
   \mathbf{1}_4 &  \mathbf{0}_4
  \end{pmatrix}, \qquad
 S^{\al \be}_{(1,2)} = \frac{i}{2}
 \begin{pmatrix}
  \mathbf{0}_4 & S^{\al \be} \\
  - S^{\al \be} & \mathbf{0}_4
 \end{pmatrix},
\end{align}
\begin{align}
 S^{11} =&
  \begin{pmatrix}
   -1 & \Z & \Z &    \\
      &  1 &    &    \\
      &    &  1 &    \\
      &    &    & -1
  \end{pmatrix},&
 S^{12} =&
  \begin{pmatrix}
      &  1 &    &    \\
    1 &    &    &    \\
   \Z & \Z &    & -1 \\
      &    & -1 &   
  \end{pmatrix},&
 S^{13} =&
  \begin{pmatrix}
      &    & -1 &    \\
      &    &    & -1 \\
   -1 &    &    &    \\
      & -1 &    &   
  \end{pmatrix},\\
 S^{21} =&
  \begin{pmatrix}
      &  1 &    &    \\
    1 &    & \Z & \Z \\
   \Z & \Z &    &  1 \\
      &    &  1 &   
  \end{pmatrix},&
 S^{22} =&
  \begin{pmatrix}
    1 &    &    &    \\
      & -1 &    &    \\
      &    &  1 &    \\
   \Z &    & \Z & -1
  \end{pmatrix},&
 S^{23} =&
  \begin{pmatrix}
      & \Z & \Z & -1 \\
      &    &  1 &    \\
      &  1 &    &    \\
   -1 &    &    &   
  \end{pmatrix},\\
 S^{31} =&
  \begin{pmatrix}
      &    &  1 &    \\
   \Z &    & \Z & -1 \\
    1 &    &    &    \\
      & -1 &    &   
  \end{pmatrix},&
 S^{32} =&
  \begin{pmatrix}
      &    &    & -1 \\
      &    & -1 &    \\
      & -1 &    &    \\
   -1 &    &    &   
  \end{pmatrix},&
 S^{33} =&
  \begin{pmatrix}
   -1 &    &    &    \\
      & -1 & \Z & \Z \\
      &    &  1 &    \\
      &    &    &  1
  \end{pmatrix},
\end{align}
where blanks are filled by zero and 
$\mathbf{0}_4$ and $\mathbf{1}_4$ are the empty matrix and 
the unit matrix of four dimensions.
\section{Bidoublet notation} 
\label{app:bidoublet}

The bidoublet notation is useful to see the
$SU(2)_{L} \times SU(2)_{R} \simeq SO(4)$ symmetry. 
We use the following $4\times 4$ matrix as bidoublet:
\begin{align}
 \Ph_{ii} =
 \begin{pmatrix}
	i \si^{2} H^\ast_i & H_i
 \end{pmatrix}.
\end{align}
Under the $SU(2)_L \times SU(2)_R$ symmetry, 
the transformation low of the bidoublet is
\begin{align}
 \Ph_{ii} \to L \Ph_{ii} R^\dag ,
\end{align}
where $L \in SU(2)_L$ and $R \in SU(2)_R$.

The correspondences between the $SU(2)_L$ doublets and the bidoublets are
\begin{align}
  H_i^\dag H_j &= 
   \frac{1}{2} \Tr \left[ 
	  \Ph_{ii}^\dag \Ph_{jj}  -\si^3 \Ph_{ii}^\dag \Ph_{jj} 
	\right], \\
  H_j^\dag H_i &= 
   \frac{1}{2} \Tr \left[ 
	  \Ph_{ii}^\dag \Ph_{jj}  +\si^3 \Ph_{ii}^\dag \Ph_{jj} 
	\right], \\
  \del_\mu H_i^\dag \del_\nu H_j &= 
   \frac{1}{2} \Tr \left[ 
	  \del_\mu \Ph_{ii}^\dag \del_\nu \Ph_{jj}  
	 -\si^3 \del_\mu \Ph_{ii}^\dag \del_\nu \Ph_{jj} 
	\right], \\
  \del_\mu H_j^\dag \del_\nu H_i &= 
   \frac{1}{2} \Tr \left[ 
	  \del_\nu \Ph_{ii}^\dag \del_\mu \Ph_{jj}  
	 +\si^3 \del_\nu \Ph_{ii}^\dag \del_\mu \Ph_{jj} 
	\right], \\
  H_i^\dag \delfb_\mu  H_j &= 
   \frac{1}{2} \Tr \left[ 
	  \Ph_{ii}^\dag \delfb_\mu \Ph_{jj} -\si^3 \Ph_{ii}^\dag \delfb \Ph_{jj} 
	\right], \\
  H_j^\dag \delfb_\mu H_i &= 
   \frac{1}{2} \Tr \left[ 
	  \Ph_{ii}^\dag \delfb_\mu \Ph_{jj} +\si^3 \Ph_{ii}^\dag \delfb \Ph_{jj} 
	\right],
\end{align}
where the following relations are used:
\begin{align}
  \Tr \left[ 
    \Ph_{ii}^\dag \Ph_{jj} 
  \right] &= 
  \Tr \left[ 
    \Ph_{jj}^\dag \Ph_{ii} 
  \right], \\
  \Tr \left[ 
    \si^3 \Ph_{ii}^\dag \Ph_{jj} 
  \right] &= 
  -\Tr \left[ 
    \si^3 \Ph_{jj}^\dag \Ph_{ii} 
  \right], \\
  \Tr \left[ 
    \Ph_{ii}^\dag \fb{\del}_\mu \Ph_{jj} 
  \right] &= 
  -\Tr \left[ 
    \Ph_{jj}^\dag \fb{\del}_\mu \Ph_{ii} 
  \right], \\
  \Tr \left[ 
    \si^3 \Ph_{ii}^\dag \fb{\del}_\mu \Ph_{jj} 
  \right] &= 
  \Tr \left[ 
    \si^3 \Ph_{jj}^\dag \fb{\del}_\mu \Ph_{ii} 
  \right].
\end{align}

Any potential terms and dimension-six derivative interactions can be
described using the relations.
In the potential and the derivative interaction,
terms including $\si^3$ violate the $SO(4)$ symmetry.
\section{Amplitudes for 2HDM without the $SO(4)$ symmetry}
\label{app:amp}

Amplitudes of the longitudinal modes and the Higgs bosons 
generated by the dimension-six derivative interactions are displayed below
for 2HDM.
We here consider the amplitudes based on the following Lagrangian:
\begin{align}
 {\cal L}^6_\text{2HDM} =& 
   \frac{c^H_{1111}}{2f^2}  O^H_{1111} 
  +\frac{c^H_{1112}}{ f^2}( O^H_{1112} +O^H_{1121} ) 
  +\frac{c^H_{1122}}{ f^2}  O^H_{1122}
  +\frac{c^H_{1221}}{ f^2}  O^H_{1221} \n &
  +\frac{c^H_{1212}}{2f^2}( O^H_{1212} +O^H_{2121} )
  +\frac{c^H_{2221}}{ f^2}( O^H_{2212} +O^H_{2221} )
  +\frac{c^H_{2222}}{2f^2}  O^H_{2222} \n &
  +\frac{c^T_{1111}}{2f^2}  O^T_{1111} 
  +\frac{c^T_{1112}}{ f^2}( O^T_{1112} +O^T_{1121} ) 
  +\frac{c^T_{1122}}{ f^2}  O^T_{1122}
  +\frac{c^T_{1221}}{ f^2}  O^T_{1221} \n &
  +\frac{c^T_{1212}}{2f^2}( O^T_{1212} +O^T_{2121} )
  +\frac{c^T_{2221}}{ f^2}( O^T_{2212} +O^T_{2221} )
  +\frac{c^T_{2222}}{2f^2}  O^T_{2222},
\end{align}
where, for simplicity, we assume all coefficients are real and 
the spontaneous $CP$ violation is avoided.
The Lagrangian apparently violates the custodial symmetry at the tree level
due to the contributions of operators $\mcl{O}^T$.
In the following, initial states,
$V_1$ and $V_2$, are the longitudinal modes of massive gauge bosons,
$W_L^\pm$ and $Z_L$.
The definitions of the Mandelstam variables are given 
by Eqs.~\eqref{eq:mandelstam_s},~\eqref{eq:mandelstam_t} 
and~\eqref{eq:mandelstam_u}.
In order to clarify the difference from the custodial invariant case,
we also show the results 
in which amplitudes are decomposed into 
the custodial invariant part, $\mcl{M}_{\text{cust}}$, 
given in Sec.~\ref{subsec:amp} 
and the custodial symmetry violating part.

Firstly, amplitudes producing the SM particles are 
displayed i.e.~$V_1 V_2 \to X_1 X_2$ 
$(X_1, X_2 \in \{ W_L^\pm, Z_L, h \})$:
\begin{align}
 \mcl{M}(W^+_L W^-_L \to W^+_L W^-_L) =& \frac{s+t}{8f^2} \Bigl(
    (3 +4c_{2\be} +c_{4\be}) 
   (c^H_{1111} +3c^T_{1111})
  +4(2s_{2\be} +s_{4\be}) 
   (c^H_{1112} +3c^T_{1112}) \n & \qquad
  +2(1 -c_{4\be})
   (c^H_{1122} +c^H_{1221}+c^H_{1212} +3(c^T_{1122} +c^T_{1221} +c^T_{1212}))
	\n & \qquad
  +4(2s_{2\be} -s_{4\be} )
   (c^H_{2221} +3c^T_{2221})
  + (3 -4c_{2\be} +c_{4\be}) 
   (c^H_{2222} +3c^T_{2222})
 \Bigr) \\
 =& \mcl{M} (W^+_L W^-_L \to W^+_L W^-_L)_{\text{cust}}
 \n & 
 + \frac{3(s+t)}{8f^2} \Bigl(
    (3 +4c_{2\be} +c_{4\be}) 
   c^T_{1111}
  +4(2s_{2\be} +s_{4\be}) 
   c^T_{1112} \n & \qquad  \quad
  +2(1 -c_{4\be})
   (c^T_{1122} +c^T_{1221} +c^T_{1212})
	\n & \qquad \quad
  +4(2s_{2\be} -s_{4\be} )
   c^T_{2221}
  + (3 -4c_{2\be} +c_{4\be}) 
   c^T_{2222}
 \Bigr), \\
 \mcl{M}(W^+_L W^-_L \to h h ) =& \frac{s}{8f^2} \Bigl(
    (2 +2c_{2\be} +(1 +2c_{2\be} +c_{4\be})c_{2(\al -\be)} 
	 -(2s_{2\be} +s_{4\be})s_{2(\al -\be)})
	c^H_{1111} \n & \quad
  +4(s_{2\be} +(s_{2\be} +s_{4\be})c_{2(\al -\be)} 
    +(c_{2\be} +c_{4\be})s_{2(\al -\be)}) 
   c^H_{1112} \n & \quad
  +2(2 -(1 +c_{4\be})c_{2(\al -\be)} +s_{4\be}s_{2(\al -\be)})
   c^H_{1122} \n & \quad
  +2((1 -c_{4\be})c_{2(\al -\be)} +s_{4\be}s_{2(\al -\be)})
   (c^H_{1221} +c^H_{1212}) \n & \quad
  +4(s_{2\be} +(s_{2\be} -s_{4\be})c_{2(\al -\be)} 
    +(c_{2\be} -c_{4\be})s_{2(\al -\be)}) 
	c^H_{2221} \n & \quad
  + (2 -2c_{2\be} +(1 -2c_{2\be} +c_{4\be})c_{2(\al -\be)} 
	 +(2s_{2\be} -s_{4\be})s_{2(\al -\be)})
   c^H_{2222}
 \Bigr) \\
  =& \mcl{M}(W^+_L W^-_L \to hh)_{\text{cust}}, \\
 \mcl{M}(W^+_L W^-_L \to Z_L Z_L) =& \frac{s}{8f^2} \Bigl(
    (3 +4c_{2\be} +c_{4\be})
	c^H_{1111}
  +4(2s_{2\be} +s_{4\be})
   c^H_{1112} \n & \quad
  +2(1 -c_{4\be})
   (c^H_{1122} +c^H_{1221} +c^H_{1212}) \n & \quad
  +4(2s_{2\be} -s_{4\be})
   c^H_{2221}
  + (3 -4c_{2\be} +c_{4\be}) c^H_{2222}
 \Bigr) \\
  =& \mcl{M}(W^+_L W^-_L \to Z_L Z_L)_{\text{cust}}, \\
 \mcl{M}(W^+_L W^-_L \to h Z_L) =&  i\frac{s+2t}{8f^2} \Bigl(
    ((3 +4c_{2\be} +c_{4\be})c_{\al -\be} -(2s_{2\be} +s_{4\be})s_{\al -\be})
	c^T_{1111} \n & \qquad
  +4((2s_{2\be} +s_{4\be})c_{\al -\be} +(c_{2\be} +c_{4\be})s_{\al -\be})
   c^T_{1112} \n & \qquad
  +2((1 -c_{4\be})c_{\al -\be} +s_{4\be}s_{\al -\be})
   (c^T_{1122} +c^T_{1221} +c^T_{1212}) \n & \qquad
  +4((2s_{2\be} -s_{4\be})c_{\al -\be} +(c_{2\be} -c_{4\be})s_{\al -\be})
   c^T_{2221} \n & \qquad
  + ((3 -4c_{2\be} +c_{4\be})c_{\al -\be} +(2s_{2\be} -s_{4\be})s_{\al -\be})
   c^T_{2222}
 \Bigr),
\end{align}
\begin{align}
 \mcl{M}(Z_L Z_L \to W^+_L W^-_L) =& \frac{s}{8f^2} \Bigl(
    (3 +4c_{2\be} +c_{4\be}) 
   c^H_{1111}
  +4(2s_{2\be} +s_{4\be}) 
   c^H_{1112} \n & \quad
  +2(1 -c_{4\be})
   (c^H_{1122} +c^H_{1221}+c^H_{1212}) \n & \quad
  +4(2s_{2\be} -s_{4\be} )
   c^H_{2221}
  + (3 -4c_{2\be} +c_{4\be}) 
   c^H_{2222}
 \Bigr) \\
  =& \mcl{M}(Z_L Z_L \to W^+_L W^-_L )_{\text{cust}}, \\
%
 \mcl{M}(Z_L Z_L \to h h) =& \frac{s}{8f^2} \Bigl(
    (2 +2c_{2\be} +(1 +2c_{2\be} +c_{4\be})c_{2(\al -\be)} 
    -(2s_{2\be} +s_{4\be})s_{2(\al -\be)})
   (c^H_{1111} +3c^T_{1111}) \n & \quad
  +4(s_{2\be} +(s_{2\be} +s_{4\be})c_{2(\al -\be)} 
    +(c_{2\be} +c_{4\be})s_{2(\al -\be)})
   (c^H_{1112} +3c^T_{1112}) \n & \quad
  + (1 +(1 -2c_{4\be})c_{2(\al -\be)} +2s_{4\be}s_{2(\al -\be)})
   (c^H_{1122} +c^H_{1221} +3(c^T_{1122} +c^T_{1221})) \n & \quad
  +3(1 -c_{2(\al -\be)})
   (c^H_{1122} -c^H_{1221} -c^T_{1122} +c^T_{1221}) \n & \quad
  +2(1 -c_{4\be}c_{2(\al -\be)} +s_{4\be}s_{2(\al -\be)})
   (c^H_{1212} +3c^T_{1212}) \n & \quad
  +4(s_{2\be} +(s_{2\be} -s_{4\be})c_{2(\al -\be)} 
    +(c_{2\be} -c_{4\be})s_{2(\al -\be)})
   (c^H_{2221} +3c^T_{2221}) \n & \quad
  + (2 -2c_{2\be} +(1 -2c_{2\be} +c_{4\be})c_{2(\al -\be)} 
    +(2s_{2\be} -s_{4\be})s_{2(\al -\be)})
   (c^H_{2222} +3c^T_{2222})
 \Bigr) \\
  =& \mcl{M}(Z_L Z_L \to hh )_{\text{cust}} \n & 
  + \frac{3s}{8f^2} \Bigl(
    (2 +2c_{2\be} +(1 +2c_{2\be} +c_{4\be})c_{2(\al -\be)} 
    -(2s_{2\be} +s_{4\be})s_{2(\al -\be)})
   c^T_{1111} 
  \n & \qquad
  +4(s_{2\be} +(s_{2\be} +s_{4\be})c_{2(\al -\be)} 
    +(c_{2\be} +c_{4\be})s_{2(\al -\be)})
   c^T_{1112} 
  \n & \qquad
  + (c_{2(\al - \be)}(1 -c_{4\be}) +s_{2(\al -\be)} s_{4\be})
   (c^T_{1122} +c^T_{1221} +c^T_{1212})
  \n & \qquad
  +4(s_{2\be} +(s_{2\be} -s_{4\be})c_{2(\al -\be)} 
    +(c_{2\be} -c_{4\be})s_{2(\al -\be)})
   c^T_{2221} 
  \n & \qquad
  + (2 -2c_{2\be} +(1 -2c_{2\be} +c_{4\be})c_{2(\al -\be)} 
    +(2s_{2\be} -s_{4\be})s_{2(\al -\be)})
   c^T_{2222}
 \Bigr)
   \n & + \frac{s}{4f^2}(1-c_{2(\al -\be)}) \Bigl( 
   3(c^T_{1221} + c^T_{1212}) 
  - (c^H_{1221} -c^H_{1212}) 
  \Bigr), \\
 \mcl{M}(Z_L Z_L \to Z_L Z_L) =& 0,\\
 \mcl{M}(Z_L Z_L \to h Z_L) =& 0,
\end{align}
\begin{align}
 \mcl{M}(W^+_L Z_L \to W^+_L h) =& - i\frac{2s+t}{8f^2} \Bigl(
    ((3 +4c_{2\be} +c_{4\be})c_{\al -\be} -(2s_{2\be} +s_{4\be})s_{\al -\be})
	c^T_{1111} \n & \qquad
  +4((2s_{2\be} +s_{4\be})c_{\al -\be} +(c_{2\be} +c_{4\be})s_{\al -\be})
   c^T_{1112} \n & \qquad
  +2((1 -c_{4\be})c_{\al -\be} +s_{4\be}s_{\al -\be})
   (c^T_{1122} +c^T_{1221} +c^T_{1212}) \n & \qquad
  +4((2s_{2\be} -s_{4\be})c_{\al -\be} +(c_{2\be} -c_{4\be})s_{\al -\be})
   c^T_{2221} \n & \qquad
  + ((3 -4c_{2\be} +c_{4\be})c_{\al -\be} +(2s_{2\be} -s_{4\be})s_{\al -\be})
   c^T_{2222}
 \Bigr),\\
 \mcl{M}(W^+_L Z_L \to W^+_L Z_L) =& \frac{t}{8f^2} \Bigl(
    (3 +4c_{2\be} +c_{4\be})
	c^H_{1111}
  +4(2s_{2\be} +s_{4\be})
   c^H_{1112} \n & \quad
  +2(1 -c_{4\be})
   (c^H_{1122} +c^H_{1221} +c^H_{1212}) \n & \quad
  +4(2s_{2\be} -s_{4\be})
   c^H_{2221}
  + (3 -4c_{2\be} +c_{4\be}) c^H_{2222}
 \Bigr) \\
  =& \mcl{M}(W^+_L Z_L \to W^+_L Z_L)_{\text{cust}}, \\
 \mcl{M}(W^+_L W^+_L \to W^+_L W^+_L) =& -\frac{s}{8f^2} \Bigl(
    (3 +4c_{2\be} +c_{4\be}) 
   (c^H_{1111} +3c^T_{1111})
  +4(2s_{2\be} +s_{4\be}) 
   (c^H_{1112} +3c^T_{1112}) \n & \qquad
  +2(1 -c_{4\be})
   (c^H_{1122} +c^H_{1221}+c^H_{1212} +3(c^T_{1122} +c^T_{1221} +c^T_{1212}))
   \n & \qquad
  +4(2s_{2\be} -s_{4\be} )
   (c^H_{2221} +3c^T_{2221})
  + (3 -4c_{2\be} +c_{4\be}) 
   (c^H_{2222} +3c^T_{2222})
 \Bigr) \\
  =& \mcl{M}(W^+_L W^+_L \to W^+_L W^+_L)_{\text{cust}}
  \n &
  -\frac{3 s}{f^2} \Bigl(
    (3 +4c_{2\be} +c_{4\be}) 
   c^T_{1111}
  +4(2s_{2\be} +s_{4\be}) 
   c^T_{1112} \n & \qquad \quad
  +2(1 -c_{4\be})
   (c^T_{1122} +c^T_{1221} +c^T_{1212})
   \n & \qquad \quad
  +4(2s_{2\be} -s_{4\be} )
   c^T_{2221}
  + (3 -4c_{2\be} +c_{4\be}) 
   c^T_{2222}
 \Bigr).
\end{align}

Secondly, in the following amplitudes, 
one of the emitted particle is a heavy Higgs boson, 
$X_1 \in \{ W_L^\pm, Z_L, h \}$ and
$X_2 \in \{ H^\pm, A, H \}$:
\begin{align}
 \mcl{M}(W^+_L W^-_L \to W^+_L H^-) =& \frac{s+t}{8f^2} \Bigl(
  - (2s_{2\be} +s_{4\be})
   (c^H_{1111} +3c^T_{1111}) 
  +4(c_{2\be} +c_{4\be})
   (c^H_{1112} +3c^T_{1112}) \n & \qquad
  +2s_{4\be}
   (c^H_{1122} +c^H_{1221} +c^H_{1212} +3(c^T_{1122} +c^T_{1221} +c^T_{1212}))
   \n & \qquad
  +4(c_{2\be} -c_{4\be})
   (c^H_{2221} +3c^T_{2221}) 
  + (2s_{2\be} -s_{4\be})
   (c^H_{2222} +3c^T_{2222})
 \Bigr) \\
  =& \mcl{M}(W^+_L W^-_L \to W^+_L H^-)_{\text{cust}}
  \n &
  + \frac{3(s+t)}{8f^2} \Bigl(
  - (2s_{2\be} +s_{4\be})
   c^T_{1111} 
  +4(c_{2\be} +c_{4\be})
   c^T_{1112} \n & \qquad \quad
  +2s_{4\be}
   (c^T_{1122} +c^T_{1221} +c^T_{1212})
   \n & \qquad \quad
  +4(c_{2\be} -c_{4\be})
   c^T_{2221} 
  + (2s_{2\be} -s_{4\be})
   c^T_{2222}
 \Bigr), \\
 \mcl{M}(W^+_L W^-_L \to h H) =& \frac{s}{8f^2} \Bigl(
  - ((2s_{2\be} +s_{4\be})c_{2(\al -\be)} 
    +(1 +2c_{2\be} +c_{4\be})s_{2(\al -\be)})
   c^H_{1111} \n & \quad
  +4((c_{2\be} +c_{4\be})c_{2(\al -\be)} -(s_{2\be} +s_{4\be})s_{2(\al -\be)})
   c^H_{1112} \n & \quad
  +2(s_{4\be}c_{2(\al -\be)} +(1 +c_{4\be})s_{2(\al -\be)})
   c^H_{1122} \n & \quad
  +2(s_{4\be}c_{2(\al -\be)} -(1 -c_{4\be})s_{2(\al -\be)})
   (c^H_{1221} +c^H_{1212}) \n & \quad
  +4((c_{2\be} -c_{4\be})c_{2(\al -\be)} -(s_{2\be} -s_{4\be})s_{2(\al -\be)}) 
   c^H_{2221} \n & \quad
  + ((2s_{2\be} -s_{4\be})c_{2(\al -\be)} 
    -(1 -2c_{2\be} +c_{4\be})s_{2(\al -\be)})
   c^H_{2222}
 \Bigr) \\
  =& \mcl{M}(W^+_L W^-_L \to h H)_{\text{cust}}, \\
 \mcl{M}(W^+_L W^-_L \to h A) =& i\frac{s+2t}{8f^2} \Bigl(
  - ((2s_{2\be} +s_{4\be})c_{\al -\be} + (1 -c_{4\be})s_{\al -\be})
   c^T_{1111} \n & \qquad
  +4((c_{2\be} +c_{4\be})c_{\al -\be} -s_{4\be}s_{\al -\be})
   c^T_{1112} \n & \qquad
  +2(s_{4\be}c_{\al -\be} +(3 +c_{4\be})s_{\al -\be})
   c^T_{1122} \n & \qquad
  +2(s_{4\be}c_{\al -\be} -(1 -c_{4\be})s_{\al -\be})
   (c^T_{1221} +c^T_{1212}) \n & \qquad
  +4((c_{2\be} -c_{4\be})c_{\al -\be} +s_{4\be}s_{\al -\be})
   c^T_{2221} \n & \qquad
  + ((2s_{2\be} -s_{4\be})c_{\al -\be} +(1 -c_{4\be})s_{\al -\be})
   c^T_{2222} 
 \Bigr) \\
  =& \mcl{M}(W^+_L W^-_L \to h A)_{\text{cust}} 
  \n & 
  + i\frac{s+2t}{8f^2} \Bigl(
  - ((2s_{2\be} +s_{4\be})c_{\al -\be} + (1 -c_{4\be})s_{\al -\be})
   c^T_{1111} \n & \qquad
  +4((c_{2\be} +c_{4\be})c_{\al -\be} -s_{4\be}s_{\al -\be})
   c^T_{1112} \n & \qquad
  +2 ( s_{4\be} c_{\al - \be} + (3 +c_{4\be}) s_{\al - \be} ) 
   (c^T_{1122} + c^T_{1221} +c^T_{1212}) \n & \qquad
  +4((c_{2\be} -c_{4\be})c_{\al -\be} +s_{4\be}s_{\al -\be})
   c^T_{2221} \n & \qquad
  + ((2s_{2\be} -s_{4\be})c_{\al -\be} +(1 -c_{4\be})s_{\al -\be})
   c^T_{2222} 
 \Bigr)
  \n & - i \frac{s +2t}{3 f^2} s_{\al -\be} \Bigl(
   3(c^T_{1221} +c^T_{1212}) -(c^H_{1221} - c^H_{1212})
   \Bigr), 
\end{align}
\begin{align}
 \mcl{M}(W^+_L W^-_L \to Z_L H ) =& i\frac{s+2t}{8f^2} \Bigl(
  ((2s_{2\be} +s_{4\be})c_{\al -\be} +(3 +4c_{2\be} +c_{4\be})s_{\al -\be})
   c^T_{1111} \n & \qquad
  -4((c_{2\be} +c_{4\be})c_{\al -\be} -(2s_{2\be} +s_{4\be})s_{\al -\be})
   c^T_{1112} \n & \qquad
  -2(s_{4\be}c_{\al -\be} -(1 -c_{4\be})s_{\al -\be})
   (c^T_{1122} +c^T_{1221} +c^T_{1212}) \n & \qquad
  -4((c_{2\be} -c_{4\be})c_{\al -\be} -(2s_{2\be} -s_{4\be})s_{\al -\be})
   c^T_{2221} \n & \qquad
  - ((2s_{2\be} -s_{4\be})c_{\al -\be} -(3 -4c_{2\be} +c_{4\be})s_{\al -\be})
   c^T_{2222}
 \Bigr),\\
 \mcl{M}(W^+_L W^-_L \to Z_L A ) =& \frac{s}{8f^2} \Bigl(
  - (2s_{2\be} +s_{4\be})
   c^H_{1111}
  +4(c_{2\be} +c_{4\be})
   c^H_{1112} \n & \quad
  +2s_{4\be}
   (c^H_{1122} +c^H_{1221} +c^H_{1212}) \n & \quad
  +4(c_{2\be} -c_{4\be})
   c^H_{2221}
  + (2s_{2\be} -s_{4\be})
   c^H_{2222}
 \Bigr) \\
 =& \mcl{M}(W^+_L W^-_L \to Z_L A )_{\text{cust}}, 
\end{align}
\begin{align}
 \mcl{M}(Z_L Z_L \to W_L^+ H^-) =& \frac{s}{8f^2} \Bigl(
  - (2s_{2\be} +s_{4\be})
   c^H_{1111}
  +4(c_{2\be} +c_{4\be})
   c^H_{1112} \n & \quad
  +2s_{4\be}
   (c^H_{1122} +c^H_{1221} +c^H_{1212}) \n & \quad
  +4(c_{2\be} -c_{4\be})
   c^H_{2221}
  + (2s_{2\be} -s_{4\be})
   c^H_{2222}
 \Bigr) \\
 =& \mcl{M}(Z_L Z_L \to W_L^+ H^-)_{\text{cust}}, \\
 \mcl{M}(Z_L Z_L \to h H) =& \frac{s}{8f^2} \Bigl(
  - ((2s_{2\be} +s_{4\be})c_{2(\al -\be)} 
    +(1 +2c_{2\be} +c_{4\be})s_{2(\al -\be)})
   (c^H_{1111} +3c^T_{1111}) \n & \quad
  +4((c_{2\be} +c_{4\be})c_{2(\al -\be)} -(s_{2\be} +s_{4\be})s_{2(\al -\be)})
   (c^H_{1112} +3c^T_{1112}) \n & \quad
  + (2s_{4\be}c_{2(\al -\be)} -(1 -2c_{4\be})s_{2(\al -\be)})
   (c^H_{1122} +c^H_{1221} +3(c^T_{1122} +c^T_{1221})) \n & \quad
  +3s_{2(\al -\be)}
   (c^H_{1122} -c^H_{1221} -c^T_{1122} +c^T_{1221}) \n & \quad
  +2(s_{4\be}c_{2(\al -\be)} +c_{4\be}s_{2(\al -\be)})
   (c^H_{1212} +3c^T_{1212}) \n & \quad
  +4((c_{2\be} -c_{4\be})c_{2(\al -\be)} -(s_{2\be} -s_{4\be})s_{2(\al -\be)})
   (c^H_{2221} +3c^T_{2221}) \n & \quad
  + ((2s_{2\be} -s_{4\be})c_{2(\al -\be)} 
    -(1 -2c_{2\be} +c_{4\be})s_{2(\al -\be)})
  (c^H_{2222} +3c^T_{2222})
 \Bigr) \\
 =& \mcl{M}(Z_L Z_L \to h H)_{\text{cust}}
 \n & 
 + \frac{3s}{8f^2} \Bigl(
  - ((2s_{2\be} +s_{4\be})c_{2(\al -\be)} 
    +(1 +2c_{2\be} +c_{4\be})s_{2(\al -\be)})
   c^T_{1111} \n & \qquad
  +4((c_{2\be} +c_{4\be})c_{2(\al -\be)} -(s_{2\be} +s_{4\be})s_{2(\al -\be)})
   c^T_{1112} \n & \qquad
  + ((-1 + c_{4\be})s_{2(\al -\be)} +s_{4\be} c_{2(\al -\be)})
   (c^T_{1122} + c^T_{1221} + c^T_{1212}) \n & \qquad
  +4((c_{2\be} -c_{4\be})c_{2(\al -\be)} -(s_{2\be} -s_{4\be})s_{2(\al -\be)})
   c^T_{2221} \n & \qquad
  + ((2s_{2\be} -s_{4\be})c_{2(\al -\be)} 
    -(1 -2c_{2\be} +c_{4\be})s_{2(\al -\be)})
   c^T_{2222}
 \Bigr) 
  \n & + \frac{s}{4 f^2} s_{2(\al -\be)} \Bigl(
  3(c^T_{1221} +c^T_{1212}) - (c^H_{1221} - c^H_{1212})
  \Bigr), \\
 \mcl{M}(Z_L Z_L \to h A) =& 0,\\
 \mcl{M}(Z_L Z_L \to Z_L H) =& 0,\\
 \mcl{M}(Z_L Z_L \to Z_L A) =& 0.
\end{align}
\begin{align}
 \mcl{M}(W^+_L Z_L \to H^+ h) =& i\frac{t}{2f^2}
   s_{\al -\be}
  (-c^H_{1221} +c^H_{1212}) \n &
 +i\frac{2s+t}{8f^2} \Bigl(
    ((2s_{2\be} +s_{4\be})c_{\al -\be} -(1 -c_{4\be})s_{\al -\be})
	c^T_{1111} \n & \qquad
  -4((c_{2\be} +c_{4\be})c_{\al -\be} -s_{4\be}s_{\al -\be})
   c^T_{1112} \n & \qquad
  -2(s_{4\be}c_{\al -\be} -(1 -c_{4\be})s_{\al -\be})
   c^T_{1122} \n & \qquad
  -2(s_{4\be}c_{\al -\be} +(1 +c_{4\be})s_{\al -\be})
   (c^T_{1221} +c^T_{1212}) \n & \qquad
  -4((c_{2\be} -c_{4\be})c_{\al -\be} +s_{4\be}s_{\al -\be})
   c^T_{2221} \n & \qquad
  - ((2s_{2\be} -s_{4\be})c_{\al -\be} +(1 -c_{4\be})s_{\al -\be})
   c^T_{2222}
 \Bigr) \\
  =& \mcl{M}(W^+_L Z_L \to H^+ h)_{\text{cust}} 
  \n & 
  +i\frac{2s+t}{8f^2} \Bigl(
    ((2s_{2\be} +s_{4\be})c_{\al -\be} -(1 -c_{4\be})s_{\al -\be})
	c^T_{1111} \n & \qquad
  -4((c_{2\be} +c_{4\be})c_{\al -\be} -s_{4\be}s_{\al -\be})
   c^T_{1112} \n & \qquad
  -2(s_{4\be}c_{\al -\be} -(1 -c_{4\be})s_{\al -\be})
   (c^T_{1122} +c^T_{1221} +c^T_{1212})  \n & \qquad
  -4((c_{2\be} -c_{4\be})c_{\al -\be} +s_{4\be}s_{\al -\be})
   c^T_{2221} \n & \qquad
  - ((2s_{2\be} -s_{4\be})c_{\al -\be} +(1 -c_{4\be})s_{\al -\be})
   c^T_{2222}
 \Bigr) 
  \n & - i \frac{2s +t}{6 f^2} s_{\al -\be} \Bigl(
  3 (c^T_{1221} + c^T_{1212}) - (c^H_{1221} -c^H_{1212})
  \Bigr),  \\
 \mcl{M}(W^+_L Z_L \to W^+_L H) =& i\frac{2s+t}{8f^2} \Bigl(
  ((2s_{2\be} +s_{4\be})c_{\al -\be} +(3 +4c_{2\be} +c_{4\be})s_{\al -\be})
   c^T_{1111} \n & \qquad
  -4((c_{2\be} +c_{4\be})c_{\al -\be} -(2s_{2\be} +s_{4\be})s_{\al -\be})
   c^T_{1112} \n & \qquad
  -2(s_{4\be}c_{\al -\be} -(1 -c_{4\be})s_{\al -\be})
   (c^T_{1122} +c^T_{1221} +c^T_{1212}) \n & \qquad
  -4((c_{2\be} -c_{4\be})c_{\al -\be} -(2s_{2\be} -s_{4\be})s_{\al -\be})
   c^T_{2221} \n & \qquad
  - ((2s_{2\be} -s_{4\be})c_{\al -\be} -(3 -4c_{2\be} +c_{4\be})s_{\al -\be})
   c^T_{2222}
 \Bigr),\\
 \mcl{M}(W^+_L Z_L \to W^+_L A) =& \frac{t}{8f^2} \Bigl(
  - (2s_{2\be} +s_{4\be})
   c^H_{1111}
  +4(c_{2\be} +c_{4\be})
   c^H_{1112} \n & \quad
  +2s_{4\be}
   (c^H_{1122} +c^H_{1221} +c^H_{1212}) \n & \quad
  +4(c_{2\be} -c_{4\be})
   c^H_{2221}
  + (2s_{2\be} -s_{4\be})
   c^H_{2222}
 \Bigr) \\
  =& \mcl{M}(W^+_L Z_L \to W^+_L A)_{\text{cust}}, \\
 \mcl{M}(W^+_L Z_L \to H^+ Z_L) =& \frac{t}{8f^2} \Bigl(
  - (2s_{2\be} +s_{4\be})
   c^H_{1111}
  +4(c_{2\be} +c_{4\be})
   c^H_{1112} \n & \quad
  +2s_{4\be}
   (c^H_{1122} +c^H_{1221} +c^H_{1212}) \n & \quad
  +4(c_{2\be} -c_{4\be})
   c^H_{2221}
  + (2s_{2\be} -s_{4\be})
   c^H_{2222}
 \Bigr) \\
  =& \mcl{M}(W^+_L Z_L \to H^+ Z_L)_{\text{cust}}, 
\end{align}
\begin{align}
 \mcl{M}(W^+_L W^+_L \to W^+_L H^+) =& -\frac{s}{8f^2} \Bigl(
  - (2s_{2\be} +s_{4\be})
   (c^H_{1111} +3c^T_{1111})
  +4(c_{2\be} +c_{4\be})
   (c^H_{1112} +3c^T_{1112}) \n & \qquad
  +2s_{4\be}
   (c^H_{1122} +c^H_{1221} +c^H_{1212} +3(c^T_{1122} +c^T_{1221} +c^T_{1212}))
   \n & \qquad
  +4(c_{2\be} -c_{4\be})
   (c^H_{2221} +3c^T_{2221})
  + (2s_{2\be} -s_{4\be})
   (c^H_{2222} +3c^T_{2222})
 \Bigr) \\
 =& \mcl{M}(W^+_L W^+_L \to W^+_L H^+)_{\text{cust}} 
 \n &
 - \frac{3s}{8f^2} \Bigl(
  - (2s_{2\be} +s_{4\be})
   c^T_{1111}
  +4(c_{2\be} +c_{4\be})
   c^T_{1112} \n & \qquad \quad
  +2s_{4\be}
   (c^T_{1122} +c^T_{1221} +c^T_{1212})
   \n & \qquad \quad
  +4(c_{2\be} -c_{4\be})
   c^T_{2221}
  + (2s_{2\be} -s_{4\be})
   c^T_{2222}
 \Bigr). 
\end{align}

Finally, we show the amplitudes of double heavy Higgs boson productions 
i.e.~$X_1, X_2 \in \{ H^\pm, A, H\}$:
\begin{align}
 \mcl{M}(W^+_L W^-_L \to H^+ H^-) =& \frac{s+t}{8f^2} \Bigl(
    (1 -c_{4\be})
	(c^H_{1111} -2c^H_{1212} +c^H_{2222} 
	+3(c^T_{1111} -2c^T_{1212} +c^T_{2222})) \n & \qquad
  -4s_{4\be}
   (c^H_{1112} -c^H_{2221} +3(c^T_{1112} -c^T_{2221})) \n & \qquad
  +2(1 +c_{4\be})
   (c^H_{1122} +c^H_{1221} +3(c^T_{1122} +c^T_{1221}))
 \Bigr) \n &
 +\frac{s-t}{2f^2}
  (c^H_{1122} -c^H_{1221} -c^T_{1122} +c^T_{1221}) \\
  =& \mcl{M}(W^+_L W^-_L \to H^+ H^-)_{\text{cust}} 
  \n & 
  + \frac{s+t}{8f^2} \Bigl(
    3(1 -c_{4\be})
	(c^T_{1111} +c^T_{2222}) 
  -12s_{4\be}
   (c^T_{1112} -c^T_{2221}) \n & \qquad \quad
   + 2 (s +5t + 3(s+t)c_{4\be})(c^T_{1122} +c^T_{1221} +c^T_{1212})
 \Bigr), \\
 \mcl{M}(W^+_L W^-_L \to H H) =& \frac{s}{8f^2} \Bigl(
    (2(1 +c_{2\be}) -(1 +2c_{2\be} +c_{4\be})c_{2(\al -\be)} 
	 +(2s_{2\be} +s_{4\be})s_{2(\al -\be)})
	c^H_{1111} \n & \quad
  +4(s_{2\be} -(s_{2\be} +s_{4\be})c_{2(\al -\be)} 
    -(c_{2\be} +c_{4\be})s_{2(\al -\be)})
   c^H_{1112} \n & \quad
  +2(2 +(1 +c_{4\be})c_{2(\al -\be)} -s_{4\be}s_{2(\al -\be)})
   c^H_{1122} \n & \quad
  -2((1 -c_{4\be})c_{2(\al -\be)} +s_{4\be}s_{2(\al -\be)})
   (c^H_{1221} +c^H_{1212}) \n & \quad
  +4(s_{2\be} -(s_{2\be} -s_{4\be})c_{2(\al -\be)} 
    -(c_{2\be} -c_{4\be})s_{2(\al -\be)})
	c^H_{2221} \n & \quad
  + (2(1 -c_{2\be}) -(1 -2c_{2\be} +c_{4\be})c_{2(\al -\be)} 
	 -(2s_{2\be} -s_{4\be})s_{2(\al -\be)}) 
   c^H_{2222}
 \Bigr) \\
  =& \mcl{M}(W^+_L W^-_L \to H H)_{\text{cust}}, \\
 \mcl{M}(W^+_L W^-_L \to A A) =& \frac{s}{8f^2} \Bigl(
    (1 -c_{4\be})
	(c^H_{1111} -2c^H_{1221} -2c^H_{1212} +c^H_{2222}) \n & \quad
  -4s_{4\be}
   (c^H_{1112} -c^H_{2221})
  +2(3 +c_{4\be})
   c^H_{1122}
 \Bigr) \\
  =& \mcl{M}(W^+_L W^-_L \to A A)_{\text{cust}}, \\
 \mcl{M}(W^+_L W^-_L \to H A) =& i\frac{s+2t}{8f^2} \Bigl(
    ((1 -c_{4\be})c_{\al -\be} +(2s_{2\be} +s_{4\be})s_{\al -\be})
	 c^T_{1111} \n & \qquad
  -4(s_{4\be}c_{\al -\be} +(c_{2\be} +c_{4\be})s_{\al -\be})
   c^T_{1112} \n & \qquad
  +2((3 +c_{4\be})c_{\al -\be} -s_{4\be}s_{\al -\be})
   c^T_{1122} \n & \qquad
  -2((1 -c_{4\be})c_{\al -\be} +s_{4\be}s_{\al -\be})
   (c^T_{1221} +c^T_{1212}) \n & \qquad
  +4(s_{4\be}c_{\al -\be} -(c_{2\be} -c_{4\be})s_{\al -\be})
   c^T_{2221} \n & \qquad
  + ((1 -c_{4\be})c_{\al -\be} -(2s_{2\be} -s_{4\be})s_{\al -\be})
   c^T_{2222}
 \Bigr) \\
  =& \mcl{M}(W^+_L W^-_L \to H A)_{\text{cust}}
  \n & 
  + i\frac{s+2t}{8f^2} \Bigl(
    ((1 -c_{4\be})c_{\al -\be} +(2s_{2\be} +s_{4\be})s_{\al -\be})
	 c^T_{1111} \n & \qquad
  -4(s_{4\be}c_{\al -\be} +(c_{2\be} +c_{4\be})s_{\al -\be})
   c^T_{1112} \n & \qquad
  +2((3 +c_{4\be})c_{\al -\be} -s_{4\be}s_{\al -\be})
   ( c^T_{1122} + c^T_{1221} + c^T_{1212} )
   \n & \qquad
  +4(s_{4\be}c_{\al -\be} -(c_{2\be} -c_{4\be})s_{\al -\be})
   c^T_{2221} \n & \qquad
  + ((1 -c_{4\be})c_{\al -\be} -(2s_{2\be} -s_{4\be})s_{\al -\be})
   c^T_{2222}
 \Bigr)
  \n & -i \frac{s +2t}{3 f^2} c_{\al -\be} \Bigl(
  3 (c^T_{1221} + c^T_{1212}) - (c^H_{1221} - c^H_{1212})
  \Bigr),
\end{align}
\begin{align}
 \mcl{M}(Z_L Z_L \to H^+ H^- ) =& \frac{s}{8f^2} \Bigl(
    (1 -c_{4\be})
	(c^H_{1111} -2c^H_{1221} -2c^H_{1212} +c^H_{2222}) \n & \quad
  -4s_{4\be}
   (c^H_{1112} -c^H_{2221})
  +2(3 +c_{4\be})
   c^H_{1122}
 \Bigr) \\
  =& \mcl{M}(Z_L Z_L \to H^+ H^- )_{\text{cust}}, \\
 \mcl{M}(Z_L Z_L \to H H) =& \frac{s}{8f^2} \Bigl(
    (2(1 +c_{2\be}) -(1 +2c_{2\be} +c_{4\be})c_{2(\al -\be)} 
	 +(2s_{2\be} +s_{4\be})s_{2(\al -\be)})
	(c^H_{1111} +3c^T_{1111}) \n & \quad
  +4(s_{2\be} -(s_{2\be} +s_{4\be})c_{2(\al -\be)} 
    -(c_{2\be} +c_{4\be})s_{2(\al -\be)})
   (c^H_{1112} +3c^T_{1112}) \n & \quad
  + (1 -(1 -2c_{4\be})c_{2(\al -\be)} -2s_{4\be}s_{2(\al -\be)})
   (c^H_{1122} +c^H_{1221} +3(c^T_{1122} +c^T_{1221})) \n & \quad
  +3(1 +c_{2(\al -\be)})
   (c^H_{1122} -c^H_{1221} -c^T_{1122} +c^T_{1221}) \n & \quad
  +2(1 +c_{4\be}c_{2(\al -\be)} -s_{4\be}s_{2(\al -\be)})
   (c^H_{1212} +3c^T_{1212}) \n & \quad
  +4(s_{2\be} -(s_{2\be} -s_{4\be})c_{2(\al -\be)} 
    -(c_{2\be} -c_{4\be})s_{2(\al -\be)})
   (c^H_{2221} +3c^T_{2221}) \n & \quad
  + (2(1 -c_{2\be}) -(1 -2c_{2\be} +c_{4\be})c_{2(\al -\be)} 
	  -(2s_{2\be} -s_{4\be})s_{2(\al -\be)})
   (c^H_{2222} +3c^T_{2222})
 \Bigr) \\
  =& \mcl{M}(Z_L Z_L \to H H)_{\text{cust}} 
  \n & 
  + \frac{3s}{8f^2} \Bigl(
    (2(1 +c_{2\be}) -(1 +2c_{2\be} +c_{4\be})c_{2(\al -\be)} 
	 +(2s_{2\be} +s_{4\be})s_{2(\al -\be)})
	c^T_{1111} \n & \quad
  +4(s_{2\be} -(s_{2\be} +s_{4\be})c_{2(\al -\be)} 
    -(c_{2\be} +c_{4\be})s_{2(\al -\be)})
   c^T_{1112} \n & \quad
  - 2 ((1- c_{4\be}) c_{2(\al -\be)} + s_{4\be} s_{2(\al - \be)}) 
  (c^T_{1122} + c^T_{1221} + c^T_{1212}) \n & \quad
  +4(s_{2\be} -(s_{2\be} -s_{4\be})c_{2(\al -\be)} 
    -(c_{2\be} -c_{4\be})s_{2(\al -\be)})
   c^T_{2221} \n & \quad
  + (2(1 -c_{2\be}) -(1 -2c_{2\be} +c_{4\be})c_{2(\al -\be)} 
	  -(2s_{2\be} -s_{4\be})s_{2(\al -\be)})
   c^T_{2222}
 \Bigr)
  \n & + \frac{s}{4 f^2} (1 + c_{2(\al -\be)}) \Bigl(
  3 (c^T_{1221} + c^T_{1212}) - (c^H_{1221} -c^H_{1212})
  \Bigr), \\
 \mcl{M}(Z_L Z_L \to A A) =& \frac{s}{2f^2} \Bigl(
   2c^H_{1122} -c^H_{1221} +c^H_{1212} -3(c^T_{1221} -c^T_{1212})
 \Bigr) \\
  =& \mcl{M}(Z_L Z_L \to A A)_{\text{cust}}
  + \frac{s}{2f^2} \Bigl(
  3 (c^T_{1221} + c^T_{1212}) - (c^H_{1221} -c^H_{1212})
  \Bigr), \\
 \mcl{M}(Z_L Z_L \to H A) =& 0,
\end{align}
\begin{align}
 \mcl{M}(W^+_L Z_L \to H^+ H) =& i\frac{t}{2f^2}
   c_{\al -\be}
  (-c^H_{1221} +c^H_{1212}) \n &
 +i\frac{2s+t}{8f^2} \Bigl(
  - ((1 -c_{4\be})c_{\al -\be} +(2s_{2\be} +s_{4\be})s_{\al -\be})
   c^T_{1111} \n & \qquad
  +4(s_{4\be}c_{\al -\be} +(c_{2\be} +c_{4\be})s_{\al -\be})
   c^T_{1112} \n & \qquad
  +2((1 -c_{4\be})c_{\al -\be} +s_{4\be}s_{\al -\be})
   c^T_{1122} \n & \qquad
  -2((1 +c_{4\be})c_{\al -\be} -s_{4\be}s_{\al -\be})
   (c^T_{1221} +c^T_{1212}) \n & \qquad
  -4(s_{4\be}c_{\al -\be} -(c_{2\be} -c_{4\be})s_{\al -\be})
   c^T_{2221} \n & \qquad
  - ((1 -c_{4\be})c_{\al -\be} -(2s_{2\be} -s_{4\be})s_{\al -\be})
   c^T_{2222}
 \Bigr) \\
  =& \mcl{M}(W^+_L Z_L \to H^+ H)_{\text{cust}} 
  \n & 
 +i\frac{2s+t}{8f^2} \Bigl(
  - ((1 -c_{4\be})c_{\al -\be} +(2s_{2\be} +s_{4\be})s_{\al -\be})
   c^T_{1111} \n & \qquad
  +4(s_{4\be}c_{\al -\be} +(c_{2\be} +c_{4\be})s_{\al -\be})
   c^T_{1112} \n & \qquad
  +2((1 -c_{4\be})c_{\al -\be} +s_{4\be}s_{\al -\be})
   (c^T_{1122} +c^T_{1221} + c^T_{1212} ) \n & \qquad
  -4(s_{4\be}c_{\al -\be} -(c_{2\be} -c_{4\be})s_{\al -\be})
   c^T_{2221} \n & \qquad
  - ((1 -c_{4\be})c_{\al -\be} -(2s_{2\be} -s_{4\be})s_{\al -\be})
   c^T_{2222}
 \Bigr)
  \n & + i \frac{s +2t}{3 f^2} c_{\al -\be} \Bigl(
  3 (c^T_{1221} + c^T_{1212}) - (c^H_{1221} -c^H_{1212})
  \Bigr), \\
 \mcl{M}(W^+_L Z_L \to H^+ A) =& \frac{t}{8f^2} \Bigl(
    (1 -c_{4\be})
   (c^H_{1111} -2c^H_{1122} +c^H_{2222})
  -4s_{4\be}
   (c^H_{1112} -c^H_{2221}) \n & \quad
  +2(1 +c_{4\be})
   (c^H_{1221} +c^H_{1212})
 \Bigr) \n &
 +\frac{2s+t}{2f^2} (
  c^T_{1221} -c^T_{1212}
 ) \\
  =& \mcl{M}(W^+_L Z_L \to H^+ A)_{\text{cust}}
  - \frac{2s+t}{6 f^2} \Bigl(
  3 (c^T_{1221} + c^T_{1212}) - (c^H_{1221} -c^H_{1212})
  \Bigr), \\
 \mcl{M}(W^+_L W^+_L \to H^+ H^+) =& \frac{s}{8f^2} \Bigl(
    (1 -c_{4\be})
	(-c^H_{1111} +2c^H_{1122} +2c^H_{1221} -c^H_{2222} 
   \n & \qquad \qquad \quad
	+3(-c^T_{1111} +2c^T_{1122} +2c^T_{1221} -c^T_{2222})) \n & \quad
  +4s_{4\be}
   (c^H_{1112} -c^H_{2221} +3(c^T_{1112} -c^T_{2221})) \n & \quad
  -2(3 +c_{4\be})
   (c^H_{1212} +3c^T_{1212})
 \Bigr) \\
  =& \mcl{M}(W^+_L W^+_L \to H^+ H^+)_{\text{cust}} 
  \n & 
  + \frac{3s}{8f^2} \Bigl(
    (1 -c_{4\be})
	(-c^T_{1111} -c^T_{2222}))
   +4s_{4\be}
    ((c^T_{1112} -c^T_{2221})) \n & \qquad
   +2 (1 - c_{4\be}) (c^T_{1122} + c^T_{1221} + c^T_{1212})
 \Bigr).
\end{align}

\section{Elimination of the $\mcl{O}^r$ and $\mcl{O}^{HT}$} 
\label{app:eom}

There are four kinds of operators,
$\mcl{O}^H$, $\mcl{O}^r$, $\mcl{O}^T$ and $\mcl{O}^{HT}$ 
in the dimension-six derivative interactions of the NHDM.
$\mcl{O}^H$, $\mcl{O}^r$, $\mcl{O}^T$ and $\mcl{O}^{HT}$.
The operators, $\mcl{O}^r$ and $\mcl{O}^{HT}$, can be eliminated 
by field redefinition.
We consider the following redefinition:
\begin{align}
 H_i \to H_i +\frac{a_{ijkl}}{f^2} H_l ( H_j^\dag H_k ),
\end{align}
where $a_{ijkl}$ are complex numbers.
By the field redefinition on the kinetic term, 
we introduce the following terms:
\begin{align}
 (\del_\mu H_i)^\dag (\del^\mu H_i) \to 
 (\del_\mu H_i)^\dag (\del^\mu H_i)
 +\left(
  \frac{a_{ijkl}}{f^2} (\del_\mu H_i)^\dag \del^\mu ( H_l ( H_j^\dag H_k ) )  
 +\hc \right)
 +{\cal O} \left( ( H/f)^4 \right),
\end{align}
where indices $i$, $j$, $k$ and $l$ are summed over 
the species of the Higgs doublets.
The second and the third terms can be written 
using $\mcl{O}^H$, $\mcl{O}^r$ and $\mcl{O}^{HT}$:
\begin{align}
 &\frac{a_{ijkl}}{f^2} \left(
    \del_\mu H_i^\dag \del^\mu H_l ( H_j^\dag H_k ) 
   +\del_\mu H_i^\dag H_l \del^\mu( H_j^\dag H_k ) \right)
	+\hc \\
 = &
  \frac{a_{ijkl}}{f^2} \left( 
    O^r_{jkil} +\frac{1}{2}\left( O^H_{jkil} -O^{HT}_{jkil} \right)
  \right) 
  +\frac{a_{ijkl}^\ast}{f^2} \left( 
    O^r_{kjli} +\frac{1}{2}\left( O^H_{kjli} +O^{HT}_{kjli} \right)
  \right) \\ 
 = &
   \frac{ a_{kijl} +a_{ljik}^\ast}{ f^2} O^r_{ijkl}
  +\frac{ a_{kijl} +a_{ljik}^\ast}{2f^2} O^H_{ijkl}
  +\frac{-a_{kijl} +a_{ljik}^\ast}{2f^2} O^{HT}_{ijkl}.
\end{align}
Accordingly,
we can choose the conditions so as to eliminate $\mcl{O}^r$ and $\mcl{O}^{HT}$:
\begin{align}
  a_{kijl} +a_{ljik}^\ast =& -\la^r_{ijkl}, \n
 -a_{kijl} +a_{ljik}^\ast =& -\la^{HT}_{ijkl}.
\end{align}
It is clear that the elimination of $\mcl{O}^r$ affects the coefficients 
of $\mcl{O}^H$. 
On the other hand,
no contribution of $\mcl{O}^{HT}$ arises in the derivative interactions 
by the prescription.

The same result can be derived using the equation of motion.
We can rewrite $O^r_{ijkl}$ and $O^{HT}_{ijkl}$ as
\begin{align}
 O^r_{ijkl} =&
   \frac{1}{2} \del_\mu (H_i^\dag H_j \del^\mu (H_k^\dag H_l))
  -\frac{1}{2} O^H_{ijkl} 
  -\frac{1}{2} \left(
    H_i^\dag H_j ( H_k^\dag \del^2 H_l +(\del^2 H_k)^\dag H_l )
  \right),\\
 O^{HT}_{ijkl} =& 
  -H_i^\dag H_j ( H_k^\dag \del^2 H_l +(\del^2 H_k)^\dag H_l ).
\end{align}
Ignoring total derivative terms and using the equation of motion,
$\mcl{O}^r$ can be written in terms of $\mcl{O}^H$, 
whereas $\mcl{O}^{HT}$ does not contribute to the derivative interactions.
\section{Cross sections of the central region}
\label{SecCentral}
The cross section limited in the region of $-1/2 < \cos \th <1/2$ is
\begin{align}
 \si_{1/2} = \frac{s}{32 \pi f^4} \left(
   \frac{(2C_s -C_t)^2}{4} +\frac{C_t^2}{48}
 \right)
\end{align}
for the amplitude given by Eq.~\eqref{EqAmp}.
If same particles are included in the final state,
the above cross section should be divided by two.

With the formula, cross sections are 
\begin{align}
  \si_{1/2} (W^+_L W^-_L \to W^+_L W^-_L)_{\text{cust}} =&
    \frac{s}{32\pi f^4} \frac{13}{48} C_1 (\be)^2 \n 
  =& 
    \frac{13}{32} \si (W^+_L W^-_L \to W^+_L W^-_L)_{\text{cust}} ,\\
  \si_{1/2} (W^+_L W^-_L \to h h)_{\text{cust}} =&
    \frac{s}{32\pi f^4} \frac{1}{2} C_2 (\al, \be)^2 \n
  =& 
    \frac{1}{2} \si (W^+_L W^-_L \to h h)_{\text{cust}} , \\
  \si_{1/2} (W^+_L W^-_L \to Z_L Z_L)_{\text{cust}} =& 
    \frac{24}{13} \si_{1/2} (W^+_L W^-_L \to W^+_L W^-_L)_{\text{cust}}, \\
  \si_{1/2} (Z_L Z_L \to W^+_L W^-_L)_{\text{cust}} =&
    \frac{48}{13} \si_{1/2} (W^+_L W^-_L \to W^+_L W^-_L)_{\text{cust}}, \\
  \si_{1/2} (Z_L Z_L \to h h)_{\text{cust}} =&
    \si_{1/2} (W^+_L W^-_L \to h h)_{\text{cust}}, \\
  \si_{1/2} (W^+_L Z_L \to W^+_L Z_L)_{\text{cust}} =&  
    \si_{1/2} (W^+_L W^-_L \to W^+_L W^-_L)_{\text{cust}}, \\
  \si_{1/2} (W^+_L W^+_L \to W^+_L W^+_L)_{\text{cust}} =&  
    \frac{24}{13} \si_{1/2} (W^+_L W^-_L \to W^+_L W^-_L)_{\text{cust}},
\end{align}
\begin{align}
  \si_{1/2} (W^+_L W^-_L \to W^+_L H^-)_{\text{cust}} =&
    \frac{s}{32\pi f^4} \frac{13}{48} C_3 (\be)^2 \n
  =& 
    \frac{13}{32} \si (W^+_L W^-_L \to W^+_L H^-)_{\text{cust}},\\
  \si_{1/2} (W^+_L W^-_L \to h H )_{\text{cust}} =&
    \frac{s}{32\pi f^4} C_4 (\al, \be)^2 \n
  =& 
    \frac{1}{2} \si (W^+_L W^-_L \to h H )_{\text{cust}},\\
  \si_{1/2} (W^+_L W^-_L \to h A )_{\text{cust}} =&
    \frac{s}{32\pi f^4} \frac{1}{108} 
	 \sin^2 (\al -\be) (c^H_{1221} -c^H_{1212} )^2 \n
  =& 
    \frac{1}{8} \si (W^+_L W^-_L \to h A )_{\text{cust}},\\
  \si_{1/2} (W^+_L W^-_L \to Z_L A)_{\text{cust}} =& 
    \frac{48}{13} \si_{1/2} (W^+_L W^-_L \to W^+_L H^-)_{\text{cust}} ,\\
  \si_{1/2} (Z_L Z_L \to W^+_L H^-)_{\text{cust}} =&  
    \frac{48}{13} \si_{1/2} (W^+_L W^-_L \to W^+_L H^-)_{\text{cust}}, \\
  \si_{1/2} (Z_L Z_L \to h H)_{\text{cust}} =&  
    \si_{1/2} (W^+_L W^-_L \to h H)_{\text{cust}}, \\
  \si_{1/2} (W^+_L Z_L \to H^+ h)_{\text{cust}} =&  
    \si_{1/2} (W^+_L W^-_L \to h A)_{\text{cust}}, \\
  \si_{1/2} (W^+_L Z_L \to W^+_L A)_{\text{cust}} =&  
    \si_{1/2} (W^+_L W^-_L \to W^+_L H^-)_{\text{cust}}, \\
  \si_{1/2} (W^+_L Z_L \to H^+ Z_L)_{\text{cust}} =&  
    \si_{1/2} (W^+_L W^-_L \to W^+_L H^-)_{\text{cust}},  \\
  \si_{1/2} (W^+_L W^+_L \to W^+_L H^+)_{\text{cust}} =&  
    \frac{48}{13} \si_{1/2} (W^+_L W^-_L \to W^+_L H^-)_{\text{cust}}, 
\end{align}
\begin{align}
  \si_{1/2} (W^+_L W^-_L \to H^+ H^-)_{\text{cust}} =&
    \frac{s}{32\pi f^4} \frac{13}{48} \left( 
	   C_5(\be)^2 -\frac{22}{13} C_5(\be) (c^H_{1221} -c^H_{1122} ) 
	  +(c^H_{1221} -c^H_{1122} )^2 
	 \right), \\
  \si_{1/2} (W^+_L W^-_L \to H H)_{\text{cust}} =&
    \frac{s}{32\pi f^4} \frac{1}{2} C_6 (\al, \be)^2 \n
  =& 
    \frac{1}{2} \si_{1/2} (W^+_L W^-_L \to H H)_{\text{cust}},\\
  \si_{1/2} (W^+_L W^-_L \to A A)_{\text{cust}} =&
    \frac{s}{32\pi f^4} \frac{1}{2} C_5 (\be)^2 \n
  =& 
    \frac{1}{2} \si (W^+_L W^-_L \to A A)_{\text{cust}},\\
  \si_{1/2} (W^+_L W^-_L \to H A )_{\text{cust}} =&
    \frac{s}{32\pi f^4} \frac{1}{108} 
	 \cos^2 (\al -\be) (c^H_{1221} -c^H_{1212} )^2 \n
  =&
    \frac{1}{8} \si (W^+_L W^-_L \to H A )_{\text{cust}},\\
  \si_{1/2} (Z_L Z_L \to H^+ H^- )_{\text{cust}} =&
    2 \si_{1/2} (W^+_L W^-_L \to A A )_{\text{cust}}, \\
  \si_{1/2} (Z_L Z_L \to H H )_{\text{cust}} =&
    \si_{1/2} (W^+_L W^-_L \to H H )_{\text{cust}}, \\
  \si_{1/2} (Z_L Z_L \to A A )_{\text{cust}} =&
     \frac{s}{32\pi f^4} \frac{1}{2} (c^H_{1122} -3c^T_{1221})^2 \n
  =& 
    \frac{1}{2} \si (Z_L Z_L \to A A )_{\text{cust}},\\
  \si_{1/2} (W^+_L Z_L \to H^+ H )_{\text{cust}} =&
    \si_{1/2} (W^+_L W^-_L \to H A )_{\text{cust}}, \\ 
  \si_{1/2} (W^+_L Z_L \to H^+ A )_{\text{cust}} =&
     \frac{s}{32\pi f^4} \frac{1}{4} \biggl(
	    ( C_5 (\be) -c^H_{1122} +c^H_{1221} -3c^T_{1221} )^2 \n & \qquad \quad
		+\frac{1}{12} \left( 
		  C_5 (\be) +\frac{c^H_{1221} +2c^H_{1212}}{3} -c^H_{1122} +c^T_{1221}
		\right)^2
	  \biggr), \\
  \si_{1/2} (W^+_L W^+_L \to H^+ H^+ )_{\text{cust}} =&
    \si_{1/2} (W^+_L W^-_L \to A A )_{\text{cust}}.
\end{align}
\bibliographystyle{JHEP} 

\begin{thebibliography}{99}

\bibitem{Nisati:2011}
  A.~Nisati, for ATLAS collaboration,
  talk presented at Lepton-Photon Conference 2011,
  Mumbai, India, August 2011; 
  V.~Sharma, for CMS collaboration, 
  talk presented at Lepton-Photon Conference 2011,
  Mumbai, India, August 2011; 
  M.~Verzocchi, for Tevatron collaboration, 
  talk presented at Lepton-Photon Conference 2011,
  Mumbai, India, August 2011.

\bibitem{Giudice:2007fh}
  G.~F.~Giudice, C.~Grojean, A.~Pomarol, R.~Rattazzi,
  JHEP {\bf 0706}, 045 (2007).

\bibitem{ArkaniHamed:2001nc}
  N.~Arkani-Hamed, A.~G.~Cohen, H.~Georgi,
  Phys.\ Lett.\  {\bf B513}, 232-240 (2001);
  N.~Arkani-Hamed, A.~G.~Cohen, T.~Gregoire, J.~G.~Wacker,
  JHEP {\bf 0208}, 020 (2002).

\bibitem{Contino:2003ve}
  R.~Contino, Y.~Nomura, A.~Pomarol,
  Nucl.\ Phys.\  {\bf B671}, 148-174 (2003);
  K.~Agashe, R.~Contino, A.~Pomarol,
  Nucl.\ Phys.\  {\bf B719}, 165-187 (2005).

\bibitem{Low:2009di}
  I.~Low, R.~Rattazzi, A.~Vichi,
  JHEP {\bf 1004}, 126 (2010).

\bibitem{Coleman:1969sm}
  S.~R.~Coleman, J.~Wess, B.~Zumino,
  Phys.\ Rev.\  {\bf 177}, 2239-2247 (1969);  
  C.~G.~Callan, Jr., S.~R.~Coleman, J.~Wess, B.~Zumino,
  Phys.\ Rev.\  {\bf 177}, 2247-2250 (1969).

\bibitem{Chanowitz:1985hj}
  M.~S.~Chanowitz, M.~K.~Gaillard,
  Nucl.\ Phys.\  {\bf B261}, 379 (1985).

\bibitem{Han:2009em}
  T.~Han, D.~Krohn, L.~-T.~Wang, W.~Zhu,
  JHEP {\bf 1003}, 082 (2010).
\bibitem{Contino}
  R.~Contino, C.~Grojean, M.~Moretti, F.~Piccinini, R.~Rattazzi,
  JHEP {\bf 1005}, 089 (2010).
\bibitem{Grober}
  R.~Grober, M.~Muhlleitner,
  JHEP {\bf 1106}, 020 (2011).

\bibitem{Contino:2011np} 
  R.~Contino, D.~Marzocca, D.~Pappadopulo and R.~Rattazzi,
  JHEP {\bf 1110}, 081 (2011).

\bibitem{ArkaniHamed:2002qx}
  N.~Arkani-Hamed, A.~G.~Cohen, E.~Katz, A.~E.~Nelson, T.~Gregoire, J.~G.~Wacker,
  JHEP {\bf 0208}, 021 (2002);
  D.~E.~Kaplan, M.~Schmaltz,
  JHEP {\bf 0310}, 039 (2003);
  W.~Skiba, J.~Terning,
  Phys.\ Rev.\  {\bf D68}, 075001 (2003);
  B.~Gripaios, A.~Pomarol, F.~Riva, J.~Serra,
  JHEP {\bf 0904}, 070 (2009);
  M.~Redi, B.~Gripaios,
  JHEP {\bf 1008}, 116 (2010);
  M.~Schmaltz, D.~Stolarski, J.~Thaler,
  JHEP {\bf 1009}, 018 (2010).

\bibitem{Mrazek:2011iu}
  J.~Mrazek, A.~Pomarol, R.~Rattazzi, M.~Redi, J.~Serra, A.~Wulzer,
  Nucl.\ Phys.\  {\bf B853}, 1-48 (2011).

\bibitem{Grzadkowski:2010dj} 
  B.~Grzadkowski, M.~Maniatis and J.~Wudka,
  JHEP {\bf 1111}, 030 (2011);
  C.~C.~Nishi,
  Phys.\ Rev.\  {\bf D83}, 095005 (2011).


 \bibitem{Gunion}
   J.~F.~Gunion, H.~E.~Haber, G.~L.~Kane and S.~Dawson, 
  (Addison-Wesley, Redwood City, CA, 1990). 

 \bibitem{Haber:1995}
  H.~E.~Haber,
  arXiv:hep-ph/9501320.

\end{thebibliography}

\end{document}